%% file: Draft.tex
\documentclass[onecolumn,preprint,preprintnumbers,nofootinbib,floats,12ptm,tightenlines]{revtex4}

\usepackage{amsmath,amssymb,color,mathrsfs,verbatim,epsfig, wasysym}
\usepackage[hyperfootnotes=false]{hyperref}
\usepackage{slashed}
\usepackage{multirow}
\usepackage{xspace}
\allowdisplaybreaks

\pdfoutput=1

\def\DeltaR{\Delta R}

\newcommand{\LHC}{LHC$_\text{14}$\xspace}
\newcommand{\HLLHC}{HL-LHC\xspace}
\newcommand{\FCC}{FCC$_\text{100}$\xspace}

\begin{document}

\baselineskip=18pt


\thispagestyle{empty}
\vspace{20pt}
\font\cmss=cmss10 \font\cmsss=cmss10 at 7pt

\begin{flushright}
\small CERN-PH-TH-2015-015
\end{flushright}

\hfill
\vspace{20pt}

\begin{center}
{\Large \textbf
{
Effective field theory analysis of double Higgs \\[0.15cm] production via gluon fusion
}}
\end{center}

\vspace{15pt}
\begin{center}
{ Aleksandr Azatov$^{\, a}$, Roberto Contino$^{\, a, b}$, Giuliano Panico$^{\, a, c}$ and
Minho Son$^{\, b}$}
\vspace{30pt}

\centerline{$^{a}$ {\small \it Theory Division, Physics Department, CERN, Geneva, Switzerland
}}
\vskip 3pt
\centerline{$^{b}$ {\small \it Institut de Th\' eorie des Ph\' enom\` enes Physiques, EPFL, Lausanne, Switzerland
}}
\vskip 3pt
\centerline{$^{c}$ {\small \it IFAE, Universitat Aut\` onoma de Barcelona, Bellaterra, Barcelona, Spain
}}

\end{center}

\vspace{20pt}
\begin{center}
\textbf{Abstract}
\end{center}
\vspace{5pt} {\small
We perform a detailed study of double Higgs production via gluon fusion in the Effective Field Theory (EFT) framework where effects from new  physics
are parametrized by local operators. 
Our analysis provides a perspective broader than the one followed in most of the previous analyses, where this process was merely considered as a way to extract the
Higgs trilinear coupling.
We focus on the $hh \to b\bar b\gamma\gamma$ channel
and perform a thorough simulation of signal and background at the $14\,$TeV LHC and a 
future $100\,$TeV proton-proton collider. 
We make use of invariant mass distributions to enhance the sensitivity on the EFT coefficients and give a first assessment of
 the impact of jet substructure techniques on the results.
The range of validity of the EFT description is estimated, as required to consistently exploit the high-energy range of distributions, 
pointing out the potential relevance of dimension-8  operators.
Our analysis contains
a few important improvements over previous studies and  identifies some inaccuracies 
there appearing in connection with the estimate of signal and background rates.
The estimated precision on the Higgs trilinear coupling that follows from our results is less optimistic than previously claimed in the literature. 
We find that a $\sim 30\%$ accuracy can be reached on the trilinear coupling at a future $100\,$TeV collider
with $3\,\text{ab}^{-1}$. Only an $O(1)$ determination seems instead possible at the LHC with the same amount of integrated luminosity.  
}

\vfill\eject
\noindent


\section{Introduction}
\input{Introduction}

\section{Effective parametrization and power counting}
\input{Parametrization}

\section{Double Higgs phenomenology at 14$\,$TeV and 100$\,$TeV}
\input{HHpheno}

\section{Analysis of $pp\to hh \to \gamma\gamma b\bar{b}$}
\input{Analysis}

\section{Sensitivity on the EFT coefficients}
\input{Results}

\label{sec:results}

\section{Comparison with previous works}
\input{comparison}
\section{Summary and outlook}
\input{Conclusions}

\acknowledgments
We would like to thank D.~Del Re, F.~Micheli and P.~Meridiani for their collaboration in the early stage of this work and for many  discussions.
We are indebted with F.~Maltoni, S.~Frixione and M.~Zaro
for their precious help with Madgraph5\_aMC$@$NLO and for important suggestions and discussions.
We would also like to acknowledge useful conversations with J\'er\'emie Quevillon,  Riccardo Rattazzi, Juan Rojo and Andrea Wulzer.
The work of R.C. was partly supported by the ERC Advanced Grant No. 267985 \textit{Electroweak Symmetry Breaking, Flavour and Dark Matter: 
One Solution for Three Mysteries (DaMeSyFla)}. 
The work of G.~P. was partially supported by the Spanish Ministerio de Econom\' ia y Competitividad (MINECO) under projects AYA2009-13936, AYA2012-39559, AYA2012-39620, FPA2012-39684, Consolider-Ingenio 2010 CSD2007-00060, and Centro de Excelencia Severo Ochoa SEV-2012-0234.

\appendix
\input{Appendix}

\bibliography{lit}

\end{document}

%% file: Introduction.tex
\label{sec:introduction}

Unveiling the dynamics at the origin of electroweak symmetry breaking (EWSB) is a primary goal of the experiments  performed
at the Large Hadron Collider (LHC) at CERN.
The discovery at Run1 of a new boson with mass $m_h \simeq 125\,$GeV and properties similar to those predicted for the Standard Model (SM) Higgs boson
has been a major leap forward in this direction~\cite{Aad:2012tfa,Chatrchyan:2012ufa}.
The many direct searches and precision measurements performed by the ATLAS and CMS collaborations all indicate the existence of
a gap between the electroweak~(EW) scale and the scale of New Physics (NP), unless the latter is very weakly coupled to the SM sector.
This justifies the use of Effective Field Theory (EFT) to give a low-energy  parametrization of NP effects in terms of a series of local operators.
Measuring the coefficients of these effective operators would give access to a wealth of information on the UV dynamics, most importantly whether it is strongly 
or weakly coupled. Higgs studies at Run1 have mostly focused on on-shell single-production and decay processes, thus probing the strength of the
underlying EWSB dynamics at the scale $Q = m_h$. 
They set limits on possible modifications of the Higgs couplings, whose naive expected size is of order $\delta c/c \sim (g_*^2/g_{SM}^2) m_h^2/m_*^2$,
where $m_*$ is the mass of the new states, $g_*$ is their coupling strength to the Higgs boson and $g_{SM}$ a SM coupling.
The improved performances in energy and luminosity which will characterize the LHC Run2,
on the other hand,  give the opportunity to directly probe the EWSB dynamics at much higher energies ($Q \sim E \gg m_h$)
through the study of $2\to 2$ scattering processes. For a typical scattering energy $E$,   effects from New Physics are expected to be of order $\sim (g_*^2/g_{SM}^2) E^2/m_*^2$,
hence enhanced by a factor $E^2/m_h^2$ compared to those entering on-shell Higgs processes.
Exploring higher energies thus gives access to potentially larger corrections, but at the same time poses the issue of assessing the validity of the EFT description.
Determining at which point this latter breaks down in fact requires adopting a power counting to estimate the size of the local operators in terms of the parameters
(masses and couplings) characterizing the UV dynamics. At the same time, the power counting puts the limits on the effective operators 
into perspective, and helps inferring how much theoretical space is being probed.

Double Higgs production via gluon fusion is one example of scattering process which can disclose key information
on the EWSB dynamics, its underlying symmetries and strength. It is the only process potentially observable at the LHC that can 
give access to the quartic couplings among two Higgs bosons and a pair of gluons or of top quarks, as well as to the
Higgs trilinear self-coupling. Previous studies of this reaction mostly focused on the extraction of the trilinear coupling in the context 
of the SM~\cite{Baur:2003gpa,Baur:2003gp,Dolan:2012rv,Papaefstathiou:2012qe,Baglio:2012np,Yao:2013ika,Barr:2013tda,Barger:2013jfa,deLima:2014dta}.
An analysis based on the wider EFT perspective, however, can give access to a much richer spectrum of information on the UV dynamics.
In this work we provide such analysis at the $14\,$TeV LHC and  a future $100\,$TeV proton-proton collider by focusing on the $hh\to b\bar b\gamma\gamma$ final state.
We perform a detailed Montecarlo (MC) simulation of signal and background, and use the kinematic distributions to maximize the sensitivity on different effective operators.
Previous studies in the EFT context appeared in Refs.~\cite{Contino:2012xk,Goertz:2014qta}, 
and a first analysis of the impact of New Physics on the kinematic distributions can be found in Ref.~\cite{Chen:2014xra}.

The paper is organized as follows. Section~\ref{sec:parametrization} contains a detailed discussion of the parametrization of double Higgs production within
the framework of EFT. The power counting of the relevant coefficients is reviewed and a few scenarios are examined where 
NP gives large modifications to the Higgs trilinear coupling whereas other couplings stay close to their SM values. 
The relevance of dimension-8 operators is also investigated and the validity of the EFT description is assessed. 
Section~\ref{sec:HHpheno} reviews the phenomenology of double Higgs production at the LHC and a future $100\,$TeV collider, giving
a first estimate of the sensitivity on the kinematic tail at large invariant masses of the Higgs pair.
Our Montecarlo analysis of the $b\bar b\gamma\gamma$ decay mode is illustrated in Section~\ref{sec:analysis}, while results on the 
sensitivity to the coefficients of the effective Lagrangian are collected in Section~\ref{sec:results}.
We compare with previous analyses of the $b\bar b\gamma\gamma$  final state in Section~\ref{sec:comparison}.
Section~\ref{sec:conclusion} puts our study into context by briefly discussing existing studies of other final states and gives an outlook on
possible future developments. We collect useful formulas and further details on our simulation of the $b\bar b\gamma\gamma$
background into the Appendices.

%% file: Parametrization.tex
\label{sec:parametrization}

Corrections due to the exchange of new heavy states can be conveniently parametrized by means of low-energy
effective Lagrangians. There are two formulations which are suited to the study of Higgs physics. The first one assumes
that the Higgs boson is part of a weak doublet, as in the SM, and that $SU(2)_L \times U(1)_Y$ is linearly realized at high energies.
It thus referred to as the ``linear'' Lagrangian. 
In the second, more general formulation, $SU(2)_L \times U(1)_Y$ is non-linearly realized, hence the name of
``non-linear'' Lagrangian, and the physical Higgs boson is a singlet of the custodial symmetry, not necessarily part of a weak doublet.
Both parametrizations have been reviewed in Ref.~\cite{Contino:2013kra}, to which we refer the reader for a more in-depth description.
The experimental data collected at the LHC during Run1 seem to indicate that the couplings of the newly discovered boson have values close 
to those predicted for the SM Higgs.  Although such experimental information is still preliminary and 
awaits the confirmation of Run2, it motivates the use of the linear Lagrangian for future studies.
Indeed, small deviations from the SM can be obtained if the Higgs boson belongs to a doublet,
provided the new states are much heavier than the weak scale. The non-linear formulation is still useful, however, in those cases where large deviations 
in the Higgs couplings are allowed. As we will see later in our analysis, this is especially true for double Higgs production, where additional couplings
not accessible via single Higgs processes can be extracted.

In the linear Lagrangian, the operators can be organized according to their dimension:
\begin{equation}
{\cal L}_{lin} = {\cal L}_{SM} + \Delta {\cal L}_6 + \Delta {\cal L}_8 + \dots
\end{equation}
The lowest-order terms coincide with the usual
SM Lagrangian ${\cal L}_{SM}$, whereas ${\cal L}_{n}$ contains the deformations due
to operators of dimension $n$, with $n > 4$.~\footnote{Here and in the following we assume that baryon and lepton numbers are conserved
by the NP dynamics and thus by the series of higher-dimensional operators. The only dimension-5 operator  invariant under
the SM gauge symmetry violates lepton number and can  be omitted.}
For our purposes it is sufficient to focus on the operators involving the Higgs boson.
The relevant ones in ${\cal L}_6$ are~\footnote{For simplicity we focus on CP-conserving operators,  CP-violating ones can be included in a straightforward way.
We omit the operator $| H^\dagger {\overleftrightarrow { D_\mu}} H |^2$ since it violates the custodial symmetry and  is strongly constrained
by LEP data. Its inclusion has no impact on our analysis.
}
\begin{equation}
\label{eq:linearL}
\begin{split}
\Delta {\cal L}_6  \supset
& \, \frac{\overline c_H}{2 v^2} \partial_\mu \big(H^\dagger H\big) \partial^\mu \big(H^\dagger H\big)
   + \frac{\overline c_u}{v^2} y_t \left(H^\dagger H\, \overline q_L H^c t_R + \textrm{h.c.}\right)  \\
& - \frac{\overline c_6}{v^2}\frac{m_h^2}{2 v^2} \big(H^\dagger H\big)^3 
   + \overline c_g \frac{g_s^2}{m_W^2} H^\dagger H G^a_{\mu\nu} G^{a\, \mu\nu}\, ,
\end{split}
\end{equation}
where $v = 1/(\sqrt{2} G_F)^{1/2} = 246\,$GeV and $m_h = 125\,$GeV is the physical Higgs mass.
In order to classify the various operators in our effective Lagrangian and estimate their coefficients we adopt the SILH 
power counting~\cite{Giudice:2007fh}. This is based on the assumption that the NP dynamics can be broadly characterized
by one mass scale $m_*$, at which new states appear, and by one coupling strength $g_*$. The latter in particular
describes the interactions  between the Higgs boson and the new states. When building higher-order operators starting 
from the SM Lagrangian, each extra insertion of the Higgs doublet is weighted by a factor $1/f \equiv g_*/m_*$, 
while each additional covariant derivative  is  suppressed by $m_*$.
If the Higgs is a composite Nambu-Goldstone (NG) boson~\cite{Kaplan:1983fs}, the operators $O_6$, $O_u$ and $O_g$ 
can  be generated only through some small explicit breaking of the global invariance since they violate 
the Higgs shift symmetry~\cite{Giudice:2007fh}.~\footnote{We 
denote as $O_i$ the operator in the linear Lagrangian whose coefficient  is $\bar c_i$.}
In this case the SILH estimate of the coefficients in Eq.(\ref{eq:linearL}) is:
\begin{equation}
\label{eq:estimatedim6}
\bar c_H , \bar c_u , \bar c_6 \sim  \frac{v^2}{f^2} \, , \qquad
\bar c_g \left( \frac{4\pi}{\alpha_2} \right) \sim  \frac{v^2}{f^2} \times \frac{\lambda^2}{g_*^2} \, ,
\end{equation}
where $\lambda$ is a weak spurion coupling suppressing $\bar c_g$, while the suppression of $\bar c_6$ and $\bar c_u$ is controlled respectively
by $(m_h^2/2v^2)$ and $y_t$.
In realistic models, the minimum amount of explicit breaking entering $\bar c_g$ is given by the top Yukawa coupling $y_t$, so that one expects 
$y_t \lesssim \lambda \lesssim g_*$. For example, one has $\lambda = y_t$ in models with fully
composite $t_R$, whereas partial compositeness of both $t_L$ and $t_R$ leads to $\lambda = \sqrt{g_* y_t}$~\cite{Giudice:2007fh}.

In the case of the non-linear Lagrangian, the terms relevant for our analysis are
\begin{equation}
\label{eq:nonlinearL}
{\cal L}_{non-lin} \supset - m_t \,\overline t t \left(c_t \frac{h}{v} + c_{2t} \frac{h^2}{v^2}\right)
- c_3 \frac{m_h^2}{2 v} \, h^3
+ \frac{g_s^2}{4 \pi^2}\left(c_g \frac{h}{v} + c_{2g} \frac{h^2}{2v^2}\right)
G^a_{\mu\nu} G^{a\, \mu\nu}\, ,
\end{equation}
where $h$ denotes the physical Higgs field (defined to have vanishing vacuum expectation value).
Compared to the linear Lagrangian, the couplings $c_i$ of Eq.~(\ref{eq:nonlinearL}) effectively resum all the corrections of order $(v^2/f^2)$.
Therefore, the non-linear Lagrangian only relies on the derivative expansion, in which higher-order terms are
suppressed by additional powers of $(E^2/m_*^2)$. The linear formulation further requires $(v/f) < 1$ for the expansion in powers
of the Higgs doublet to be under control.
When both parametrizations are valid, the coefficients of the two Lagrangians are related by the following simple formulas:
\begin{equation}
\label{eq:dictionary}
c_t = 1 - \frac{\overline c_H}{2} - \overline c_u\,, \quad
c_{2t} = -\frac{1}{2}\left( \bar c_H + 3\bar c_u \right) \,,
\quad  c_3 = 1  - \frac{3}{2}\overline c_H + \overline c_6\,,
\quad  c_g = c_{2g} = \overline c_g \left( \frac{4\pi}{\alpha_2} \right)  , 
\end{equation}
where $\alpha_2 \equiv g^2/4\pi$.
Notice in particular that the same operator $O_u$ which gives a (non-universal) modification of the top Yukawa coupling
also generates the new quartic interaction~$\bar tt hh$.

It is worth at this point to comment about the normalization of the operator $O_6$ in Eq.~(\ref{eq:linearL}).
This differs from the one of Ref.~\cite{Contino:2013kra}, where it was defined $\Delta {\cal L} \supset - \bar c_6'\, \lambda_4 (H^\dagger H)^3/v^2$
(here we introduce the symbol $\bar c_6'$ to distinguish it from $\bar c_6$  in Eq.~(\ref{eq:linearL})), with $\lambda_4$ denoting the coefficient of the $(H^\dagger H)^2$
operator in the SM Lagrangian. 
We find $c_3 = (1 + 5 \bar c_6'/2)/(1 + 3\bar c_6'/2)$, which should be compared with the relation between $c_3$ and $\bar c_6$
(valid at all orders in $\bar c_6$) appearing in Eq.~(\ref{eq:dictionary}).
For small values of $\bar c_6'$ one has $\lambda_4 \simeq (m_h^2/2v^2)$, hence $\bar c_6 = \bar c_6' + O(\bar c_6'^2)$,
so that the dependence of $c_3$ upon $\bar c_6'$ is the same as in Eq.~(\ref{eq:dictionary}) up to small corrections. 
The parametrization of  Eq.~(\ref{eq:linearL}) is thus more convenient than that of Ref.~\cite{Contino:2013kra} when $\bar c_6$ (or $\bar c_6'$) becomes large,
since the formula for the trilinear coupling $c_3$ remains linear in $\bar c_6$. 
This is relevant for this analysis, since we anticipate that the results of Section~\ref{sec:results} will constrain values of $\bar c_6$ larger than 1 at the LHC.

\subsection{Modified power counting for the Higgs trilinear coupling}
\label{sec:modifiedpowercounting}

The estimates of Eq.~(\ref{eq:estimatedim6}) shows that  assuming the SILH power counting the modification of the Higgs trilinear coupling is expected
to be of order $(v/f)^2$, i.e. of the same size as the shifts to other Higgs couplings. These latter however are already constrained from single-Higgs
measurements to be close to their SM value, in particular the coupling of the Higgs to two vector bosons.
It is thus interesting to ask whether there are scenarios, characterized by a power counting different from the SILH one, where the Higgs trilinear
coupling can get a large modification from NP effects while all the other Higgs couplings are close to their SM values, in agreement with the 
current LHC limits.
Interestingly the answer to this question is affirmative: it is possible to imagine at least a few scenarios where the largest NP effects are in the trilinear coupling.

A first possibility is a scenario where the Higgs is a generic bound state --i.e. not a Nambu-Goldstone boson--  of some new strong dynamics.
In this case no weak spurion suppression is required to generate $O_6$, and the naive estimate of $\bar c_6$ is enhanced by a factor $g_*^2/\bar \lambda_4$ compared
to the SILH case, where we conveniently defined $\bar\lambda_4 \equiv m_h^2/2v^2\simeq 0.13$. 
The Higgs trilinear coupling (in SM units) is thus expected to be of order $c_3 = 1 + O(g_*^2 v^4/f^2 m_h^2)$, and large modifications are possible 
for $g_*$ large even if $(v/f)^2 \ll 1$. For example, $(v/f)^2 =0.05$ gives $\bar c_H \sim 0.05$ and $\bar c_6 \sim 3.5 \, (g_*/3)^2$, corresponding to 
$c_3 \simeq 0.93 + 3.5 \, (g_*/3)^2$.
The price to pay for this enhancement, however, is the tuning which is now required to keep the Higgs mass light, 
since naively one would expect $m_h^2 \sim m_*^2$. This is in addition to the $O(v^2/f^2)$ tuning which must occur in the vacuum alignment
even for a NG boson Higgs. 
Notice that similarly to~$O_6$, other operators which previously required a breaking of the Goldstone symmetry to be generated
will now be unsuppressed. In particular, $\bar c_g$ and the coefficient of the operator $H^\dagger H B_{\mu\nu}B^{\mu\nu}$ get enhanced by a factor $(g_*^2/\lambda^2)$
compared to their SILH estimate, thus leading to  modifications of the Higgs couplings to gluons and  photons of order $v^2/f^2$ times their (loop-induced)
SM value. However, for $(v/f)^2\ll 1$  the current LHC constraints on these couplings can be easily satisfied.

It is possible to avoid the tuning of the Higgs mass while keeping an enhanced trilinear coupling by envisaging a different scenario. 
Consider a theory where a new strongly-interacting sector,
characterized by a mass scale~$m_*$ and a coupling strength $g_*$, couples to the SM sector only through the Higgs mass term portal: 
${\cal L}_{int} = \lambda H^\dagger H O$. The Higgs field is thus elementary, and couples to the composite operator $O$, made of  fields of the
strong sector, with coupling $\lambda$. This leads to the following estimates:
\begin{equation} \label{eq:estimatesforportal}
\Delta \lambda_4 \sim \frac{\lambda^2}{g_*^2}\, ,\qquad
\bar c_H \sim \frac{\lambda^2}{g_*^4} \, \frac{v^2}{f^2}\, ,  \qquad
\bar c_u \sim \frac{\lambda^2}{g_*^4}\, \frac{v^2}{f^2}\, \frac{y_t^2}{16\pi^2}\, , \qquad 
\bar c_6 \sim \frac{\lambda^3}{g_*^4\bar\lambda_4} \, \frac{v^2}{f^2}\, ,
\end{equation}
where $\Delta \lambda_4$ is the correction to the coefficient of the operator $(H^\dagger H)^2$ induced by the strong dynamics.
The contribution to $\bar c_u$ follows from a Higgs-dependent modification of the top quark kinetic term arising at the 1-loop level, and it is subleading
compared to $\bar c_H$. Similar effects are also generated for the lighter quarks proportional to their Yukawa coupling squared.
Corrections to Eq.~(\ref{eq:estimatesforportal}) of higher order in $\lambda$  are suppressed by powers of $(\lambda v^2/m_*^2)$.
An additional factor $g_*^2/16\pi^2$ should be included in the above estimates if they arise from diagrams involving a loop of strongly-coupled 
particles.~\footnote{This is what happens, for example, when the  strong dynamics consists of a single real scalar field $S$, neutral under the SM gauge group.
By requiring invariance under the parity $S\to -S$, the Lagrangian describing the strong dynamics is ${\cal L} = (\partial_\mu S)^2/2 - m_S^2 S^2/2 - \lambda_S S^4/4$,
and $O = S^2$. One can thus identify $m_* = m_S$ and $g_* = \sqrt{\lambda_S}$.}
Since $\bar c_6\sim (\lambda/\bar\lambda_4) \, \bar c_H $, one can have an enhanced NP correction to the Higgs trilinear coupling for $\lambda > \bar\lambda_4$.
However, requiring that no fine-tuning occurs in the coefficient of the quartic operator --i.e. in the physical Higgs mass-- 
implies $\lambda \lesssim \sqrt{\bar\lambda_4} \, g_*$. Furthermore, only for $(\lambda v^2/m_*^2) \lesssim 1$ one can make sense of the series in powers
of the Higgs field, since when this inequality is saturated all orders in $\lambda$ become equally important and perturbation theory is lost.
From the above considerations it follows the bound $\bar c_6 \lesssim O(1)$. In this scenario 
it is thus possible to avoid tuning the Higgs mass but the corrections to the Higgs trilinear are at most of $O(1)$; larger enhancements require fine-tuning.
For example, $g_* = 4\pi$, $m_* = 500\,$GeV and $\lambda = 4\pi \sqrt{\bar\lambda_4} \simeq 4.5$ gives $\bar c_H \sim 0.03$ and $\bar c_6 \sim 1$.

These examples show that it is possible, in scenarios characterized by a power counting different from the SILH one, to have large modifications
to the  trilinear coupling while keeping the other Higgs couplings close to their SM values.
In some cases, like that of a generic composite Higgs, $\bar c_6$ can be of order 1 or larger without invalidating the perturbative expansion in powers of the Higgs field.

\subsection{The relevance of higher-order operators}
\label{sec:higherorder}

Let us now investigate more quantitatively the validity of our effective Lagrangians and  discuss the importance of higher-order operators.
We will do so by adopting the SILH power counting.
We already mentioned that in the case of the linear Lagrangian, for a given process, operators with higher powers of the Higgs doublet
imply corrections suppressed by additional factors $(v/f)$.  This means that if $v/f \sim 1$ the linear Lagrangian cannot be consistently used anymore,
although we can still rely on the non-linear one. As regards the derivative expansion, the same considerations apply instead to both descriptions.
The perturbative parameter in that case is $(E/m_*)$, so that naively higher-order operators  become important at $E \sim m_*$, where the effective
description breaks down. However,  in certain processes it can happen that the leading operators are suppressed due to symmetry reasons, and the higher-order ones
become relevant at large energies below the  cutoff scale. This is in fact the case of double Higgs production via gluon fusion, where the dimension-6
operator $O_g$ is suppressed  by two powers of a weak spurion if the Higgs is a NG boson.

The dimension-8 operators relevant for double Higgs production via gluon fusion are:
\begin{equation} \label{eq:dim8L}
\begin{split}
\Delta {\cal L}_8 \supset \frac{g_s^2}{m_W^4} \Big[ 
& \bar c_{gD0} \, \big(D_\rho H^\dagger D^\rho H\big) G^a_{\mu\nu} G^{a\,\mu\nu} \\
& + \bar c_{gD2} \, \big(\eta^{\mu\nu} D_\rho H^\dagger D^\rho H - 4\, D^\mu H^\dagger D^\nu H\big)
G^a_{\mu\alpha} G_\nu^{a\,\alpha} \Big]\,.
\end{split}
\end{equation}
They mediate scatterings of two gluons into two Higgs bosons with total angular momentum equal to respectively 0 and 2, 
and can therefore lead to different kinematic configurations, as we will discuss later on.
Both operators, however, are similar from the power-counting viewpoint.
Neither of the two breaks the shift symmetry of a NGB Higgs, and thus no spurion suppression appears in the  estimate of their coefficients:~\footnote{This estimate 
assumes that $O_{gD0}$ and $O_{gD2}$ are generated at 1-loop by the New Physics. The same assumption has been done on $O_g$ in the estimate of 
Eq.~(\ref{eq:estimatedim6}).}
\begin{equation}
\label{eq:estimatedim8}
\bar c_{gD0,2} \left( \frac{4\pi}{\alpha_2} \right) \sim  \frac{v^2}{f^2} \times \frac{m_W^2}{m_*^2} \, .
\end{equation}

We can now compare the contributions of the various operators to double Higgs production via gluon fusion.~\footnote{R.C. would like to thank Riccardo Rattazzi
for illuminating discussions on the validity of the effective theory in scattering processes. For a related analysis of $WW$ scattering, see: R.~Rattazzi, talk 
at \textit{BSM physics opportunities at 100TeV}, CERN, February 2014.}
By using the estimates of Eqs.~(\ref{eq:estimatedim6}),(\ref{eq:estimatedim8}), the  scattering amplitude can be schematically expressed as
\begin{equation} \label{eq:amplitude}
A(gg \to hh) \sim \left(\frac{\alpha_s}{4\pi}\right) \times \!\left[\,  y_t^2 \left( 1 + O\!\left( \frac{v^2}{f^2} \right) \right) + g_6^2(E) + g_8^2(E) + \dots \right]\, ,
\end{equation}
where
\begin{equation} \label{eq:couplings}
g^2_6(E) \sim  \bar c_g \, \frac{4\pi}{\alpha_2} \, \frac{E^2}{v^2} \sim \frac{\lambda^2 E^2}{m_*^2}\, , 
\quad\qquad 
g^2_8(E)  \sim \bar c_{gD0,2} \, \frac{4\pi}{\alpha_2} \, \frac{E^4}{v^2 m_W^2} \sim \frac{g_*^2 E^4}{m_*^4}\, .
\end{equation}
The first term in the square parentheses of Eq.~(\ref{eq:amplitude}) corresponds to the SM contribution  plus the $O(v^2/f^2)$ correction implied by 
$O_H$ and $O_u$.~\footnote{Here we neglect the
logarithmic energy growth of the triangle diagram with the $\bar t thh$ quartic interaction generated by $O_u$. 
The correction implied by $O_6$ enters through the SM triangle diagram and is further suppressed at
high energies, see next section.}
The second and third terms in the parentheses denote  the contributions of respectively $O_g$ and  $O_{gD0}$, $O_{gD2}$,
and thus define the coupling strengths of the interactions mediated by these operators.
Since these couplings grow with $E$, 
at sufficiently large energies, yet below the cutoff scale, they can become larger than the top Yukawa coupling. 
It is thus possible to obtain NP corrections larger than the SM contribution within the validity of the effective Lagrangian.
This should be contrasted with the corrections from dimension-6 operators to the on-shell production and decay rates of the Higgs boson: those are at most
of order $O(v^2/f^2)$, hence always smaller than the SM term.
We thus see the potential advantage of studying $2\to 2$ scatterings at high energies compared to the single-Higgs measurements
performed during Run1 at the LHC: going off-shell at higher energies one can directly probe the strength of the New Physics dynamics in the regime
in which it gives large effects~\cite{Contino:2013gna}.
In practice, as we will discuss more in detail in the following, such regime is difficult to exploit in the process $gg \to hh \to b\bar b \gamma\gamma$ 
considered in this work. This is  due to the steep fall off of the gluon pdfs at large $x$ and to the small final-state branching ratio, which strongly limit 
the exploration of the region with large $hh$ invariant mass.
For $\lambda > \sqrt{g_* y_t}$ the coupling $g_6(E)$ is the first to become larger than $y_t$ at a scale $E \sim y_t f \, (g_*/\lambda)$. The smaller $\lambda$ is,
the higher the crossover scale becomes, until  $E \sim y_t f \, (g_*/y_t)^{1/2}$ is reached where $g_8(E) \sim y_t$. Above this energy, 
for $\lambda \lesssim \sqrt{g_* y_t}$, the largest effects 
come from the dimension-8 operators. The validity of the effective Lagrangian extends up to energies $E \sim m_*$, so that the
coupling strengths are bounded by 
\begin{equation} \label{eq:constraint1}
g_6(E) < \lambda \lesssim g_* \, , \qquad\quad g_8(E) < g_* \,.
\end{equation}

\begin{figure}[tp]
\begin{center}
\includegraphics[width=0.55\linewidth]{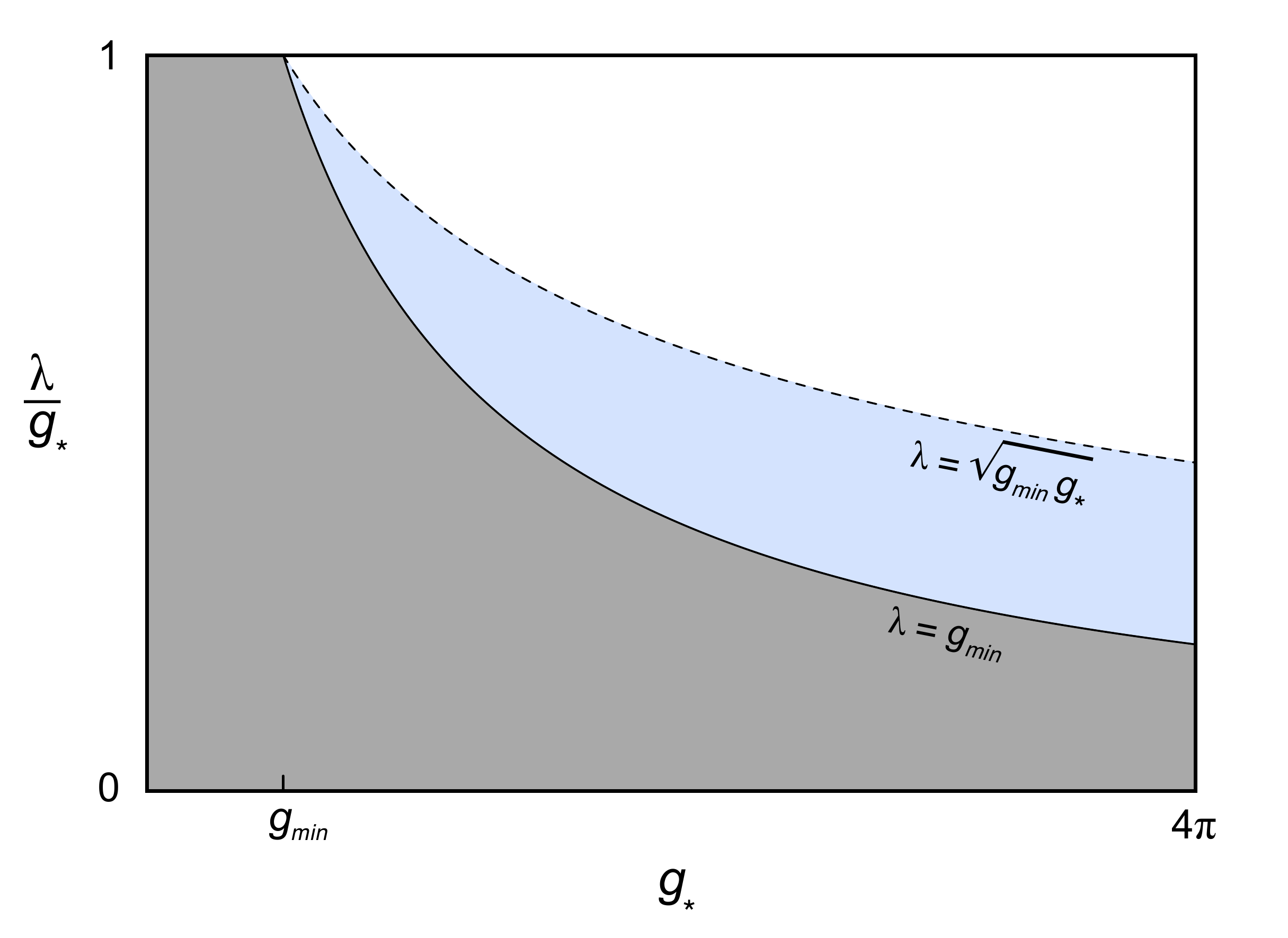}
\caption{Cartoon of the region in the plane $(g_*, \lambda/g_*)$, defined by Eqs.~(\ref{eq:ineq1}),(\ref{eq:ineq2}), that can be probed by an analysis including 
only dimension-6 operators (in white).
No sensible effective field theory description is possible in the gray area ($\lambda < g_{min}$), 
while exploration of the light blue region ($g_{min} < \lambda < \sqrt{g_* g_{min}}$) 
requires including the dimension-8 operators. 
}
\label{fig:cartoon}
\end{center}
\end{figure}

As we anticipated, studying the kinematic region where higher-order operators give a contribution larger than the SM one is challenging in double Higgs production
via gluon fusion. However, it is still interesting, and relevant for the analysis carried in this work, to ask at which scale the dimension-8 operators become
more important than the dimension-6 ones, independently of their absolute size.
From Eq.~(\ref{eq:couplings}) it follows that $g_6(E) \sim g_8(E)$ for $E \sim \lambda f$; at this scale $g_6(E) \sim \lambda^2/g_*$. Hence, a further condition to be 
satisfied in an analysis which includes only dimension-6 operators neglecting those with dimension 8 is:
\begin{equation} \label{eq:constraint2}
g_6(E) < \frac{\lambda^2}{g_*} \, .
\end{equation}
Let us then indicate with $\delta$ the precision obtained in such analysis on $c_{2g} = \bar c_g (4\pi/\alpha_2)$,  by making use of events 
with invariant masses up to $\bar E$.
This means that the smallest  value of $g_6$  probed is $g_{min} \sim \sqrt{\delta} \, (\bar E/v)$, so that 
Eqs.~(\ref{eq:constraint1}),(\ref{eq:constraint2})  can be re-cast into
the following constraints:
\begin{align} 
\label{eq:ineq1}
\lambda & > g_{min} \\[0.2cm]
\label{eq:ineq2}
\frac{\lambda^2}{g_*} & > g_{min}\, . 
\end{align}
These inequalities define the region in the $(g_*, \lambda)$ parameter space, sketched in Fig.~\ref{fig:cartoon}, which
can be sensibly probed, within the validity of the effective theory, by an analysis including only 
dimension-6 operators.
%
%
Clearly, smaller values of $g_{min}$, obtained by increasing the precision $\delta$ for a fixed energy $\bar E$, lead to a larger viable region.
Notice that the two cases $\lambda = y_t$ (fully composite $t_R$) and $\lambda = \sqrt{g_* y_t}$ (partially composite $t_L$ and $t_R$) can be probed only 
if $g_{min} < y_t$. This is not the case of course if the analysis does not have enough sensitivity to probe the SM cross section.

\subsection{Cross section of double Higgs production}
\label{sec:crosssection}

We can now discuss our parametrization of the cross section of double Higgs production via gluon fusion.
We will use the non-linear Lagrangian (\ref{eq:nonlinearL}) and start by neglecting higher-derivative terms (which correspond to dimension-8 operators
in the limit of linearly-realized EW symmetry).  The effect of the neglected derivative operators will be then studied by analyzing their impact on
angular differential distributions and shown to be small in our case due to the limited sensitivity on the high $m_{hh}$  region.

\begin{figure}[tp]
\begin{center}
\includegraphics[width=0.30\textwidth]{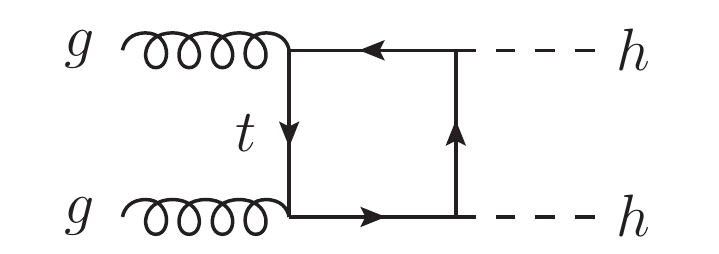}
\hspace{0.8em}
\includegraphics[width=0.335\textwidth]{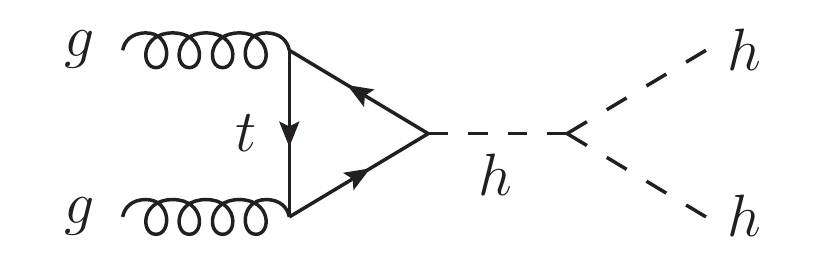}
\hspace{0.8em}
\includegraphics[width=0.28\textwidth]{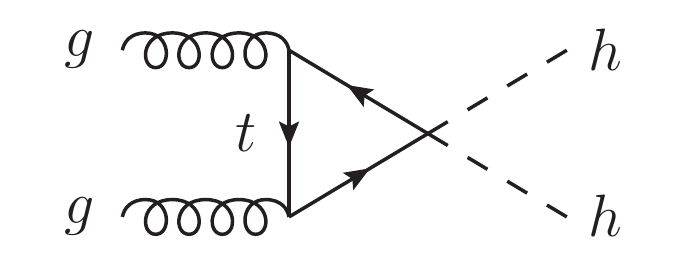}\\
\vspace{1.2em}
\includegraphics[width=0.30\textwidth]{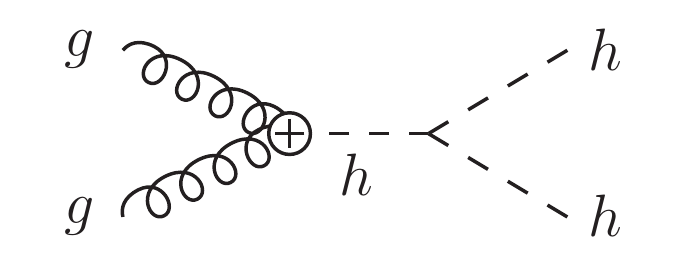}
\hspace{2.5em}
\includegraphics[width=0.233\textwidth]{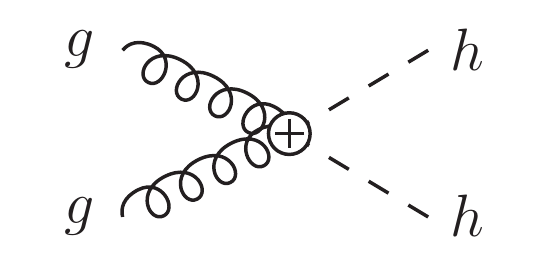}
\caption{Feyman diagrams contributing to double Higgs production via gluon fusion
(an additional contribution comes from the crossing of the box diagram).
The last diagram on the first line contains the $\bar t t hh$ coupling, while those in the second line
involve contact interactions between the Higgs and the gluons denoted with a cross.}
\label{fig:diagrams}
\end{center}
\end{figure}

The Feynman diagrams that contribute to the $gg \rightarrow hh$ process are shown in
Fig.~\ref{fig:diagrams}.
%
%
Each diagram is characterized by a different scaling at large energies $\sqrt{\hat s} = m_{hh} \gg m_t,\, m_h$. 
We find:
%
\begin{equation}
\begin{split}
{\cal A}_\square & \sim  c_t^2\, \frac{\alpha_s}{4\pi} \, y_t^2\,,\\[0.1cm]
{\cal A}_\triangle & \sim  c_t c_3\, \frac{\alpha_s}{4\pi} \, y_t^2 \,\frac{m_h^2}{\hat s}
\left(\log\frac{m_t^2}{\hat s} + i \pi\right)^2\,,\\[0.1cm]
{\cal A}_{\triangle nl} & \sim  c_{2t}\, \frac{\alpha_s}{4\pi} \, y_t^2
\left(\log\frac{m_t^2}{\hat s}+ i \pi\right)^2\,,\\[0.1cm]
{\cal A}_{3} & \sim  c_g c_3\, \frac{\alpha_s}{4\pi} \, \frac{m_h^2}{v^2}\,,\\[0.1cm]
{\cal A}_{4} & \sim c_{2g} \,\frac{\alpha_s}{4\pi} \, \frac{\hat s}{v^2}\,,
\end{split}
\end{equation}
where ${\cal A}_\square$, ${\cal A}_\triangle$ are the amplitudes of respectively the box and  triangle diagram
with Higgs exchange, ${\cal A}_{\triangle nl}$ is the amplitude of the triangle diagram
with the  $t\bar t h h$ interaction, and  ${\cal A}_3$ and ${\cal A}_4$
are the amplitudes of the diagrams with the Higgs-gluon contact interactions.
One can see that each NP contribution affects the $m_{hh}$ distribution
in a different way.  In particular, the diagrams that depend on the
Higgs trilinear coupling $c_3$ are always suppressed at large $\hat s$, and their contribution
affects the process mostly at threshold. Modified values of the top Yukawa coupling $c_t$ and the non-linear
interactions $c_{2t}$ and $c_{2g}$, instead, tend to increase the cross section at higher
invariant masses. Finally, including the dimension-8 operators would lead
to an additional contribution to ${\cal A}_4$ growing 
as $\hat s^2$ and distort the tail of the $m_{hh}$ distribution. 
A shape analysis can thus help to differentiate the different effects and break the degeneracy of the total cross section
on the Higgs couplings. This will be our strategy in the study of double Higgs production discussed in the next section,
where we will use $m_{hh}$ as the main kinematic variable to characterize signal events.

By focussing on gluon fusion, the  total cross section for the process $pp \to hh$ can be written 
as a simple polynomial of the parameters of the effective Lagrangian:~\footnote{The contribution from Vector Boson Fusion is smaller by at most one order
of magnitude, and can be isolated by selecting the number of jets in the final state. See for example Refs.~\cite{Dolan:2013rja,LesHouches} 
for up-to-date studies with $m_h = 125\,$GeV.}
\begin{equation}
\label{eq:tot_xsec}
\begin{split}
\sigma  = \, \sigma_{SM} \Big[ \, & A_1\, c_t^4 + A_2 \, c_{2t}^2  + A_3\,  c_t^2 c_3^2  + A_4 \, c_g^2 c_3^2  + A_5\,  c_{2g}^2  + A_6\, c_{2t} c_t^2 + A_7\,  c_t^3 c_3 \\[0.1cm]
& + A_8\,  c_{2t} c_t\, c_3  + A_9\, c_{2t} c_g c_3 + A_{10}\, c_{2t} c_{2g} + A_{11}\,  c_t^2 c_g c_3 + A_{12}\, c_t^2 c_{2g} \\[0.1cm]
& + A_{13}\, c_t c_3^2 c_g  + A_{14}\, c_t c_3 c_{2g} +A_{15}\, c_g c_3 c_{2g} \Big] \,.
\end{split}
\end{equation}
The LO value of the numerical coefficients $A_i$ and of the SM cross section $\sigma_{SM}$ is reported in Table~\ref{tab:tot_xsec_coefficients} for 
hadron colliders with center-of-mass energy  $\sqrt{s} = 14\,$TeV and $100\,$TeV.~\footnote{We computed the $A_i$'s by performing a corresponding
number of MC simulations, each with $\sim 10^6$ events. We estimate the statistical uncertainty on the $A_i$'s to be at the $10^{-2}$ level. This is less accurate
than one could naively deduce due to cancellations that occur when extracting some of the coefficients, but still much smaller than the theoretical error
from PDFs and higher-order QCD corrections (see footnote~\ref{foot:therror} and Ref.~\cite{Baglio:2012np}).
When we consider  different $m_{hh}$ categories later in the analysis, the fit is individually performed in each $m_{hh}$ bin through a similar procedure, 
thus maintaining the same level of statistical uncertainty.
}
They were computed with our dedicated {\small C++} code linked to QCDLoop~\cite{Ellis:2007qk} and to the LHAPDF routines~\cite{Whalley:2005nh}, by setting
$m_h = 125\,$GeV, $m_t = 173\,$GeV and using the CTEQ6ll parton distribution functions. The factorization and renormalization scales have been fixed to 
$Q = \sqrt{\hat s}= m_{hh} $.
%
\begin{table}
\centering
\small
\begin{tabular}{ccccccccc}
$\sqrt{s}$ & $ \sigma_{SM}\,[\text{fb}]$ & $A_1$ & $A_2$ & $A_3$ & $A_4$ & $A_5$ & $A_6$
& $A_7$ \\
\hline
\rule{0pt}{1.25em}
 $14\,\text{TeV}$ & $16.2$ & $2.13$ & $10.1$ & $0.300$ & $21.8$ & $188$ & $-8.62$ & $-1.43$ \\
\rule{0pt}{1.25em}
$100\,\text{TeV}$ & $874$ &  $1.95$ & $11.2$ & $0.229$ & $16.0$ & $386$ & $-8.32$ & $-1.18$ \\[0.4cm]
 $\sqrt{s}$ & $A_8$ & $A_9$ & $A_{10}$ & $A_{11}$ & $A_{12}$ & $A_{13}$ & $A_{14}$ & $A_{15}$\\
\hline
\rule{0pt}{1.25em}
 $14\,\text{TeV}$  & $2.93$
                             & $21.0$ & $59.8$ & $-9.93$ & $-23.1$ & $4.87$ & $10.5$ & $96.6$\\
\rule{0pt}{1.25em}
$100\,\text{TeV}$ & $2.55$
& $16.9$ & $52.4$ & $-7.49$ & $-17.3$ & $3.55$ & $8.46$ & $87.5$
\end{tabular}
\caption{Coefficients of the fit of the total $pp \rightarrow hh$ cross section via gluon fusion
given in Eq.~(\ref{eq:tot_xsec}). They have been computed at LO for the $14\ \mathrm{TeV}$ LHC
and for a future collider with $\sqrt{s} = 100\,$TeV.}
\label{tab:tot_xsec_coefficients}
\end{table}
%

\begin{figure}[tp]
\begin{center}
\includegraphics[width=0.47\linewidth]{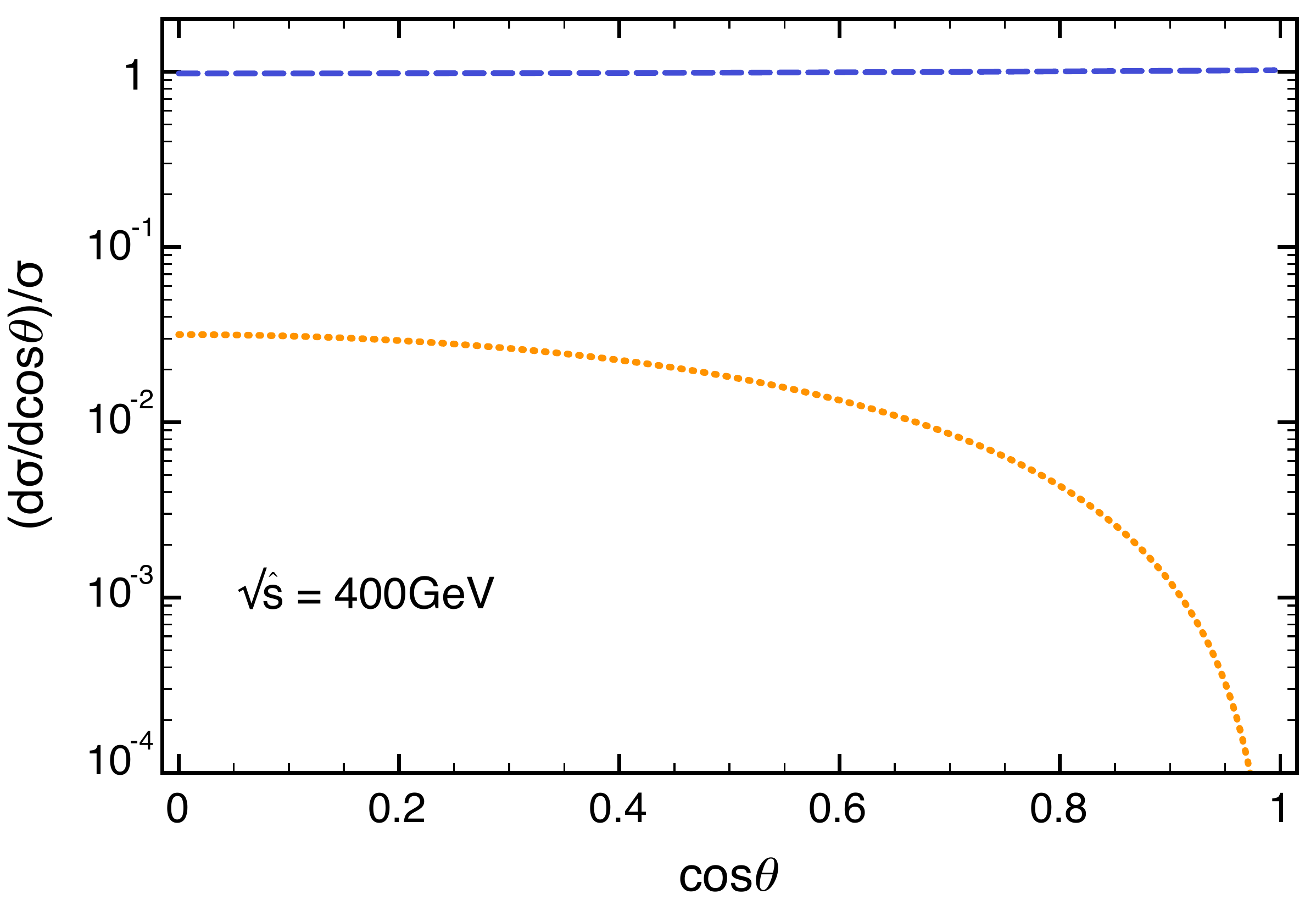}
\includegraphics[width=0.47\linewidth]{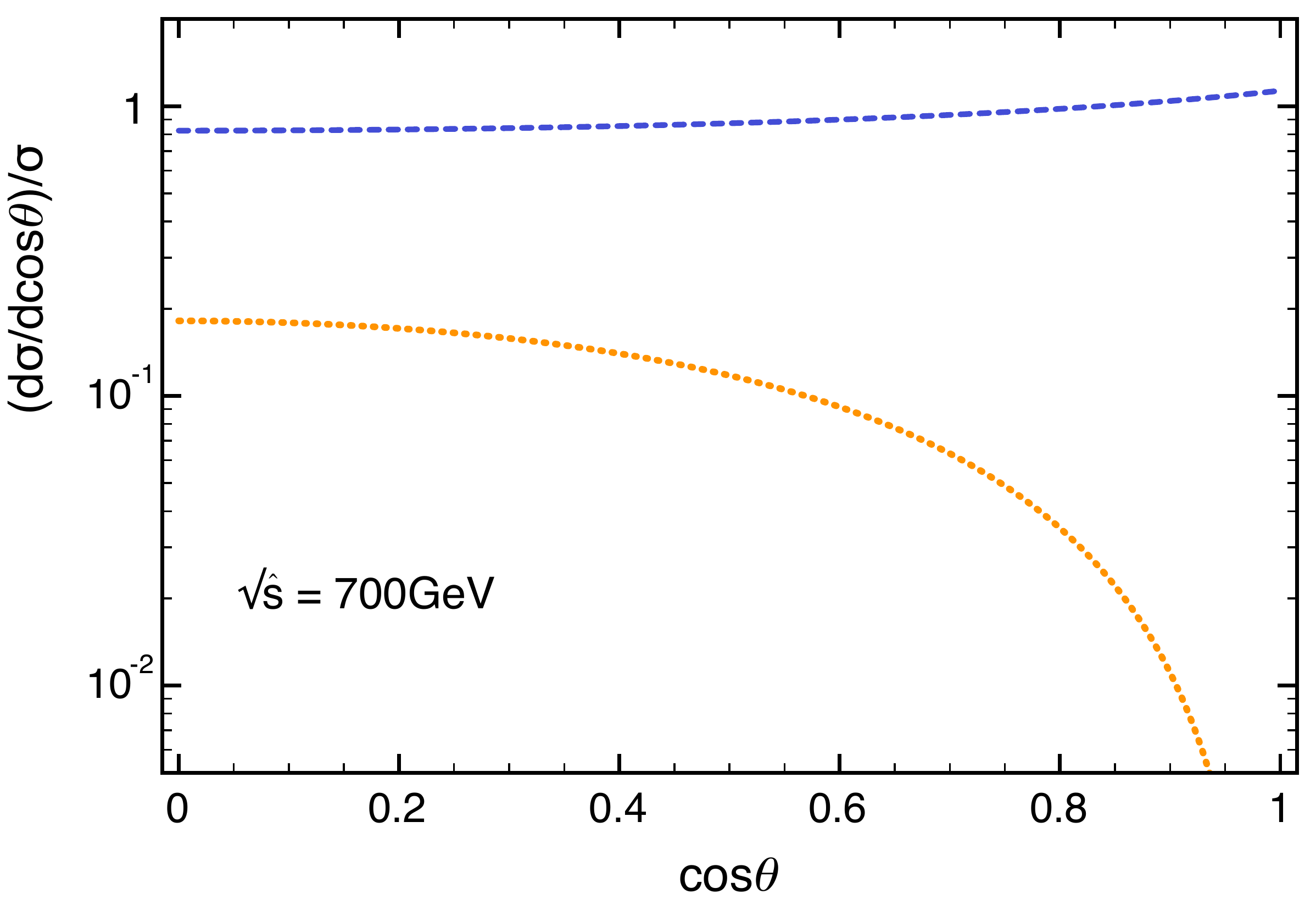}
\caption{Differential cross sections
obtained by including only the contribution of $M_0$ (dashed blue curve) or $M_2$ (dotted orange curve) in the SM, as functions of $\cos\theta$.
Both curves are normalized to the total SM cross section. The partonic center-of-mass energy has been fixed to $\sqrt{\hat s}=400\,$GeV in the left
plot and to  $\sqrt{\hat s}=700\,$GeV in the right plot.}
\label{fig:angdepSM}
\end{center}
\end{figure}

\begin{figure}[tp]
\begin{center}
\includegraphics[width=0.45\linewidth]{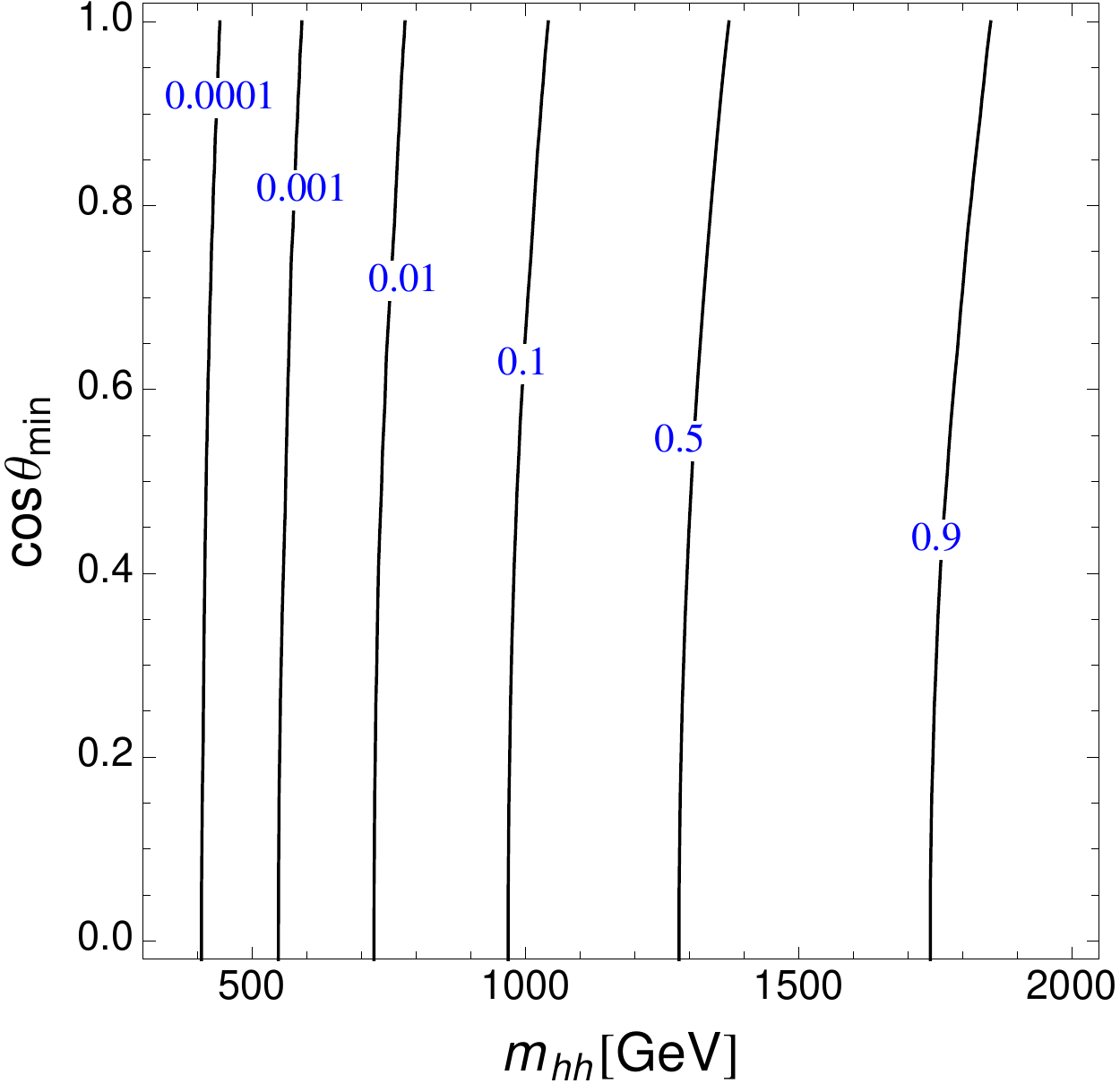}
\caption{Isocontours of the ratio $r = \sigma_{gD2}/\sigma_{tot}$,  defined in the text, in the plane $(m_{hh}, \cos\theta_{min})$.}
\label{fig:angdepO8}
\end{center}
\end{figure}

Besides $m_{hh}$, the other kinematic variable that characterizes the $gg \rightarrow hh$
events is the angle $\theta$ between either of the two Higgses and the beam axis in the center-of-mass frame. By total angular momentum conservation,
the scattering amplitude can be decomposed into two terms, $M_0$ and $M_2$,  describing transitions with respectively $J_z = 0$ and $J_z = \pm 2$,
where $J_z$ is the projection of $J$ along the beam axis.
The amplitude $M_2$ receives a contribution only from the box diagram and from the dimension-8 operator $O_{gD2}$ (through the 
last diagram of Fig.~\ref{fig:diagrams}). All the other diagrams, instead,  mediate $J_z =0$ transitions.
The explicit expression of $M_0$ and $M_2$ is reported in Appendix~\ref{app:angular} for convenience.
While transitions with all integer values of the total angular momentum $J$ can occur, the leading contributions to $M_0$ and $M_2$ come  
respectively from $J=0$ (s-wave) and $J=2$ (d-wave). A simple angular momentum decomposition then shows that, up to small corrections, 
the angular dependence is $M_0 \sim const.$,  $M_2 \sim \sin^2\theta$.~\footnote{The angular dependence of a scattering amplitude  with $J=j$
and $J_z = m$ is given by the Wigner function $d_{0m}^{(j)}(\theta)$.}
In the SM, the contribution of $M_2$ to the cross section is always small, as illustrated by Fig.~\ref{fig:angdepSM}:
it is  negligible at the peak of the $m_{hh}$ distribution ($m_{hh} \sim 400\,$GeV) and smaller than $\sim 20\%$ on its tail  ($m_{hh} \sim 700\,$GeV).
%
%
A shift in the top Yukawa coupling due to New Physics modifies the value of the box diagram, but cannot change the above conclusion (unless
extreme shifts are considered). 
It is thus interesting to ask whether the dimension-8 operator $O_{gD2}$ can give a sizable contribution to $M_2$ through the last diagram of Fig.~\ref{fig:diagrams}.
If this were the case, one could reach a better sensitivity on 
$O_{gD2}$ by performing a suitable analysis of the angular distributions of the Higgs decay  products.
Unfortunately, we find that the effect is numerically small. This is illustrated in Fig.~\ref{fig:angdepO8}, where we show the isocurves of $r = \sigma_{gD2}/\sigma_{tot}$ 
in the plane $(m_{hh}, \cos\theta_{min})$. 
We defined $r$ to be the ratio of the cross section induced by $O_{gD2}$ alone, $\sigma_{gD2}$, to the total cross section, $\sigma_{tot}$,
obtained by adding the contribution of $O_{gD2}$ to the SM.
Both cross sections are computed at the partonic level for the process $gg\to hh$ and by integrating over angles $\theta_{min} \leq \theta \leq \pi/2$.
We set the coefficient $\bar c_{gD2}$ equal to its NDA estimate (\ref{eq:estimatedim8}) and choose as benchmark values $f = 635\,$GeV, $m_* = 1.9\,$TeV, which
correspond to $g_* = 3$, $\xi \equiv (v/f)^2 = 0.15$.
%
%
For $r \sim 0.5$ the contribution of $O_{gD2}$ is as large as the SM one. 
The plot of Fig.~\ref{fig:angdepO8} shows that this occurs at the crossover scale $m_{hh} \sim 1.3\,$TeV, in agreement with the naive estimate of 
section~\ref{sec:higherorder}, $m_{hh} \sim y_t f \, (g_*/y_t)^{1/2}$. 
Increasing the value of $\theta_{min}$ (i.e. decreasing $\cos\theta_{min}$) enhances the  importance of the $J_z =2$ component of the cross section, 
hence that of $O_{gD2}$, but the effect is never large. For example, the value of the crossover scale is reduced at most by $\sim 10\%$.
We thus conclude that, although an analysis of angular distributions could in principle help disentangling the effects of dimension-8 operators,
in practice there is little leverage, and the efficacy of such strategy is further reduced in the  case of the $b\bar b\gamma\gamma$  final state
by the limited range in $m_{hh}$ which can be realistically probed.
In our analysis of $gg\to hh\to b\bar b \gamma\gamma$ we will thus make use of the $m_{hh}$ distribution as the main tool to probe the
effects induced by New Physics, neglecting for simplicity angular distributions.

%% file: HHpheno.tex
\label{sec:HHpheno}

The presence of various Higgs decay channels with a non-negligible branching ratio allows the exploration of the Higgs
properties in several final states. This is especially true in single Higgs processes, where the relatively large
production cross section compensates for the small decay probability in some of the cleanest channels, such as
$h \rightarrow \gamma \gamma$ and $h \rightarrow Z Z^* \rightarrow 4 l$. 
In the case of double Higgs production, however, 
the low signal rate forces one to consider only final states with a sizable branching ratio.
This is particularly so at the $14\,$TeV LHC, where the NLO total SM production rate is around $37\,\mathrm{fb}^{-1}$,
but remains  true even at a hypothetical future $100\,$TeV  machine, where the enhanced rates are unavoidably
accompanied by larger backgrounds.
In order to obtain a large enough branching ratio,  at least one of the two Higgses must decay into a $b\bar{b}$ pair.
For the second Higgs different choices seem possible and have been considered in the literature, namely $h \rightarrow b\bar{b}$, $h \rightarrow WW^*$,
$h \rightarrow \tau^+\tau^-$ and $h \rightarrow \gamma\gamma$.
The  channel $hh\rightarrow b\bar{b} b\bar{b}$ has the highest signal rate ($BR\simeq 33.3\%$ in the SM), but is plagued by a large QCD background. 
Even imposing four $b$-tags, it seems hard to exploit and could require a sophisticated
search strategy~\cite{Baur:2003gpa,Dolan:2012rv,deLima:2014dta}.
The channel with the second highest rate is $hh\to b\bar{b}WW^*$, with a branching ratio $BR \simeq 24.9\%$ in the SM.
Its observation is also threatened by a large background, mainly coming from $t\bar t$~\cite{Dolan:2012rv,Baglio:2012np}. It has been recently claimed that 
a good signal-to-background ratio can be obtained by using jet substructure techniques and focusing on a very specific region of the parameter space where 
the two Higgses and their decay products are highly boosted~\cite{Papaefstathiou:2012qe}. Although encouraging, the results of this analysis suggest that the $b\bar{b}WW^*$ 
final state can be observed at the $14\,$TeV LHC only with its high-luminosity extension.
The other channel which has been extensively studied in the literature is 
$hh\to b\bar{b}\tau^+\tau^-$~\cite{Baur:2003gpa,Dolan:2012rv,Barr:2013tda,Baglio:2012np,Goertz:2014qta}.
It is very promising and potentially relevant for the $14\,$TeV LHC, 
since its SM branching ratio is sizable, $BR \simeq 7.35\%$, and good signal-to-background ratios seem to be achievable while keeping a relatively large number 
of signal events. Its actual significance relies however on the ability to reconstruct the tau pair, and will have to be fully assessed by an experimental study.

For the purpose of our study we will focus on the channel $hh\to \gamma\gamma b\bar{b}$  ($BR \simeq 0.264\%$ in the SM),  which has been 
considered to be the cleanest one despite its small rate. As shown by previous theoretical studies~\cite{Baur:2003gp,Barger:2013jfa,Yao:2013ika,Baglio:2012np}
as well as a recent non-resonant experimental search at $\sqrt{s} = 8\,$TeV~\cite{Aad:2014yja} and a study for the $\sqrt{s} = 14\,$TeV high-luminosity 
LHC ~\cite{ATL-PHYS-PUB-2014-019} by ATLAS,~\footnote{See Ref.~\cite{Barger:2014taa} and Refs.~\cite{CMS:2014ipa,Aad:2014yja}  
for respectively theoretical and experimental analyses of the resonant di-Higgs production in the $b\bar b \gamma\gamma$ final state.}
the analysis strategy to observe this decay mode is relatively simple and straightforward. This allows us to
avoid unnecessary complications and to concentrate primarily
on the interpretation of the search in the context of the effective field theory description.
Clearly, exploring the other available channels from a similar perspective
could improve significantly the sensitivity. We leave this interesting follow-up for a future study.

Before proceeding with the details of our analysis it is useful to briefly summarize the properties of the kinematic
distributions and discuss the main differences between the $14\,$TeV LHC and a future $100\,$TeV collider.

\begin{figure}[tbp]
\begin{center}
\includegraphics[width=0.47\linewidth]{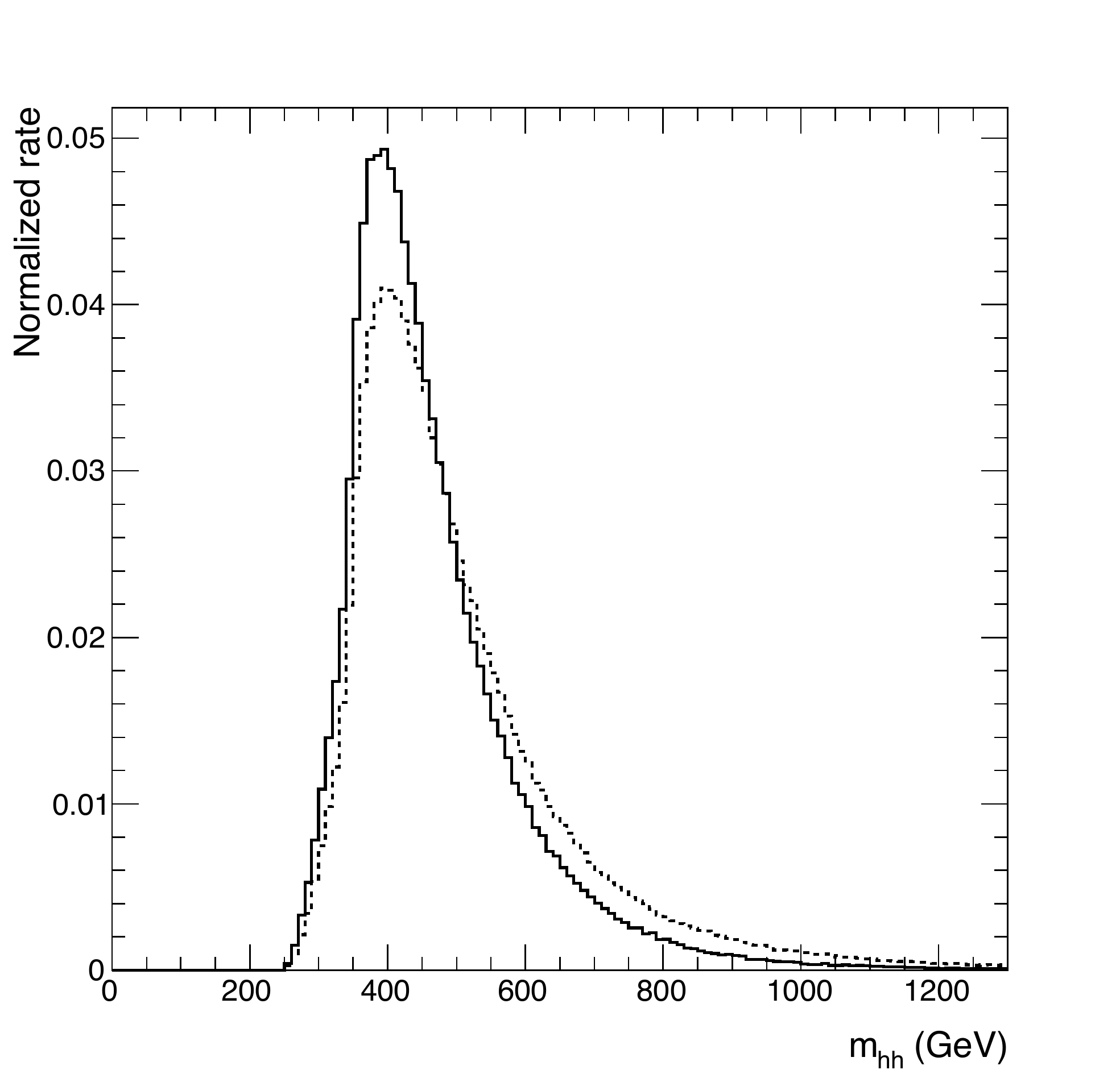}
\includegraphics[width=0.47\linewidth]{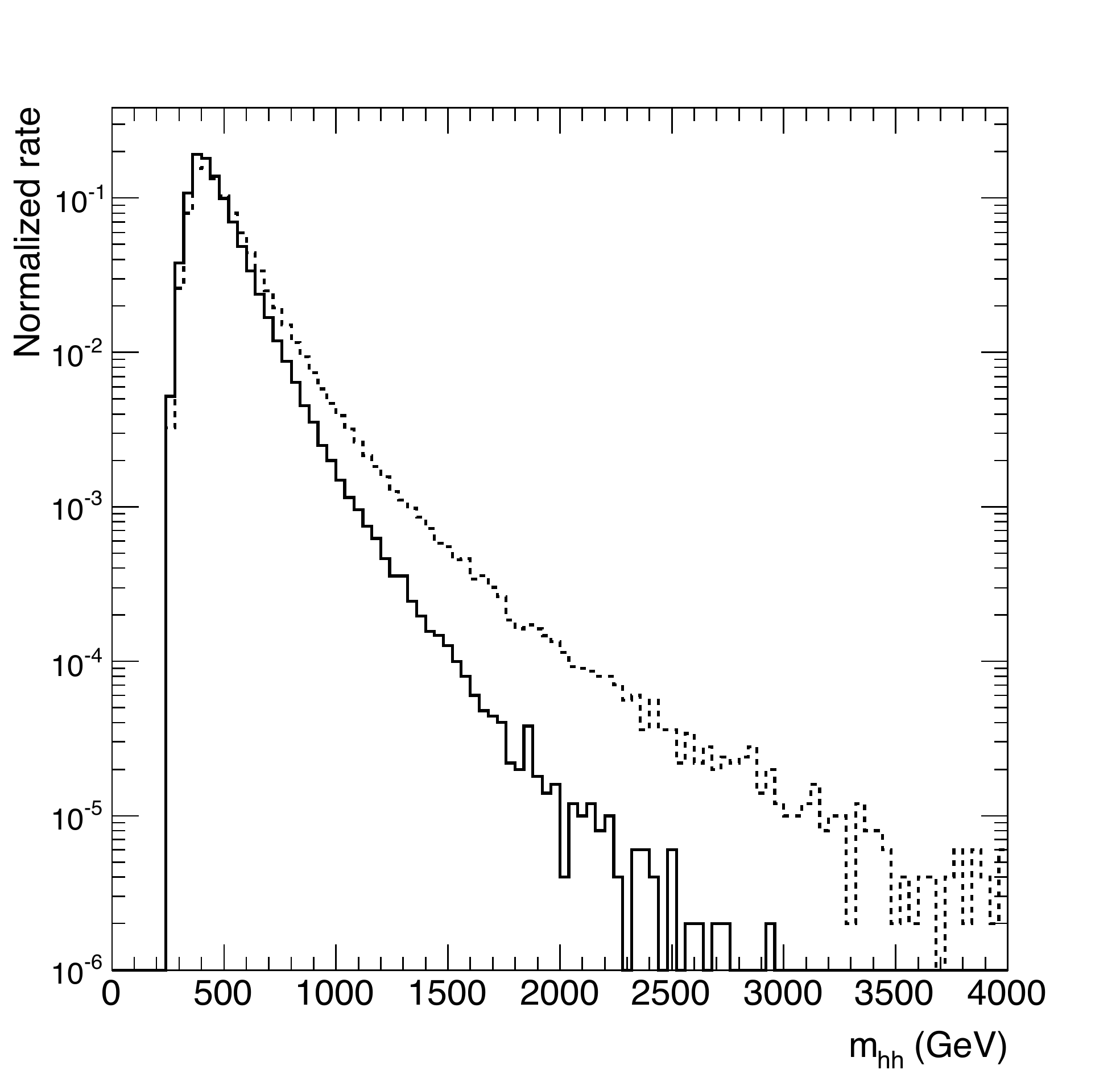}
\caption{Left plot: Normalized differential cross section for $pp  \to hh$ in the SM
as a function of the invariant mass of the two Higgs bosons. 
The solid and dotted lines correspond respectively to $\sqrt{s} = 14$ and $100\,$TeV. Right plot: Same as on the left but with  logarithmic scale.
}
\label{fig:mhhdist}
\end{center}
\end{figure}

It is well known (see for example Ref.~\cite{Contino:2012xk}) that in the SM a cancellation between the box diagram
and the triangle diagram involving the Higgs trilinear coupling leads to a depletion of the signal at threshold
($m_{hh} \gtrsim 250\ \mathrm{GeV}$). The peak of the distribution is essentially determined by the fast
decrease of the gluon parton distribution functions and is located around $m_{hh} \simeq 400\ \mathrm{GeV}$.
This conclusion is valid independently of the collider center-of-mass energy $\sqrt{s}$. The main difference between the LHC and a
higher-energy collider is a rescaling of the overall cross section (the SM cross section at a $100\,$TeV
machine is around $40$ times bigger than the one at the $14\,$TeV LHC), with small effects on the $m_{hh}$
distribution. As it can be seen from Fig.~\ref{fig:mhhdist}, this latter is modified significantly only on its tail
($m_{hh} \gtrsim 700\ \mathrm{GeV}$), which is enhanced at larger collider energies due to the higher luminosity of gluons. 
In fact, as we will discuss in section~\ref{sec:boostedanalysis},
an important change is also present in the pseudo-rapidity distribution of the Higgs bosons, which is related to the amount of boost determined by the initial parton energies.
%

Due to the above considerations, one would naively expect that an analysis similar to the one designed for $14\,$TeV
would continue to work well at $100\,$TeV. This is in fact only partially true.
Indeed, the enhanced signal rates can impact the analysis results in two ways.
First, they will improve the sensitivity on the parameters, such as the Higgs trilinear coupling, by simply reducing the statistical uncertainty 
even if the analysis strategy is not modified. Second, due to the larger cross section and enhanced tail, 
much higher values in $m_{hh}$ will be accessible, which are potentially more sensitive to higher-order
operators growing with the energy. For this reason, fully exploiting the potential of a $100\,$TeV collider  requires
some modification of our $14\,$TeV analysis strategy in order to better reconstruct events with a higher boost.
Although we will not present a fully optimized analysis for the $100\,$TeV, we will give a first assessment 
of how much our final results can improve when jet substructure techniques are used.

\begin{figure}[tbp]
\begin{center}
\includegraphics[width=0.47\linewidth]{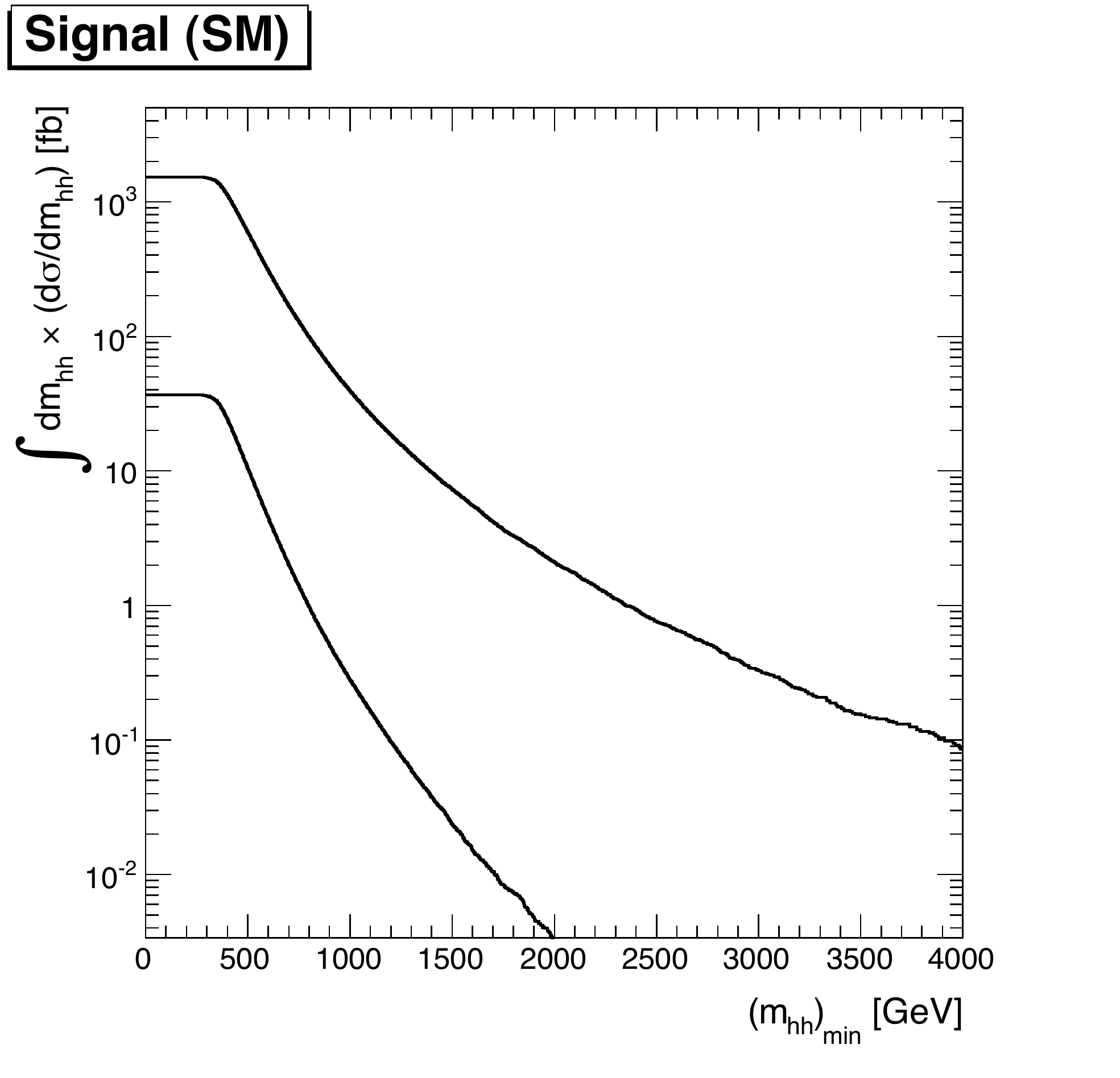}
\includegraphics[width=0.47\linewidth]{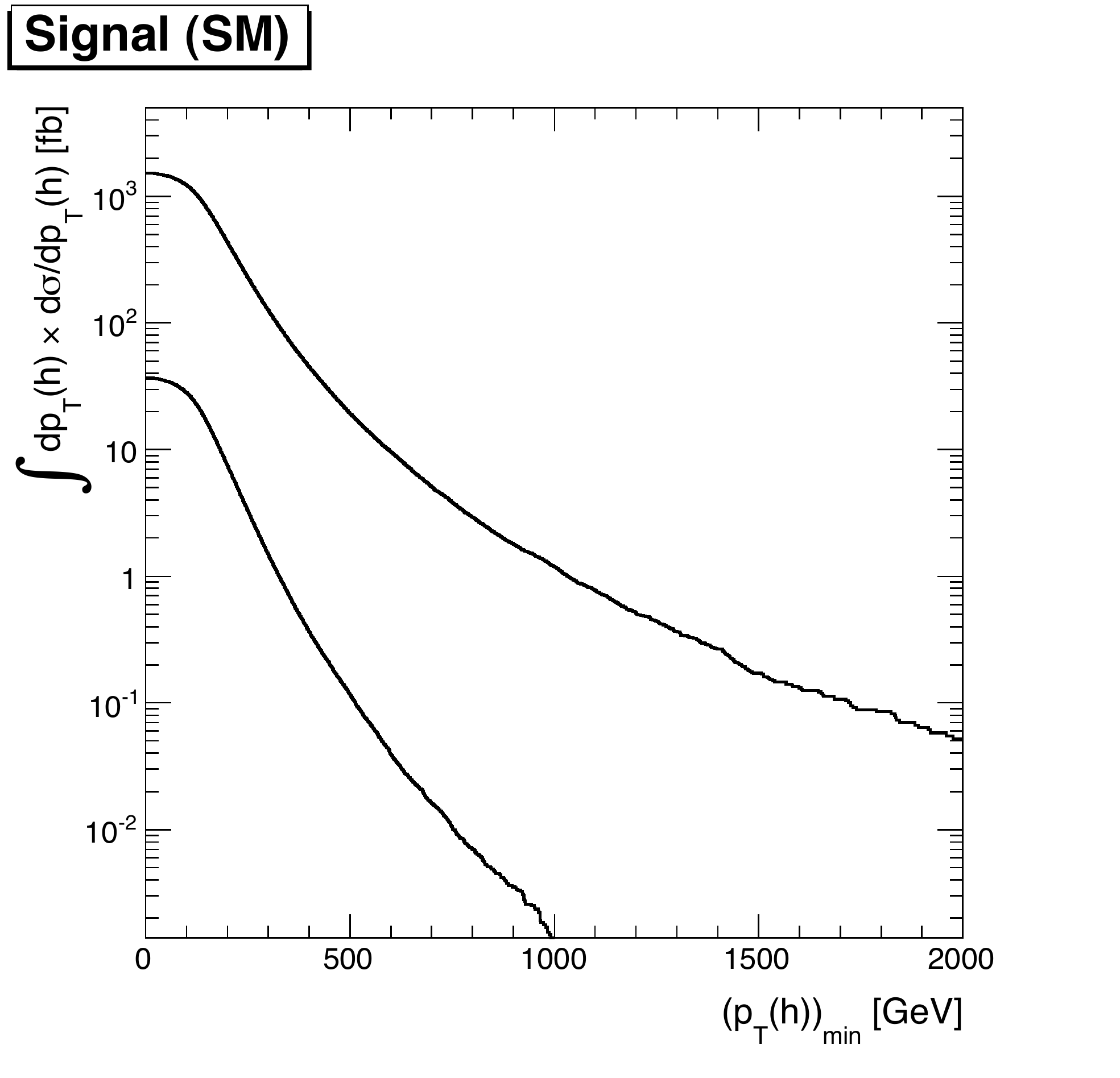}
\vspace{-0.2cm}
\caption{Integrated differential cross sections for $pp\to hh$ in the SM as a function of the minimum $hh$ invariant mass, $(m_{hh})_{\rm min}$ (left plot), and the minimum
Higgs transverse momentum, $(p_{T}(h))_{\rm min}$ (right plot). The upper (lower) curve is for $\sqrt{s} = 100\,$TeV ($14\,$ TeV).
The cross sections have been computed at NNLO using the k-factors of Eq.~(\ref{eq:kfactors}).
}
\label{fig:dsigmaOfxx}
\end{center}
\end{figure}

As a first step in this direction, it is useful 
to get an idea of the reach that can be achieved on $m_{hh}$ and $p_T(h)$ at  different colliders and for  different search channels.
A complete assessment of this point would require specifying completely the scenario we are interested
in and the size of the possible new physics contributions to the cross section.
To get a rough approximation, however, we can estimate the reach by demanding that a few events
(we choose $5$ for the estimate) are still present in the tails of the SM distributions above the considered
$m_{hh}$ (or $p_T(h)$) value.  To be more realistic we also assume a $10\%$ overall signal efficiency due to kinematic cuts. 
The estimates can be easily extracted from the plots of Fig.~\ref{fig:dsigmaOfxx}, which show the integrated differential
cross sections for the SM signal $pp\to hh$ as a function of the lower cut on $m_{hh}$ and of the minimum $p_T(h)$. 
%
%
By including the branching ratio of the most important decay modes we obtain the results shown in Table~\ref{tab:prospect}.
%
\begin{table}[tbp]
\centering
\scalebox{0.95}{
\begin{tabular}{l|c|c|c|c}
channel & $b\bar{b}b\bar{b}$ (33.3\%) & $b\bar{b}WW^*$ (24.9\%) & $b\bar{b}\tau^+\tau^-$ (7.35\%) & $\gamma\gamma b\bar{b}$ (0.264\%) \\ \hline \hline
Cross section   & $>$ 0.05 fb & $>$ 0.067 fb &$>$ 0.227 fb & $>$  6.31 fb \\
$m_{hh}$ [GeV]   & $<$ 1340 (4290) & $<$ 1280 (4170) & $<$ 1039 (3235) & $<$ 558 (1552) \\
$p_T(h)$ [GeV]   & $<$ 575 (2000)  & $<$ 550 (1890) & $<$ 440 (1430)  & $<$ 210 (664) 
\end{tabular}}
\vspace{0.2cm}
\caption{Lower limits on the $pp\to hh$ cross section in the SM and upper limits of the phase space that guarantee 5 signal events.
We assume an integrated luminosity $L=3000\,\text{fb}^{-1}$ at the $14\,$TeV LHC and a 10\% efficiency due to kinematic cuts.
The numbers in parenthesis refer to a $100\,$TeV collider, with the same integrated luminosity.
The cross sections have been computed at NNLO using the k-factors of Eq.~(\ref{eq:kfactors}). 
}
\label{tab:prospect}
\end{table}

The reach on $m_{hh}$ is rather limited in the $b\bar b\gamma\gamma$ channel due to the small signal rate, and even at a $100\,$TeV collider
it does not extend much beyond $\sim 1.5\,$TeV. Other channels with larger rates will be crucial to push further the exploration of higher invariant masses.
Similar considerations apply also for the reach on $p_T(h)$. 
The maximal value of $p_T(h)$ 
can be used  to estimate whether a given search can benefit from jet substructure techniques or a simple jet analysis is enough.
The angular separation between two partons coming from the Higgs decay scales like $\Delta R \sim 2 m_h/p_T(h)$.
In order to resolve the jets with the standard techniques we must demand $\Delta R \gtrsim R_{ATLAS,\, CMS}$,
where $R_{ATLAS} = 0.4$ and $R_{CMS} = 0.5$ are the reconstruction cones used by the ATLAS and CMS collaborations
respectively. It is easy to see that for $p_T\gtrsim 500-600$ GeV the two partons are not often resolved as two
separated objects and jet substructure techniques are likely to be helpful.
The results in Table~\ref{tab:prospect} show that the size of the boosted phase space at $14\,$TeV is
almost negligible for the $\gamma\gamma b \bar{b}$ channel and very limited for
$b\bar{b}b\bar{b}$, $\gamma\gamma WW^*$ and $b\bar{b}\tau^+\tau^-$.
The story  changes however for an $100\,$TeV collider. In this case the SM signal can be probed up to
$m_{hh} \sim 1.5\ \mathrm{TeV}$ in the $\gamma\gamma b \bar{b}$ channel and up to $m_{hh} \sim 4\ \mathrm{TeV}$
for $b \bar{b}b \bar{b}$. In these kinematic regions many boosted Higgses are  produced and jet substructure techniques are crucial
to reconstruct them. We will analyze this aspect in more details in Section~\ref{sec:boostedanalysis}.

%% file: Analysis.tex
\label{sec:analysis}

After discussing the general properties of the $pp \to gg \rightarrow hh$ process, we now focus  on the
$b \bar b \gamma \gamma$ final state and describe our analysis strategy.
We generated the signal events through
our own dedicated code described in section~\ref{sec:crosssection}.
The code includes the 1-loop diagrams of Fig.~\ref{fig:diagrams} and is based on the parametrization of Eq.~(\ref{eq:nonlinearL}).
It is available from the authors upon request.
By using the CTEQ6ll PDF's (LO PDF with LO $\alpha_s$), setting the factorization and renormalization scales to $Q = m_{hh}$ and $m_h = 125\,$GeV, $m_t=173\,$GeV,
we find the following LO cross sections for the Standard Model: $16.2\,$fb and $873.6\,$fb for respectively $14\,$TeV and $100\,$TeV collider energies.
In order to (partially) include the NLO and NNLO corrections, we rescale the cross sections by the k-factors
\begin{equation}
\label{eq:kfactors}
k_{14\,\text{TeV}} = 2.27 \, ,\qquad\quad  k_{100\,\text{TeV}} = 1.75
\end{equation}
which were computed for the SM in Ref.~\cite{deFlorian:2013jea}.~\footnote{Compared to the NLO calculation (performed in Ref.~\cite{Dawson:1998py}),
the NNLO corrections give a $\sim 20\%$ enhancement of the total cross section.
Notice that both the NLO and NNLO corrections to the total rate have been computed in the infinite top mass approximation.
It is well known that at LO this approximation leads to very inaccurate kinematic distributions, in particular the $m_{hh}$ one.
For this reason the k-factors used in our analysis must be  considered only as a crude approximation. 
Recently, a step forward towards a fully differential NLO calculation has been done in Ref.~\cite{Frederix:2014hta}, where real emission diagrams
were computed at 1-loop, while the finite part of the two-loop virtual corrections was extracted by resorting to the infinite top mass approximation.
For an estimate of the finite-top mass effects and of the dependence of the k-factor on a cut on $\sqrt{\hat s}$, see Refs.~\cite{Grigo:2013rya,Maltoni:2014eza}.
See also Ref.~\cite{Shao:2013bz} for the resummation of threshold  effects in the SM.
}

The main backgrounds that we considered are the non-resonant processes $b \bar b \gamma \gamma$ and $\gamma \gamma j j$
and the resonant processes $b\bar b h$, $Z h$ and $t\bar t h$ (with the
subsequent decays $h\to \gamma\gamma$, $Z\to b\bar b$, and $t\bar t \to b\bar b + X$). 
Further backgrounds involving fake photons from jets have been neglected
as their estimate is beyond the scope of this work. Experimental analyses of single Higgs production have shown that similar processes in that case
can be safely reduced to a subleading level, and we thus assume that this will be possible for double Higgs production as well.
We generated all the backgrounds with \textsc{MadGraph}5\_aMC$@$NLO v2.1.1~\cite{Alwall:2014hca} by switching off virtual corrections (i.e. working in LO mode).
The output has been interfaced with \textsc{PYTHIA} v6.4~\cite{Sjostrand:2006za} for parton showering and hadronization, and with {\tt{FastJet}} v3.0.6~\cite{Cacciari:2011ma} 
for jet clustering.
The factorization and renormalization scales have been set to the default dynamic scale in \textsc{MadGraph}5.
Further details about the generation can be found in Appendix~\ref{app:xsec:sig:bkg}.
The backgrounds $b\bar{b}\gamma\gamma$, $b\bar{b}h$ and $Zh$ have been matched up to one additional jet via the $k_T$-jet MLM matching~\cite{Alwall:2007fs} to 
partially account for  NLO effects.
In the case of $b\bar b \gamma\gamma$ we found that allowing for the extra jet increases the cross section by a factor $\sim 2$.
We found that this is  in good agreement with a complete NLO computation performed with \textsc{MadGraph}5\_aMC$@$NLO, indicating
that real emissions in this case represent the bulk of the NLO correction. Considering that  $b\bar{b}\gamma\gamma$ is the dominant background
after all cuts, including this effect is very important, but it has been omitted in most of the previous studies. More details about our generation of $b\bar b \gamma\gamma$
and a thorough discussion of the NLO correction are given in Appendix~\ref{app:NLOestimate}.
For the resonant $t\bar{t}h$ background we  apply instead a simple k-factor rescaling of the LO cross section to the NLO value reported in
Ref.~\cite{Dittmaier:2011ti}.
Finally, the $\gamma\gamma jj$ background was simulated without matching and a k-factor $k_{\gamma\gamma jj} =2$ was applied  (this is a more conservative choice than
the one proposed in~\cite{Alwall:2014hca} based on an NLO simulation).

To estimate the sensitivity of our analysis we should, ideally, take into account parton-shower/ha\-dro\-ni\-za\-tion and smearing
in every point of our parameter space. This procedure, however, would be  computationally too expensive.
Instead we adopt the following simplified approach. We fully extract the signal rate after cuts only for the SM point
by performing a hadron-level analysis. For the other points in the parameter space, we apply the same type of analysis
to the parton-level samples and we rescale the signal by the hadron-to-parton cross section ratio  computed for the SM:
\begin{equation}
\sigma^\text{had}_\text{BSM}\big|_\text{w/ cuts} \simeq \sigma^\text{part}_\text{BSM}\big|_\text{w/ cuts} 
\times \left(\frac{\sigma^\text{had}_\text{SM}}{\sigma^\text{part}_\text{SM}}\right)\!\Big|_\text{w/  cuts} \, .
\end{equation}
We use the fact that such ratio is approximately the same for both the SM and in a generic point of the BSM parameter space.
The rescaling is performed individually for each of the bins in $m_{hh}$ of Tables~\ref{tab:mhh:14TeV},~\ref{tab:mhh:100TeV} and~\ref{tab:mhh:100TeV:tradVSjetsub}.
Our analysis  takes correctly into account possible non-universal signal effects coming from distortions of the
kinematic distributions at the partonic level. The rescaling  then approximately includes the effect of showering and hadronization.
For the jet substructure analysis of Section~\ref{sec:boostedanalysis}, on the other hand, we follow a different procedure, since there is no parton-level cross section
we can use after cuts. We thus compute the BSM signal by starting with the SM one computed at the hadron level and multiply by the ratio of BSM over SM cross sections
before cuts at the parton level
\begin{equation}
\sigma^\text{had}_\text{BSM}\big|_\text{w/ cuts} \simeq \sigma^\text{had}_\text{SM}\big|_\text{w/ cuts} 
\times \left(\frac{\sigma^\text{part}_\text{BSM}}{\sigma^\text{part}_\text{SM}}\right)\!\Big|_\text{w/o  cuts} \quad \text{(boosted analysis)}\, .
\end{equation}
This ratio is expected to be approximately the same at the partonic level before cuts and at the hadronic level after  cuts if one considers sufficiently narrow bins  in $m_{hh}$,
since to very good accuracy the latter is the only variable that controls the kinematics of the signal.
We checked that this procedure is accurate at $\sim\text{a few }\%$ level in the high-$m_{hh}$ categories used for the boosted analysis.

We designed  different strategies for $14\,$TeV and $100\,$TeV colliders.
We begin by describing the one at $14\,$TeV.
Events are triggered by demanding exactly two isolated photons satisfying the minimal reconstruction requirements
\begin{equation}
p_T(\gamma) > 25\ \mathrm{GeV}\, ,\qquad |\eta(\gamma)| < 2.5\,.
\end{equation}
To forbid extraneous leptonic activity we  veto events with isolated leptons (electrons or muons) satisfying
\begin{equation}
p_T(l) > 20\ \mathrm{GeV}\, ,\qquad |\eta(l)| < 2.5\,.
\end{equation}
Similar isolation criteria are applied to both photons and leptons. A photon (lepton) is considered isolated
if the surrounding hadronic activity within a cone of size $R = 0.4$ (0.3) satisfies
$p_T(\gamma)/(p_T(\gamma)+p_T({\rm cone}))>0.9$ (0.85).
We have checked that more sophisticated photon isolation criteria, like the one proposed in~\cite{Frixione:1998jh}, give similar results.
We then cluster the events into $R=0.5$ anti-$k_T$ jets~\cite{Cacciari:2008gp}. We accept only jets with
\begin{equation}
p_T(j) > 25\ \mathrm{GeV}\, ,\qquad  |\eta(j)| < 2.5
\end{equation}
and require that at least two of them are $b$-tagged (the leading two $b$-jets are selected if more than two $b$-jets exist).
We assume 70\% efficiency for $b$ tagging $\epsilon_b =0.7$, corresponding to~1\% of mis-tag rate
$\epsilon_{j\rightarrow b} =0.01$~\cite{Chatrchyan:2012jua,CMS:2013vea,ATL-PHYS-PUB-2013-009},~\footnote{More specifically, we tag with 70\% probability only those jets
which contain a $b$-hadron with transverse momentum larger than $5\,$GeV$/R_{jet}$. For $R_{jet}=0.5$ this translates
into a minimum $10\,$GeV transverse momentum.  When the jet substructure is used (see section~\ref{sec:boostedanalysis}), 
$R_{jet}$ is set to $\Delta R(bb)$, corresponding to the distance between two leading subjets after BDRS declustering.}
and 80\% efficiency for photon tagging $\epsilon_\gamma =0.8$~\cite{CMS:2013aoa,ATL-PHYS-PUB-2013-009}~\footnote{Typical rejection rates for fake photons 
from jets at this working point are in the range $0.1-0.5\%$. We will not use this number however, since backgrounds from fake photons are not included in this analysis.}.
After this step, an event consists of two isolated photons, two $b$-tagged jets, and possible additional jets (whether or not $b$-tagged).

Once the reconstruction of the basic objects by the above procedure is done, we apply the
set of cuts described in the following. We first restrict the events to those with two hard photons
and two $b$-tagged jets satisfying
\begin{equation}
\label{eq:acceptanceI}
\begin{gathered}
p_{T>}(b),p_{T>}(\gamma) > 50\ {\rm GeV}\,, \quad p_{T<}(b),p_{T<}(\gamma) > 30\ {\rm GeV}\,,\\
 60 < m^{\rm reco}_{b\bar{b}} < 200\ {\rm GeV}\,,\quad 60 < m^{\rm reco}_{\gamma\gamma} < 200\ {\rm GeV}\,,
\end{gathered}
\end{equation}
where $p_{T>}(\gamma)$ ($p_{T<}(\gamma)$) denote the transverse momentum of the hardest (softest)  photon, and 
$p_{T>}(b)$, $p_{T<}(b)$ are similarly defined for the two $b$-jets.
At this stage, the broad mass windows in Eq.~(\ref{eq:acceptanceI}) are placed merely to allow for a fair comparison of the signal and  background cross sections.

\begin{figure}[tp]
\begin{center}
\includegraphics[width=0.33\linewidth]{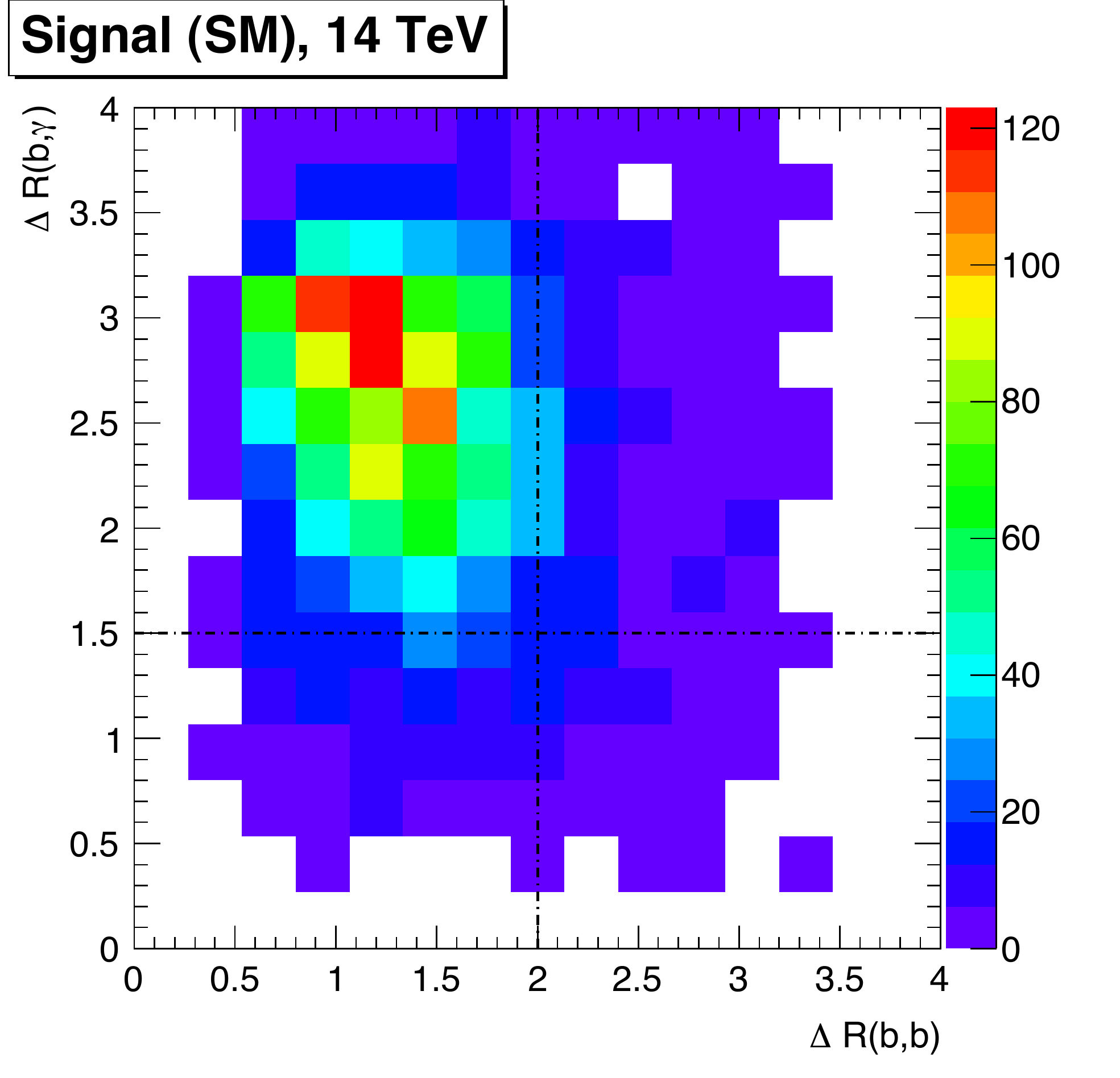}\hspace{-0.1cm}
\includegraphics[width=0.33\linewidth]{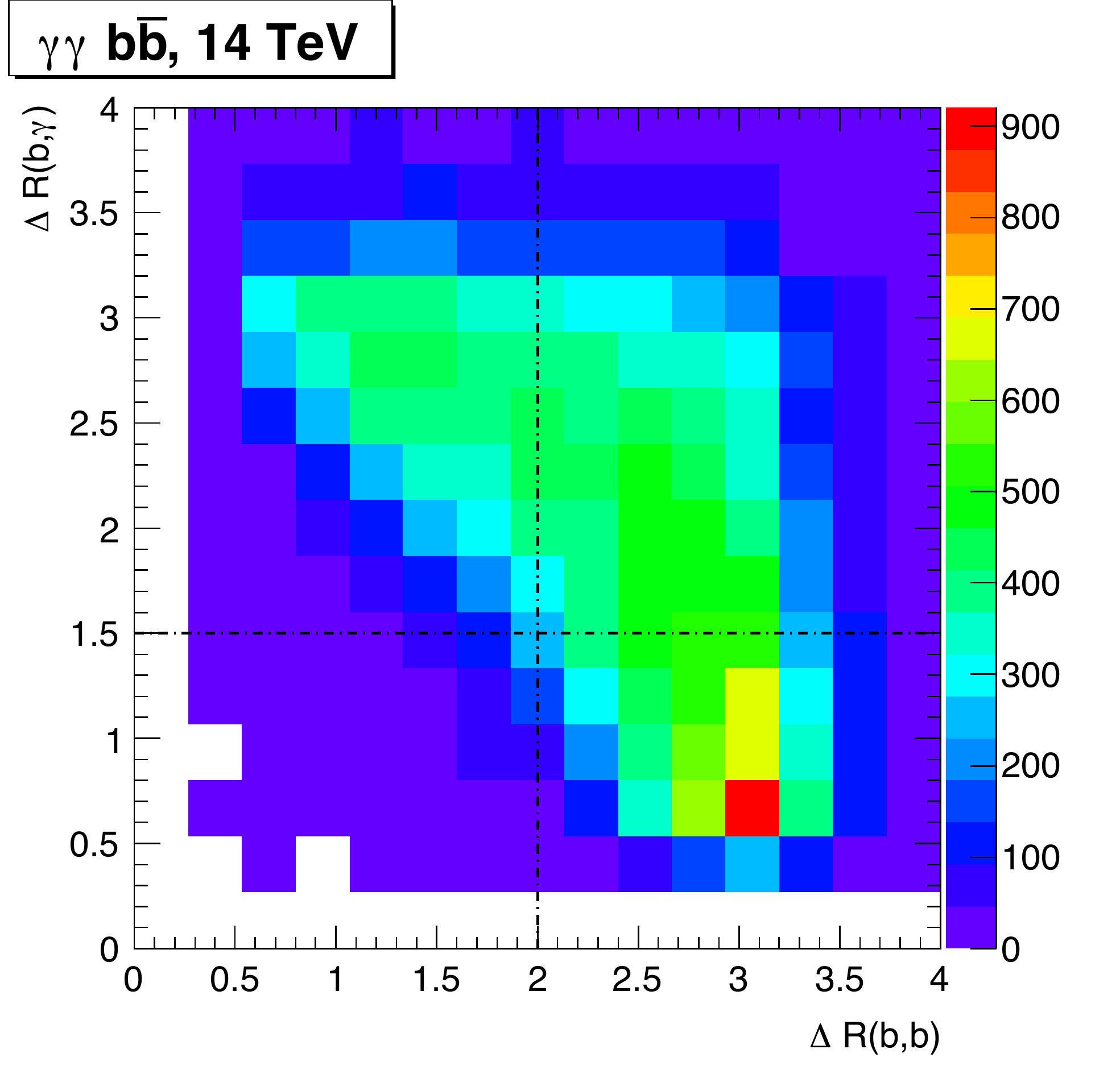}\hspace{-0.1cm}
\includegraphics[width=0.33\linewidth]{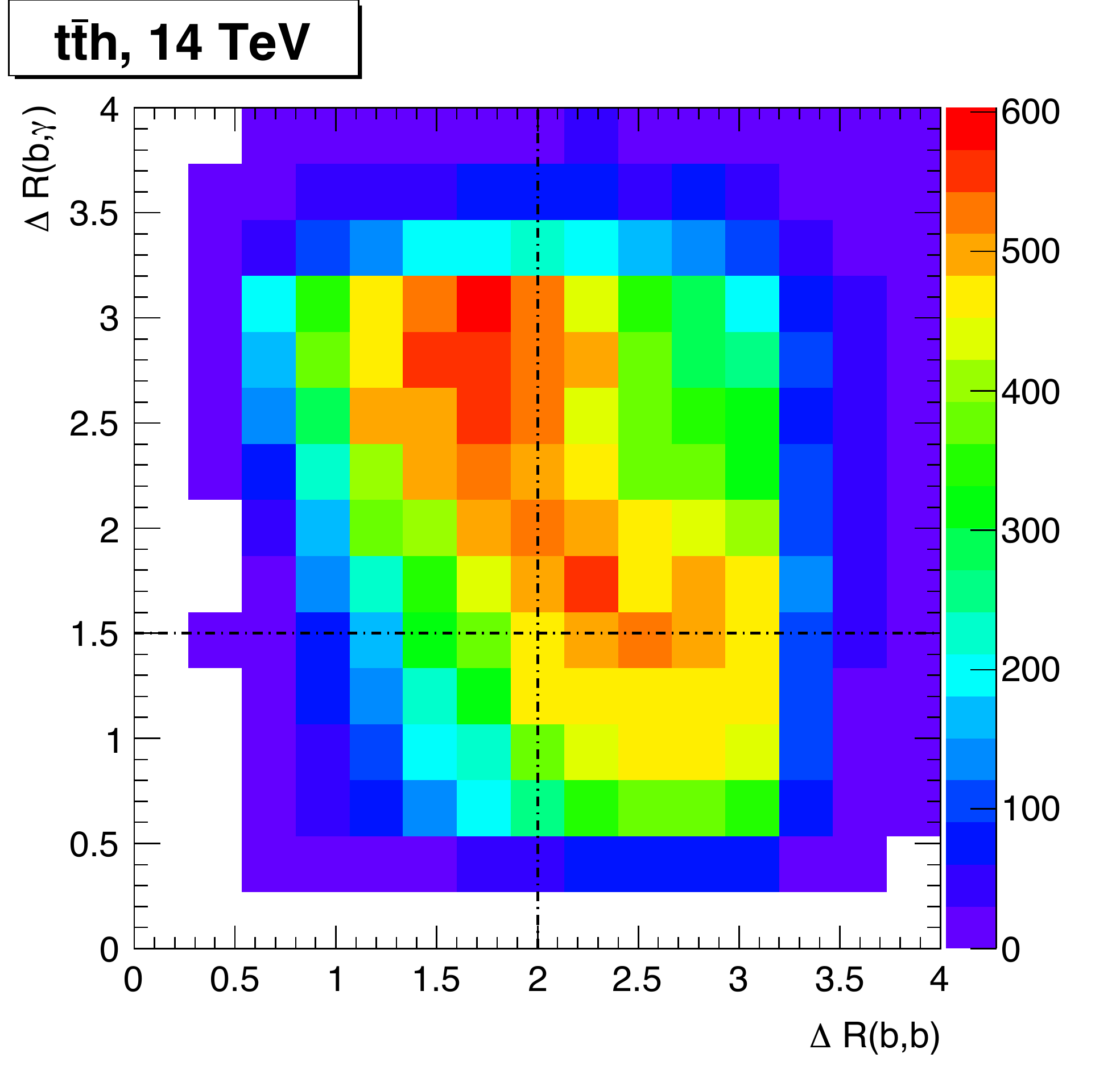}
\caption{Angular distributions (in arbitary units) after the cuts of Eq.~(\ref{eq:acceptanceI}) of the SM signal and the two dominant backgrounds, 
$\gamma\gamma b\bar{b}$ and $t\bar{t}h$, for $\sqrt{s} = 14\,$TeV. The dashed  lines indicate to the  cuts of Eq.~(\ref{eq:acceptanceII}).}
\label{fig:DRdist:14TeV}
\end{center}
\end{figure}

In the signal, the $b\bar{b}$ and $\gamma\gamma$ subsystems tend to be in a back-to-back configuration,
and the angular distance between two photons (and similarly between two $b$-jets) is of order $\sim 2 m_h/p_T(h) \sim {\mathcal O}(1)$ 
in the majority of the phase space.  While the $\gamma\gamma$ subsystem in resonant backgrounds
has a kinematics similar to the signal, the different origin of the $b\bar{b}$ pair in each process
can be used to distinguish them from the signal.
The angular distributions of the signal and of the two dominant backgrounds are shown in Fig.~\ref{fig:DRdist:14TeV}.
%
%
We find that the following cuts (indicated by the dashed lines in Fig.~\ref{fig:DRdist:14TeV})
\begin{equation}
\label{eq:acceptanceII}
 \DeltaR(b,b) < 2~,\quad \DeltaR(\gamma,\gamma) <2~, \quad \DeltaR(b,\gamma) > 1.5\,,
\end{equation}
can efficiently reduce the background, especially $b\bar b\gamma\gamma$, while retaining most of the signal.~\footnote{At leading order the
$b\bar b\gamma\gamma$ process is initiated by $q\bar{q}$ and $gg$ and proceeds through diagrams where the two $b$'s tend to 
emerge with a large relative angle. Diagrams initiated by $gq$ and $g\bar{q}$ at next-to-leading order lead to topologies which can more easily fake 
the angular configuration of the signal but account only for a minor fraction of events.}
As a final cut, we restrict the invariant masses of two $b$-jets and of the two photons to the Higgs mass window~\footnote{The width of the interval
$120 < m^{\rm reco}_{\gamma\gamma} < 130\ {\rm GeV}$ corresponds to $\sim 3$ times the experimental resolution on photon pairs, 
see~\cite{Khachatryan:2014ira,Aad:2014eha}. The same mass window was adopted by the recent CMS study of di-Higgs resonant production~\cite{CMS:2014ipa}, 
see also Ref.~\cite{Aad:2014yja}. The width of the interval  $105 < m^{\rm reco}_{b\bar{b}} < 145\ {\rm GeV}$ corresponds to $\sim 2$ times the experimental 
resolution on $b$-jet pairs from Higgs decays (after correcting for resolution effects), see~\cite{Aad:2014xzb,Chatrchyan:2013zna}.
We do not smear photons in our simulation, nor include a specific efficiency for the reconstruction of the photon pair. The mass window
on $m^{\rm reco}_{\gamma\gamma}$ is on the other hand sufficiently wide that almost all of the signal is retained, so that the efficiency would be close
to $100\%$ even including a finite energy resolution on photons. The efficiency of the cuts of Eq.~(\ref{eq:masswindow}) reported in Table~\ref{tab:cutflow:14TeV} 
corresponds, in the case of the signal, to the efficiency for the reconstruction of the $b\bar b$ pair (equal to $\sim 0.72$).
}
\begin{equation} \label{eq:masswindow}
 105 < m^{\rm reco}_{b\bar{b}} < 145\ {\rm GeV}~,\quad 120 < m^{\rm reco}_{\gamma\gamma} < 130\ {\rm GeV}\,.
\end{equation}
The resulting cut flow  is shown in Table~\ref{tab:cutflow:14TeV}. 
%
\begin{table}[tbp]
\centering
\begin{tabular}{l|c|ccccc}
\hline &&&&&& \\[-0.45cm]
$\sqrt{s} = 14$ TeV  & $hh$   &  $b\bar{b}\gamma\gamma$  & $\gamma\gamma jj$ & $t\bar{t}h$ &  $b\bar{b}h$  & $Zh$ \\[0.08cm]
\hline \hline &&&&&& \\[-0.42cm]
After selection cuts of Eq.~(\ref{eq:acceptanceI})                         & 25.8  & 6919 & 684  & 130     & 7.2    & 25.4    \\[0.08cm]
After $\Delta R$ cuts of Eq.~(\ref{eq:acceptanceII})                       & 17.8  & 1274 & 104  & 29    & 1.2     & 15.8    \\[0.08cm]
After $m^{\rm reco}_{b\bar{b}, \, \gamma\gamma}$ cuts of Eq.~(\ref{eq:masswindow}) & 12.8  & 24.2 & 2.21 & 9.9    & 0.40    & 0.41 
\\[0.16cm] \hline
\end{tabular}
\caption{Cut flow for the SM signal and the various backgrounds at $\sqrt{s} = 14\,$TeV. 
The values correspond to the number of events, assuming an integrated luminosity $L=3\,\text{ab}^{-1}$.
All reconstruction efficiencies and branching ratios to the final state under study are included.
}
\label{tab:cutflow:14TeV}
\end{table}
%
We report in Table~\ref{tab:tot_rate_coefficients} a fit of the signal rate ($r = \sigma \times BR(hh\to b\bar b\gamma\gamma)$) 
obtained after all cuts based on a  parametrization  analog to that given in Eq.~(\ref{eq:tot_xsec}) for the cross section.
%
\begin{table}[tpb]
\centering
\small
\begin{tabular}{ccccccccc}
$\sqrt{s}$ & $ r_{SM}\,[\text{ab}]$ & $A_1$ & $A_2$ & $A_3$ & $A_4$ & $A_5$ & $A_6$
& $A_7$ \\
\hline
\rule{0pt}{1.25em}
 $14\,\text{TeV}$ & $4.28$ & $1.70$ & $10.7$ & $0.117$ & $6.11$ & $217$ & $-7.56$ & $-0.819$ \\
\rule{0pt}{1.25em}
$100\,\text{TeV}$ & $92.9$ & $1.59$ & $12.8$ & $0.090$ & $5.20$ & $358$ & $-7.66$ & $-0.681$ \\[0.4cm]
 $\sqrt{s}$ & $A_8$ & $A_9$ & $A_{10}$ & $A_{11}$ & $A_{12}$ & $A_{13}$ & $A_{14}$ & $A_{15}$\\
\hline
\rule{0pt}{1.25em}
 $14\,\text{TeV}$  & $1.95$ & $10.9$ & $51.6$ & $-3.86$ & $-12.5$ & $1.46$ & $5.49$ & $58.4$\\
\rule{0pt}{1.25em}
$100\,\text{TeV}$ & $1.83$ & $9.25$ & $51.2$ & $-2.61$ & $-7.35$ & $1.03$ & $4.65$ & $65.5$
\end{tabular}
\caption{Coefficients of the fit of the total signal rate ($r = \sigma \times BR(hh\to b\bar b\gamma\gamma)$) obtained after all cuts at 
$14\,$TeV and $100\,$TeV. The fit is based on a parametrization analog to that given in Eq.~(\ref{eq:tot_xsec}) for the cross section.
The SM rates $r_{SM}$ include the NLO k-factors of Eq.~(\ref{eq:kfactors}).
}
\label{tab:tot_rate_coefficients}
\end{table}
%

\begin{figure}
\begin{center}
\includegraphics[width=0.45\linewidth]{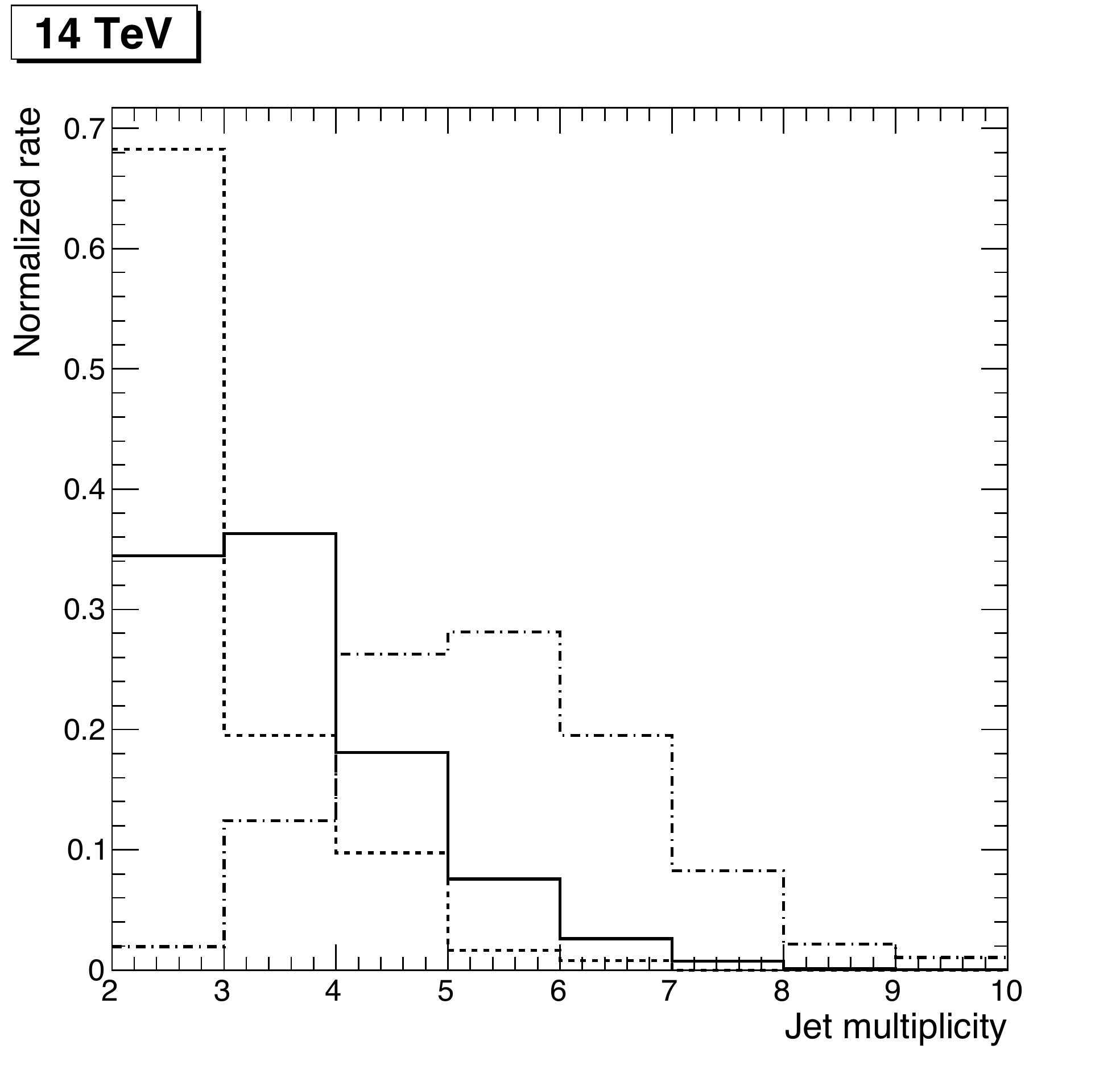}\quad
\includegraphics[width=0.45\linewidth]{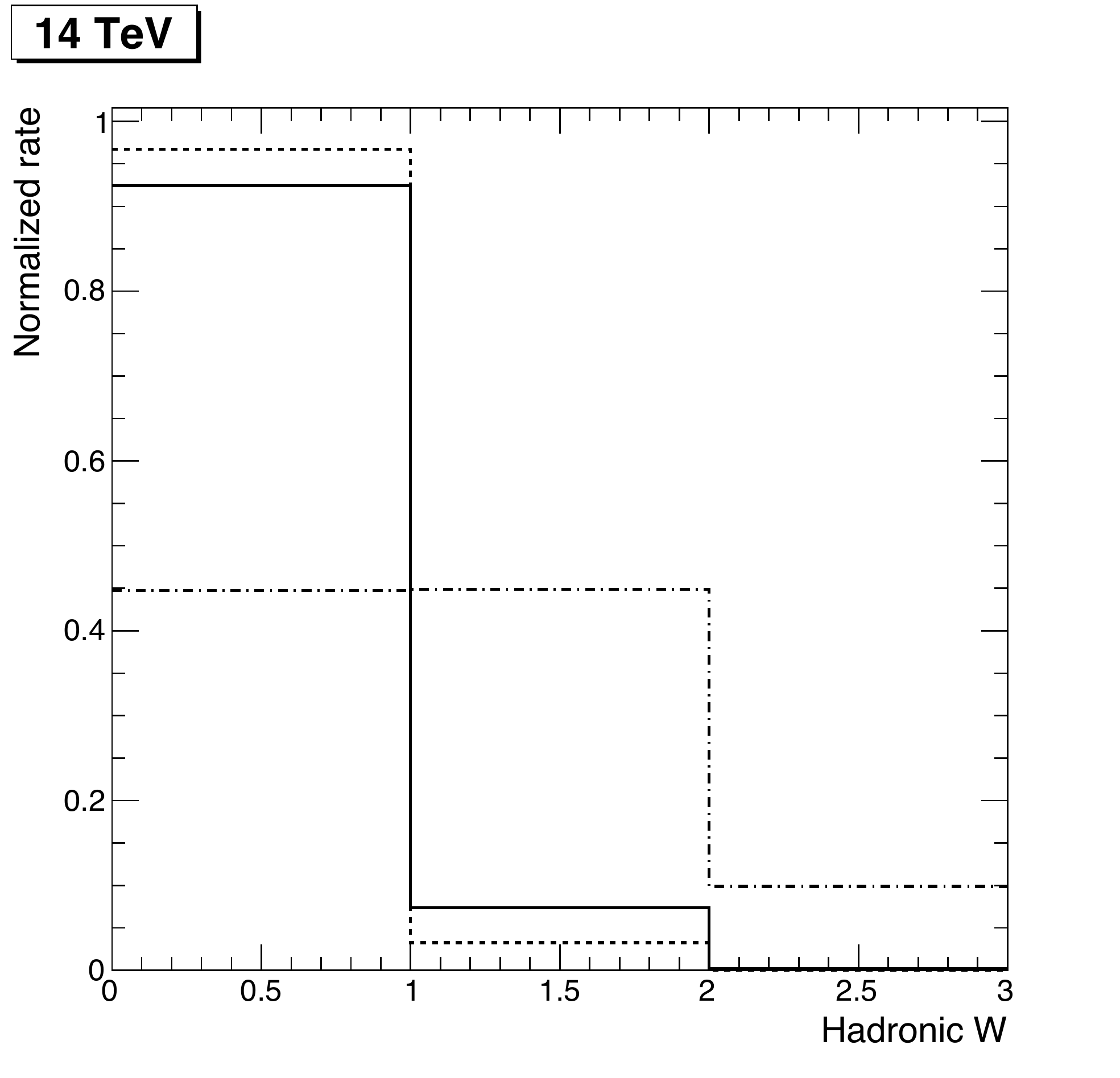}
\caption{Multiplicity of jets (left plot) and of reconstructed hadronic $W$ bosons (right plot) after all cuts at $\sqrt{s} = 14\,$TeV.
The solid, dotted and dot-dashed curves denote respectively the 
SM signal, $\gamma\gamma b\bar{b}$ and $t\bar{t}h$.}
\label{fig:Multiplicity14TeV}
\end{center}
\end{figure}

With our analysis strategy, $\gamma\gamma b\bar{b}$ and $t\bar{t}h$ are the two major backgrounds. The latter tends
to produce extra  jets from the hadronic decay of the $W$'s.
One could thus consider applying a veto on the extra hadronic activity to enhance the signal significance.
The potential impact of a jet veto can be seen in Fig.~\ref{fig:Multiplicity14TeV}.
%
%
For example, further restricting the events to the region with $N(jets) < 4$, in addition
to all the previous cuts, can remove roughly 80\% of the $t\bar{t}h$ background while keeping $\sim$ 70\% of the signal.
Alternatively, one could look for hadronic $W$'s by iteratively forming jet pairs with invariant mass lying in a given window around the $W$ mass,
for example we will use $(70, 100)\,$GeV in the following. 
As illustrated in Fig.~\ref{fig:Multiplicity14TeV}, vetoing hadronic $W$'s in addition to our cuts can reduce $\sim$ 50\% of the $t\bar{t}h$
background while retaining more than 90\% of the signal. We find that  applying either of these cuts leads to a modest
increase in the significance, at the cost of reducing the number of signal events. We thus decided not to exploit any form of extra-jet vetoing at $14\,$TeV,
motivated by the necessity of retaining as many signal events as possible.

As discussed in Section~\ref{sec:crosssection}, several diagrams with different energy scalings contribute to the signal.
We can thus use the  invariant mass of the reconstructed $hh$ system to differentiate the various effects
and improve the sensitivity on the Higgs couplings. To this purpose the events are subdivided into six different categories
in $m^\text{reco}_{hh}$. The corresponding numbers are reported in Table~\ref{tab:mhh:14TeV}  and will be used in section~\ref{sec:results} to extract
our bounds on the coefficients of the effective operators.
%
\begin{table}[tbp]
\centering
\begin{tabular}{c|cccccc}
$m^{\rm reco}_{hh}$  [GeV]   & $250-400$ & $400-550$ & $550-700$ & $700-850$ & $850-1000$ & $1000-$\\[0.07cm] 
\hline \hline &&&&&& \\[-0.47cm]
$hh$        & 2.14 & 6.34 & 2.86 & 0.99 & 0.33 & 0.17 \\[0.05cm]
 \hline &&&&&& \\[-0.47cm]
$\gamma\gamma b\bar{b}$              & 7.69 & 10.1 & 3.35  & 1.38 & 1.18 & 0.59 \\[0.05cm]
$\gamma\gamma jj$                    & 0.66 & 0.95 & 0.31 & 0.16 & 0.08 & 0.045 \\[0.05cm]
$t\bar{t}h$  & 3.33 & 4.53 & 1.41  & 0.41   & 0.16   & 0.043  \\[0.05cm]
$b\bar{b}h$  & 0.20 & 0.16 & 0.03  & 0.0054 & 0.0022 & 0.00054 \\[0.05cm]
$Zh$        & 0.13 & 0.19 & 0.067 & 0.021  & 0.009  & 0.0009
\end{tabular}
\caption{Expected number of events after all cuts at $\sqrt{s} = 14\,$TeV in each of the six $m^\text{reco}_{hh}$ categories considered, assuming an integrated
luminosity $L =3000\,$fb$^{-1}$. The last category is inclusive.}
\label{tab:mhh:14TeV}
\end{table}

\begin{figure}
\begin{center}
\includegraphics[width=0.45\linewidth]{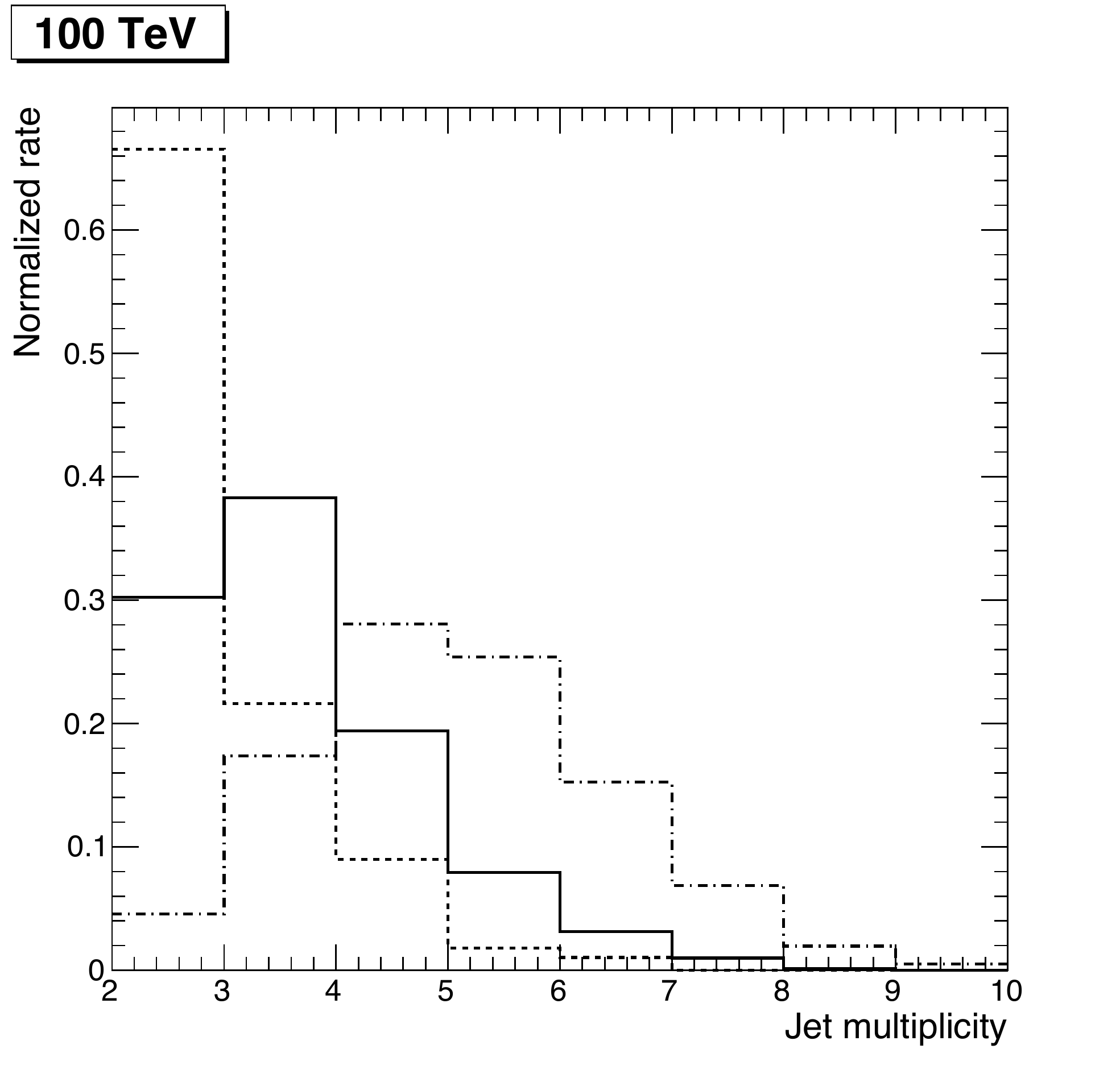}
\includegraphics[width=0.45\linewidth]{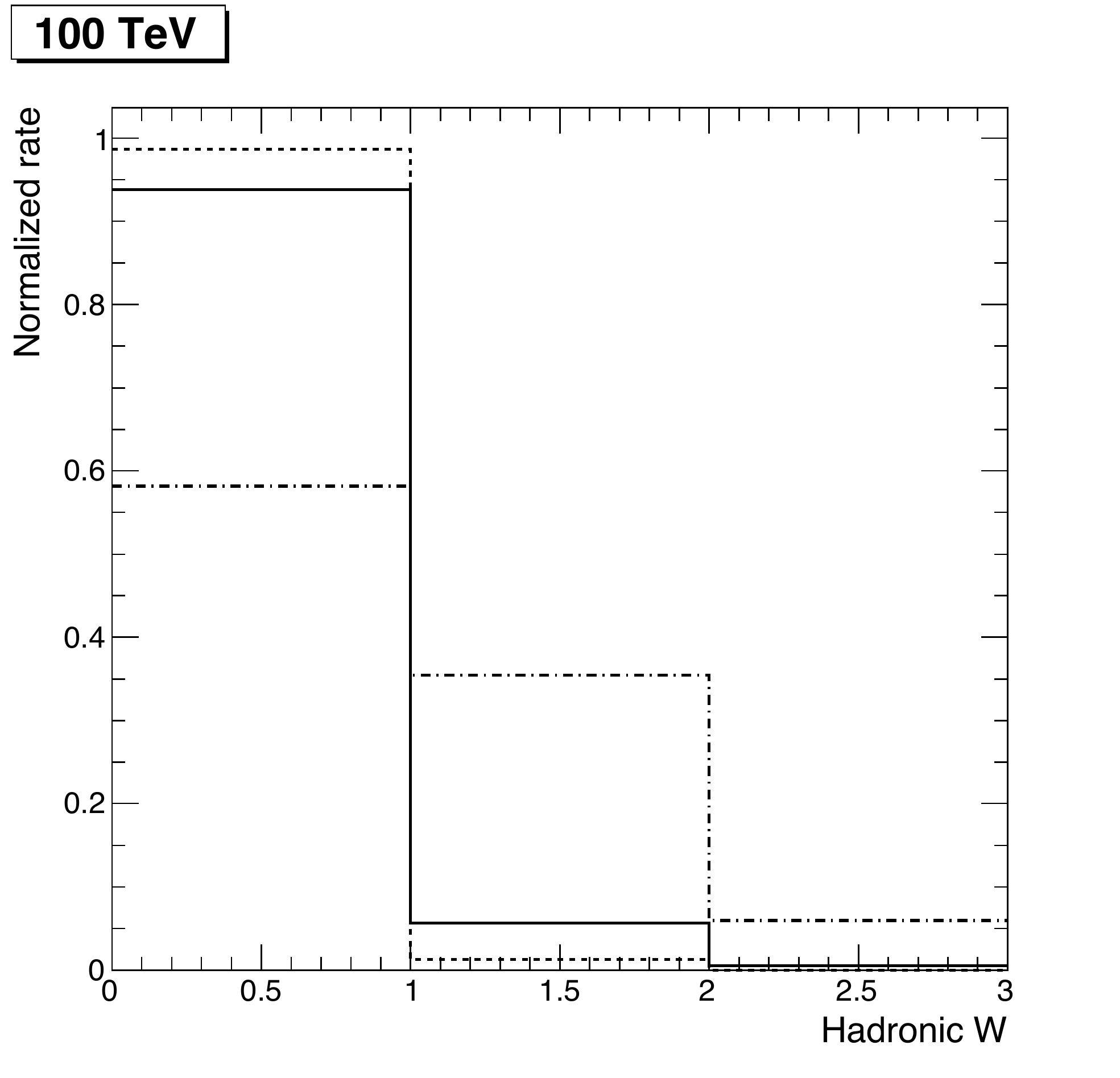}
\caption{Multiplicity of jets (left plot) and of reconstructed hadronic $W$ bosons (right plot) after all cuts at $\sqrt{s} = 100\,$TeV.
The solid, dotted and dot-dashed curves denote respectively the 
SM signal, $\gamma\gamma b\bar{b}$ and $t\bar{t}h$.}
\label{fig:Multiplicity100TeV}
\end{center}
\end{figure}

Let us now discuss the case of a $100\,$TeV collider.  We start by considering a strategy similar to the one adopted at $14\,$TeV,
and differing from this one 
 mostly for the numerical values of the cuts and selection parameters.
The nominal $p_T$ threshold on the reconstructed jets is increased to $p_T(j) > 35\,$GeV to take into account the
busier environment of the higher energy collision. We define the same pseudorapidity region for the acceptance of jets, photons, and leptons
although more signal events fall into the high-$|\eta|$ region due to the boost along the beam axes (this point will be discussed in
detail in section~\ref{sec:boostedanalysis}).
Similarly, in our initial selection of events we increase the $p_T$ cuts on the $b$-tagged jets and photons while keeping the mass windows to be the same:
\begin{equation}
\label{eq:acceptanceI:100TeV}
\begin{gathered}
p_{T>}(b),p_{T>}(\gamma) > 60\ {\rm GeV}\,, \quad p_{T<}(b), p_{T>}(\gamma) > 40\ {\rm GeV}\,,\\[0.1cm]
60 < m^{\rm reco}_{b\bar{b}} < 200\ {\rm GeV}\,,\quad 60 < m^{\rm reco}_{\gamma\gamma} < 200\ {\rm GeV}\,.
\end{gathered}
\end{equation}
After the above cuts we find angular distributions similar to those shown in Fig.~\ref{fig:DRdist:14TeV}, and we thus apply the same cuts 
as in Eq.~(\ref{eq:acceptanceII}) and Eq.~(\ref{eq:masswindow}).
The cut flow at $\sqrt{s} =100\,$TeV is reported in Table~\ref{tab:cutflow:100TeV}.
%
\begin{table}[tbp]
\centering
\begin{tabular}{l|c|ccccc}
\hline &&&&&& \\[-0.45cm]
$\sqrt{s} = 100$ TeV & $hh$ & $\gamma\gamma b\bar{b}$ & $\gamma\gamma jj$ & $t\bar{t}h$  &  $b\bar{b}h$   &  $Zh$ \\[0.08cm]
\hline \hline &&&&&& \\[-0.42cm]
After selection cuts of Eq.~\ref{eq:acceptanceI:100TeV}                           & 500 & 22319 & 2469 & 2694 & 52  & 102 \\[0.08cm]
After $\Delta R$ cuts of Eq.~\ref{eq:acceptanceII}                                & 390 & 4819  & 626  & 836  & 17  & 86   \\[0.08cm]
After $m^{\rm reco}_{b\bar{b}, \, \gamma\gamma}$ cuts of  Eq.~\ref{eq:masswindow} & 303 & 137   & 18.2 & 303  & 6.2 & 3.2  \\[0.08cm]
After hadronic $W$ vetoing                                                        & 279 & 135   & 15.5 & 179  & 5.9 & 3.2 \\[0.16cm]  
\hline
\end{tabular}
\caption{Cut flow for the SM signal and the various backgrounds at $\sqrt{s} = 100\,$TeV. 
The values correspond to the number of events, assuming an integrated luminosity $L=3000\,\text{fb}^{-1}$.
All reconstruction efficiencies and branching ratios to the final state under study are included.
}
\label{tab:cutflow:100TeV}
\end{table}
%
Differently from the $14\,$TeV case, at this level the dominant background is $t\bar th$, rather than $b\bar b\gamma\gamma$,
due to its steeper increase with the collider energy. 
Imposing a veto on  extra hadronic activity is beneficial  to reduce this background without strongly affecting the signal, as
illustrated by the distributions of Fig.~\ref{fig:Multiplicity100TeV}.
%
%
We find that a veto on hadronic $W$'s is most efficient to maximize the signal significance, and thus apply it.
The final number of signal and background events after the $W$ veto is reported in the last row of Table~\ref{tab:cutflow:100TeV},
while a fit to the total signal rate after all cuts is given  in Table~\ref{tab:tot_rate_coefficients}.
As a final step, we subdivide events into six $m^\text{reco}_{hh}$ categories and report the corresponding numbers in Table~\ref{tab:mhh:100TeV}.
%
\begin{table}[tbp]
\centering
\scalebox{0.94}{
\begin{tabular}{c|cccccc}
$m^{\rm reco}_{hh}$  [GeV]   & $250-400$ & $400-550$ & $550-700$ & $700-850$ & $850-1000$ & $1000-$\\[0.07cm] 
\hline \hline &&&&&& \\[-0.47cm]
$hh$                    & 27.1 (29.4) & 116.3 (125.8) & 74.5 (81.0) & 35.3 (38.5) & 15.9 (17.6) & 9.7 (10.8) \\[0.1cm]
\hline &&&&&& \\[-0.47cm]
$\gamma\gamma b\bar{b}$ & 56.1 (57.5) &  50.5 (51.3) & 13.1 (13.1) & 8.37 (8.38) & 3.63 (3.90) & 2.79 (3.07) \\[0.1cm]
$\gamma\gamma jj$       & 4.23 (4.96) & 6.14 (7.38) & 2.54 (2.87) & 1.34 (1.46)  & 0.46 (0.63) &  0.78 (0.90) \\[0.1cm]
$t\bar{t}h$             &  53.7 (91.3) &  77.3 (130.3) & 30.1 (51.7) & 11.5 (18.7) & 4.37 (6.99) & 2.49 (3.58) \\[0.1cm]
$b\bar{b}h$             & 2.41 (2.47) & 2.65 (2.74)  & 0.61 (0.649) & 0.18 (0.197) & 0.062 (0.065) & 0.021 (0.026)\\[0.1cm]
$Zh$                    &  0.70 (0.70) & 1.31 (1.32)  & 0.67 (0.674)&  0.34 (0.353)& 0.10 (0.10) & 0.031 (0.031)
\end{tabular} 
}
\vspace{0.01cm}
\caption{Expected number of events after all cuts at $\sqrt{s} = 100\,$TeV in each of the six $m^\text{reco}_{hh}$ categories considered, assuming an integrated
luminosity $L =3000\,$fb$^{-1}$. The last category is inclusive.
The numbers in the parenthesis are obtained by removing the veto on hadronic $W$'s.}
\label{tab:mhh:100TeV}
\end{table}

\subsection{Recovering the boosted topologies}
\label{sec:boostedanalysis}
As we anticipated in the previous discussion, a possible issue with our analysis strategy is the loss of sensitivity
for the kinematical configurations containing boosted Higgses. Although this effect is not likely to be relevant for the
$14\,$TeV LHC, it can potentially affect the $gg \rightarrow hh$ searches at a future higher-energy collider.
In this section we present a first estimate of the improvement that can be achieved in the reconstruction of highly energetic
events by the use of jet substructure techniques.
Notice that the development of a fully optimized  analysis will most likely require
a hybrid  strategy that smoothly interpolates between a traditional jet analysis and one using boosted techniques~\cite{Gouzevitch:2013qca}.
Devising such a strategy is beyond the scope of the present work and we  postpone it to a future study.
In this section we instead adopt a ``minimal'' approach and use either the standard analysis or the jet substructure technique
in each $m^\text{reco}_{hh}$ bin.

\begin{figure}
\begin{center}
\includegraphics[width=0.45\linewidth]{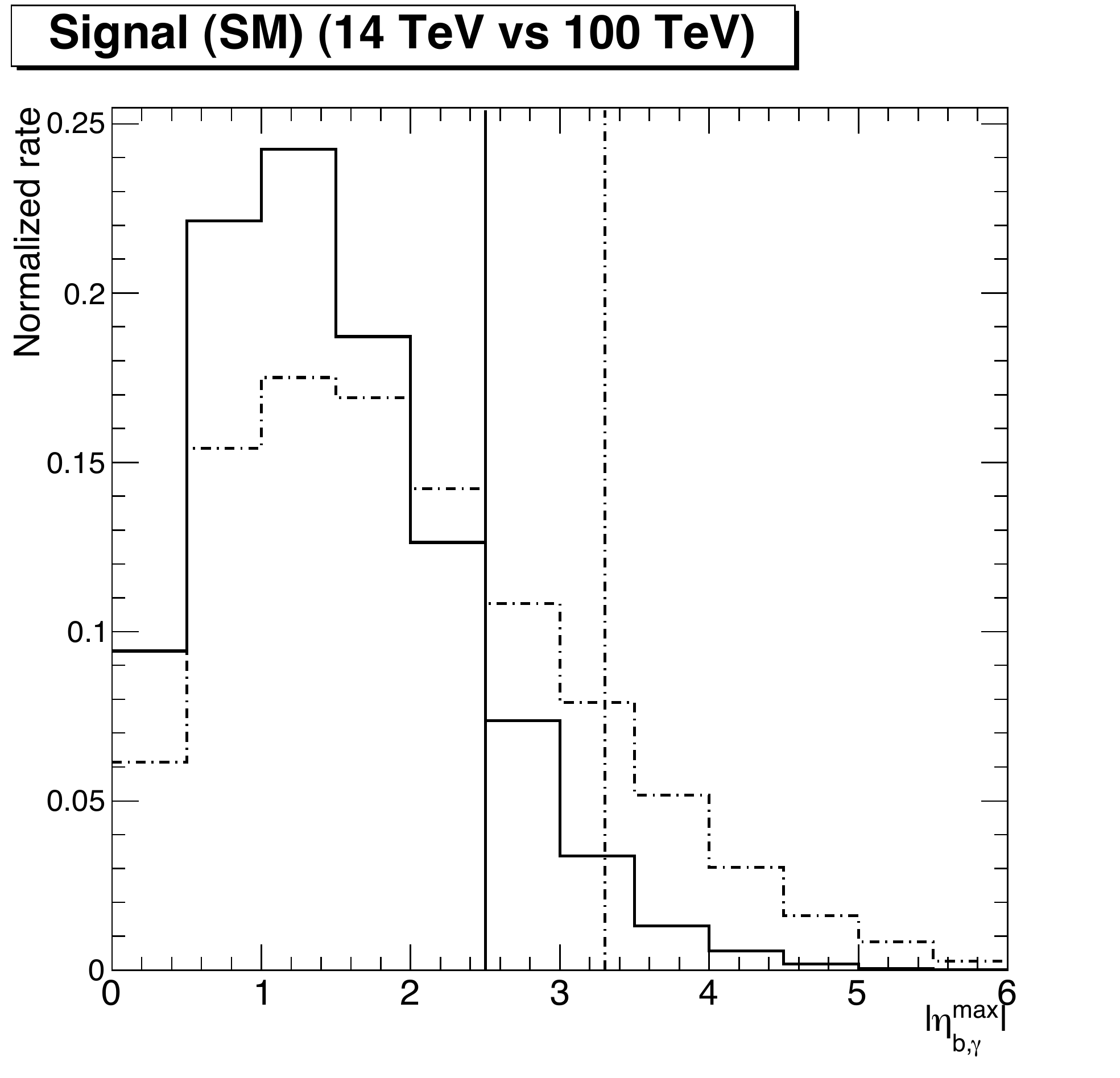}\quad
\includegraphics[width=0.45\linewidth]{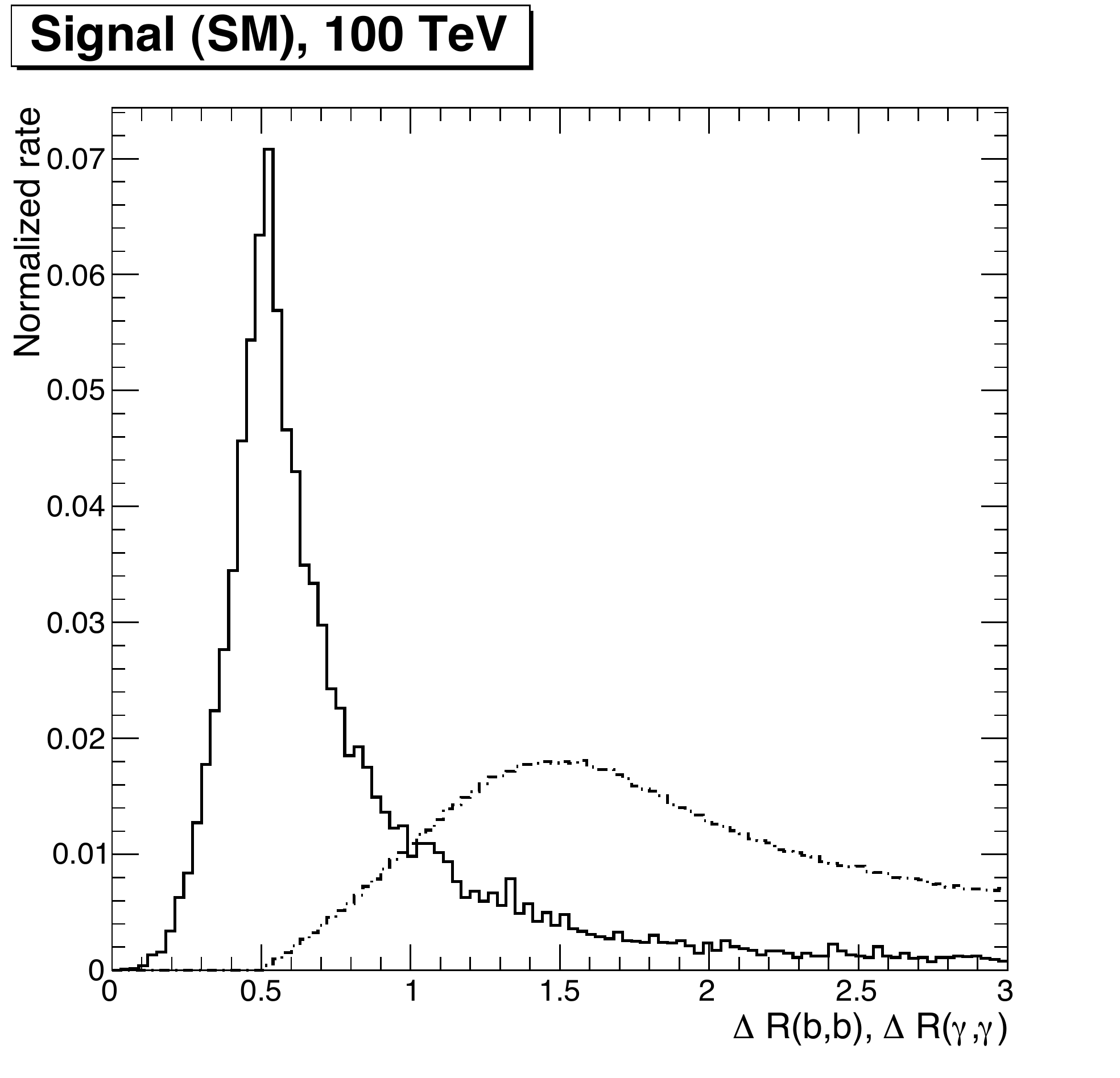}
\caption{Left: pseudorapidity distribution of the most forward bottom quark or photon in the SM $gg\rightarrow hh \to b\bar{b}\gamma\gamma$ process 
at $14\,$TeV (solid curve) and $100\,$TeV (dot-dashed curve). The vertical solid line corresponds to the acceptance region at the LHC. The dot-dashed vertical line denotes 
the acceptance region which should be adopted at $100\,$TeV to retain the same fraction of events of the $14\,$TeV case.
Right: $\Delta R$ distance between two bottom quarks or photons   in the SM $gg\rightarrow hh \to b\bar{b}\gamma\gamma$ process at $100\,$TeV.
The solid (dot-dashed) curve corresponds to the events with $m_{hh} >1000\,$GeV  ($m_{hh}<1000\,$GeV). 
Both plots are made at the partonic level.
}
\label{fig:etaDRbbdist}
\end{center}
\end{figure}

There are basically two different effects that lead to boosted Higgses, namely 
the boost of the whole $hh$ system along the beam axis and the production of a Higgs with high $p_T$.
Both effects will become relevant at a future high-energy collider.
At $100\,$TeV the di-Higgs system acquires on average a non-negligible amount of momentum along the beam axis
and, as a result, the events more frequently leak into the high-$|\eta|$ region. The situation is illustrated by the left plot of
Fig.~\ref{fig:etaDRbbdist}, which suggests that a larger pseudorapidity range for object reconstruction and flavor tagging
could improve the sensitivity to the $gg \rightarrow hh$ process.
%
%
For instance, at the $14\,$TeV LHC roughly $13\%$ of the partonic signal lays outside the acceptance region $|\eta| < 2.5$
(solid vertical line in the left panel of Fig.~\ref{fig:etaDRbbdist}).
If one assumes the same acceptance region, this fraction increases to $\sim$ 30\% at a $100\,$TeV collider.
In order to obtain the same acceptance efficiency of the $14\,$TeV LHC,  the coverage region should be extended to $|\eta| < 3.3$ at the higher-energy collider
(vertical dot-dashed line in the left panel of Fig.~\ref{fig:etaDRbbdist}).~\footnote{After imposing the cuts in Eqs.~(\ref{eq:acceptanceI}) and (\ref{eq:acceptanceII})
on  partonic signal events, these fractions reduce to $\sim 6.5$\% for the $14\,$TeV LHC
and $\sim 24$\% for the $100\,$TeV case. To recover the $14\,$TeV acceptance efficiency at the higher-energy collider
the pseudorapidity region should be extended to $|\eta| < 3.6$.}

A further problem with boosted events is that our analysis relying on a dijet-style search fails when the two $b$-jets are not well separated.
A similar argument can be applied to the di-photon system. 
The boosted Higgses are more likely to be produced in the high invariant mass tail of the $m_{hh}$ distribution.
This effect is illustrated in the right panel of Fig.~\ref{fig:etaDRbbdist}, where we plot the
$\Delta R$ separation between two $b$'s or photons at the partonic level 
for the two categories of events with $m_{hh} > 1000$ GeV and $m_{hh} < 1000$ GeV.
Most events belonging to the lower $m_{hh}$ category have a clear two-prong topology,
whereas a significant amount of  events in the higher $m_{hh}$ category fail to be resolved as two well separated partons.

For our present purposes it will be sufficient to perform a simple analysis on the $b\bar{b}$ subsystem
and compare the performance of our traditional jet-based analysis with the
substructure method.
For the $\gamma\gamma$ subsystem we consider two possibilities: a reduced isolation cone size ($R_{iso}=0.2$) and an
ad-hoc mutual photon isolation (a somewhat similar prescription was proposed in~\cite{Katz:2010iq} for a boosted di-tau system).
In the latter,  photon isolation is imposed neglecting the other photon in the $\gamma\gamma$ system and an event is retained
as long as $\Delta R(\gamma,\gamma)>0.2$. This procedure is quite similar to our traditional jet-based analysis up to the slight
modification of the photon isolation criterion. 
We do not modify instead the isolation criterion on leptons.~\footnote{While naively one would expect that leptons from the boosted tops in the resonant $t\bar{t}h$ 
background fail more frequently  the isolation requirement, we find that the efficiency drop is not significant for the $m_{hh}$ range of interest.}

After this step, the events are clustered into $R=1.5$ ``fat-jets'' with the Cambridge/Aachen (C/A)
jet algorithm~\cite{Dokshitzer:1997in,Wobisch:1998wt}.
We iteratively look over those fat-jets and apply the BDRS subjet-finding technique
of Ref.~\cite{Butterworth:2008iy} (with the same declustering parameters as in~\cite{Butterworth:2008iy}). To ensure enough boost we only look into
fat-jets with $p_T (j) > 150\,$GeV. The declustering algorithm stops when it successfully identifies three
subjets (the third subjet takes into account the leading gluon emission from either bottom quark line).
The fat-jet is identified as a Higgs-jet if the invariant mass of the three subjets falls into the Higgs mass window
$m^{\rm reco}_{b\bar{b}}=$ [105,145] GeV (see Eq.~(\ref{eq:masswindow})). If multiple candidates exist, we pick the
one whose invariant mass is closest to the Higgs mass. Two $b$-taggings are performed
at subjet-level assuming a $70\%$ $b$-tagging efficiency. The cuts on the angular separations in Eq.~(\ref{eq:acceptanceII})
are applied to the subjets and the photons. 
Finally,  we do not apply any veto on extra hadronic activity or on hadronic $W$ bosons.

\begin{figure}
\begin{center}
\includegraphics[width=0.45\linewidth]{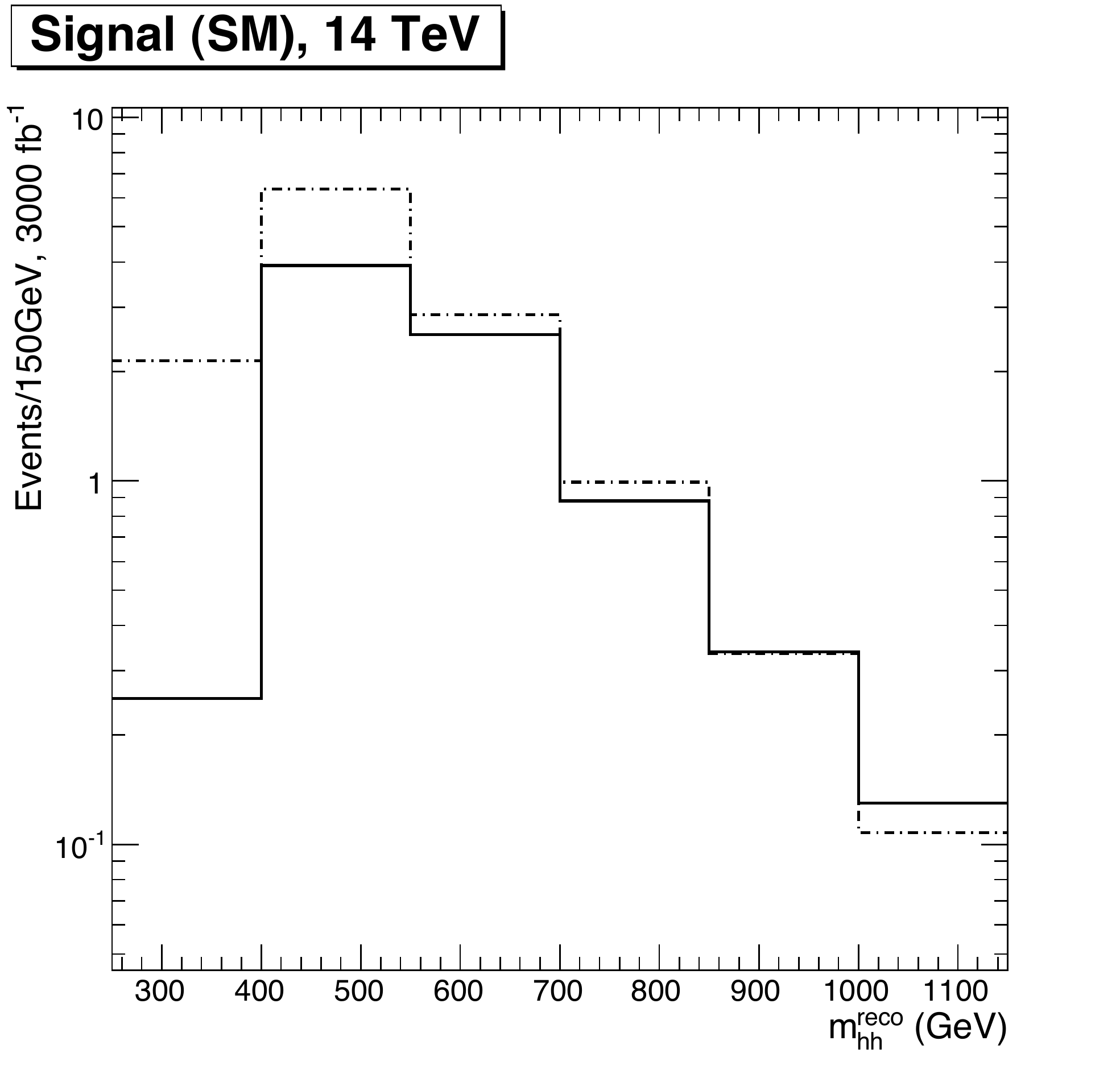}\quad
\includegraphics[width=0.45\linewidth]{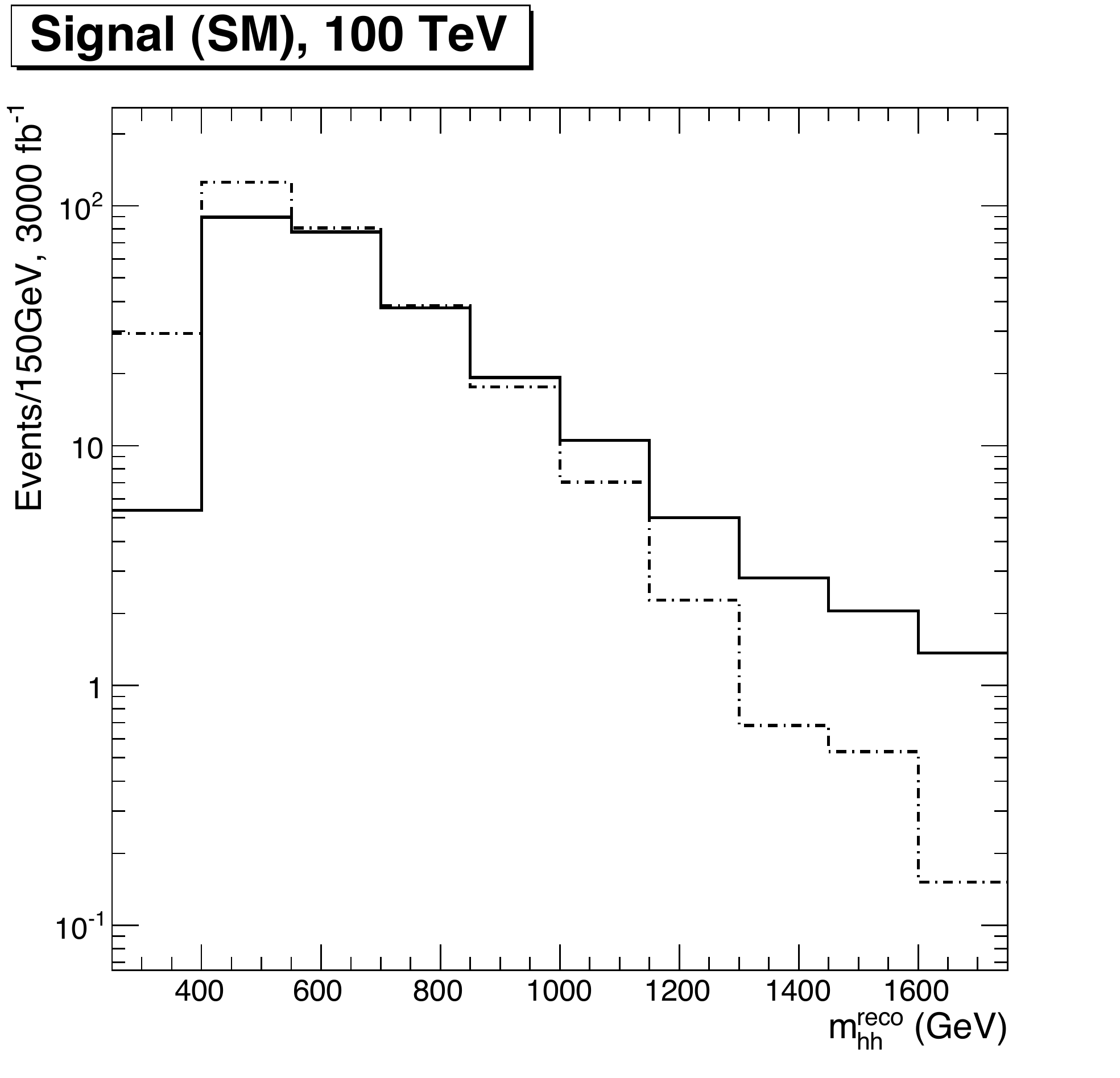}
\caption{Number of signal events as a function of $m^{\rm reco}_{hh}$ after all cuts at $14\,$TeV (left plot) and $100\,$TeV (right plot). 
The solid line is obtained by adopting the substructure technique with  reduced photon isolation cone size (scenario ``Substructure II''
described in the text), whereas the dot-dashed line corresponds to a traditional jet-based analysis.}
\label{fig:mhh:tradjetVSjetsub}
\end{center}
\end{figure}

The performances of the substructure analysis on signal events are summarized in Table~\ref{tab:mhh:100TeV:tradVSjetsub} and compared
to the traditional  jet-based analysis.
%
\begin{table}[tbp]
\centering
\setlength{\tabcolsep}{4.5pt}
\scalebox{0.76}{
\begin{tabular}{l|cccccccccc}
\multicolumn{1}{l}{} & \multicolumn{10}{l}{Signal event rate at $\sqrt{s} =14\,$TeV, assuming $L=3\,\text{ab}^{-1}$}\\[0.07cm]
\cline{1-7} &&&&&&&&&& \\[-0.44cm]
$m^{\rm reco}_{hh}$  [GeV]   & $250\!-\!400$ & $400\!-\!550$ & $550\!-\!700$ & $700\!-\!850$ & $850\!-\!1000$ & $1000\!-$ & & & & \\[0.07cm]
\cline{1-7} \cline{1-7}
Trad. anal.  & 2.14 & 6.34 & 2.86 & 0.99 & 0.33 & 0.17 & & & &  \\[0.05cm]
\cline{1-7} &&&&&&&&&& \\[-0.44cm]
Substr. I     & 0.21 & 3.53  & 2.31 & 0.81 & 0.31 & 0.22 & & & & \\[0.05cm]
Substr. II    & 0.25 & 3.91  & 2.52 & 0.88 & 0.34 & 0.24 & & & & \\[0.05cm]
Substr. III   & 0.21 & 3.53  & 2.31 & 0.81 & 0.32 & 0.23 &  &  &  &  \\[0.05cm]
\cline{1-7}
\multicolumn{11}{c}{}\\
\multicolumn{1}{l}{} & \multicolumn{10}{l}{Signal event rate at $\sqrt{s} =100\,$TeV, assuming $L=3\,\text{ab}^{-1}$}\\[0.07cm] 
\hline &&&&&&&&&& \\[-0.44cm]
$m^{\rm reco}_{hh}$  [GeV]   & $250\!-\!400$ & $400\!-\!550$ & $550\!-\!700$ & $700\!-\!850$ & $850\!-\!1000$ & $1000\!-\!1150$ & $1150\!-\!1300$ 
                                         &  $1300\!-\!1450$ & $1450\!-\!1600$ & $1600\!-$\\[0.07cm] 
\hline \hline &&&&&&&&&& \\[-0.44cm]
Trad. anal. & 29.3 & 125.6 & 80.8 & 38.4 & 17.6 & 7.04 & 2.27 & 0.68 & 0.52 & 0.23 \\[0.05cm]
 \hline &&&&&&&&&& \\[-0.44cm]
Substr. I    & 4.1  & 70.6  & 64.7 & 32.1 & 16.6 & 9.64 & 4.02 & 1.90 & 1.21 & 0.99 \\[0.05cm]
Substr. II   & 5.39 & 89.8  & 78.1 & 37.5 & 19.3 & 10.5 & 5.0 & 2.81 & 2.05 & 4.7 \\[0.05cm]
Substr. III  & 4.1  & 70.6  & 64.7 & 32.5 & 17.0 & 9.71 & 4.40 & 2.50 & 1.82 & 4.02 \\[0.05cm]
 \hline
\end{tabular}}
\vspace{0.25cm}
\caption{Comparison between the traditional  jet-based analysis (second line) and the substructure analysis 
(third, fourth and fifth lines) on SM signal events. 
The three scenarios ``Substructure I, II, III'' differ by the treatment of $\gamma\gamma$ system and are defined in the text.
}
\label{tab:mhh:100TeV:tradVSjetsub}
\end{table}
%
We consider three possible scenarios which differ by the treatment of the $\gamma\gamma$ system:  ``Substructure I'' uses a standard photon isolation criterion with 
$R_{iso}=0.4$,  whereas in ``Substructure II'' the cone size is reduced to $R_{iso}=0.2$. ``Substructure III'' corresponds instead to the mutual isolation criterion.
Not surprisingly,  the substructure analysis becomes more efficient compared to the traditional one for $m^{\rm reco}_{hh} \gtrsim 1000\,$GeV; this matches our naive expectation
based on  the $\Delta R(b,b)$ and $\Delta R(\gamma, \gamma)$ distributions of Fig.~\ref{fig:etaDRbbdist}.
For $m^{\rm reco}_{hh} \lesssim 1000\,$GeV, on the other hand, only part of the events are correctly reconstructed using the substructure technique (due to the
finite fat-jet size $R=1.5$) and the traditional analysis is more efficient.
This is also clearly illustrated by the plots of Fig.~\ref{fig:mhh:tradjetVSjetsub}, which show the $m^{\rm reco}_{hh}$ distributions for signal events with both strategies.
%
%
These results suggest that the use of  jet substructure in the $b\bar b\gamma\gamma$ final state is crucial only 
at a $100\,$TeV collider, while it has a minor impact at the $14\,$TeV LHC, where large values of $m_{hh}$ are suppressed by the fast drop of the gluon luminosity.
The relevance of jet substructure can be determined in a more quantitative way by comparing 
the reach in $m^{\rm reco}_{hh}$ at $14\,$TeV and $100\,$TeV. 
At the partonic level, Fig.~\ref{fig:dsigmaOfxx} shows that the same number of events one has at $14\,$TeV after imposing the cut $m_{hh}>850\,$GeV is obtained
at $100\,$TeV for $m_{hh}>2570\,$GeV.
At the hadronic level,  the increase in the reach is much more modest than the naive expectation when adopting a traditional jet-based analysis. 
For instance, the signal rate at 14 TeV is $N_{events} \sim 0.5$ for $m^{\rm reco}_{hh} > 850$ GeV. The same number of events is obtained at $100\,$TeV
for $m^{\rm reco}_{hh} > 1540\,$GeV using a traditional analysis, and $m^{\rm reco}_{hh} > 2560\,$GeV using jet substructure with the reduced photon isolation cone 
$R_{iso}=0.2$.  Hence, a jet substructure analysis (along with the appropriate modification of the photon isolation criteria) is essential to get close to the naive expectation.

Focusing on the $100\,$TeV case, we show in Table~\ref{tab:mhh:100TeV:jetsub} the number of signal and background events expected with 
the substructure analysis and a reduced photon isolation cone size (scenario ``Substructure II'') in six categories with high $m^\text{reco}_{hh}$. 
%
\begin{table}[tbp]
\centering
\scalebox{1.}{
\begin{tabular}{c|cccccc} \hline &&&&&& \\[-0.45cm]
$m^{\rm reco}_{hh}$  [GeV]   & $850-1000$ & $1000-1200$ & $1200-1400$ & $1400-1600$ & $1600-1800$ & $1800-$\\[0.05cm] 
\hline \hline &&&&&& \\[-0.45cm]
$hh$                         & 19.3 & 12.5 & 4.86 & 3.03 & 1.75  & 2.96  \\[0.05cm] 
\hline &&&&&& \\[-0.45cm]
$\gamma\gamma b\bar{b}$      & 3.92 & 2.35 & 1.18 & 0.59 & 0.29  & 0.098 \\[0.05cm]
$t\bar{t}h$                  & 8.63 & 4.91 & 2.32 & 1.16 & 0.51  & 1.1 \\[0.05cm] \hline
\end{tabular}}
\caption{Expected numbers of  signal and background events at $\sqrt{s} = 100\,$TeV after the jet substructure analysis 
(scenario ``Substructure II'' described in the text with photon isolation 
cone $R_{iso}=0.2$) assuming an integrated luminosity $L=3000\,$fb$^{-1}$. The last category is inclusive.}
\label{tab:mhh:100TeV:jetsub}
\end{table}
%
We include only the dominant $b\bar b\gamma\gamma$ and $t\bar th$ backgrounds for simplicity.~\footnote{We have checked that $Zh$ and $b\bar b h$ 
are negligible, as the corresponding number of events in each category is always smaller than 0.1. We expect that also the background $\gamma\gamma jj$ 
can be reduced to a subdominant level as long as the efficiency for  making two $b$-tags at the subjet level is sufficiently high.
The  $\gamma\gamma b\bar{b}$ background  was newly generated for the substructure analysis with relaxed generation cuts 
$\Delta R (b, \bar{b})>0.1$, $\Delta R (\gamma,\gamma)>0.15$; these are less stringent than those specified in Appendix~\ref{eq:gencutsLOaabb:14TeV}. }  
A further reduction of the $t\bar th$ background could in principle be possible by iteratively looking for boosted top- and $W$-jets through dedicated tagging algorithms,
and by discarding events where such objects are reconstructed.  This is however beyond the scope of the present work. On the other hand, we have checked that a simple 
veto on hadronic $W$'s reconstructed from pairs of jets, as applied in the traditional analysis, is not efficient on events with such high $m^\text{reco}_{hh}$.
The results in Table~\ref{tab:mhh:100TeV:jetsub} will be used in Section~\ref{sec:results} to assess the impact of jet substructure in setting bounds on the coefficients
of the effective operators.

%% file: Results.tex

The results of the previous section have been  processed through a Bayesian statistical analysis to extract the sensitivity
on the coefficients of the Higgs effective Lagrangian. In each case, a probability distribution of the relevant parameters is obtained
by constructing a likelihood function from the signal and background number of events in the six $m_{hh}$ categories, and  marginalizing
over (or fixing) the remaining parameters. The injected signal is the SM one for all the results presented in this section.
A flat prior is assumed for the coefficients of the Higgs effective Lagrangian, except when
they are constrained by single-Higgs measurements. In the latter case, we use the ATLAS projections for the high-luminosity LHC 
of Ref.~\cite{ATL-PHYS-PUB-2013-014}
to construct a likelihood function and use it as a prior. Reference~\cite{ATL-PHYS-PUB-2013-014}
reports the estimated precision on various Higgs decay channels expected for $\sqrt{s} = 14\,$TeV 
with integrated luminosities $L = 300\,\text{fb}^{-1}$ and $L = 3\,\text{ab}^{-1}$.
We will further study the case of a 
future proton-proton circular collider operating at $\sqrt{s} = 100\,$TeV 
with $3\,\text{ab}^{-1}$ of integrated luminosity. 
When single-Higgs measurements are required to perform marginalization in this latter scenario, we use the ATLAS projections for $L = 3\,\text{ab}^{-1}$
at $\sqrt{s} = 14\,$TeV, lacking a specific estimate of the precision reachable at a $100\,$TeV machine.  
We will thus consider the following three benchmark scenarios:
\begin{center}
\begin{tabular}{llll}
                   & \LHC                                & \HLLHC                       & \FCC \\[0.05cm]
\hline 
&&& \\[-0.37cm]
$\sqrt{s}$   & $14\,$TeV                       & $14\,$TeV                  & $100\,$TeV \\[0.08cm]
Luminosity \ & $L = 300\,\text{fb}^{-1}$ \ \ & $L = 3\,\text{ab}^{-1}$ \ \ & $L = 3\,\text{ab}^{-1}$ 
\end{tabular}
\end{center}
For simplicity, we do not include in our analysis the theoretical uncertainty on the prediction of the signal cross section nor the systematic
uncertainties which will characterize the extraction of the background from data in a realistic experimental  analysis.~\footnote{The theoretical error was estimated 
by the authors of Ref.~\cite{Baglio:2012np} to amount to a $\sim 18\% \, (12\%)$  from scale variation, $\sim 10\% \, (10\%)$ from the use
of EFT in the calculation of the NLO k-factor,  and $\sim 7\% \, (6\%)$ from PDFs at the \LHC (\FCC).
See Ref.~\cite{Goertz:2013kp} for a proposal on how
to reduce the theoretical and experimental uncertainties by considering the ratio of double and single cross sections. \label{foot:therror}}

As a first result, we derive the precision that can be obtained on the signal strength multiplier $\mu = \sigma/\sigma_{SM}$ in the three benchmark
scenarios described above. No marginalization is performed in this simple case.
We obtain the following 68\% intervals on $\mu$:
\begin{center}
\begin{tabular}{r|ccc}
& \LHC & \HLLHC & \FCC \\[0.05cm] 
\hline &&& \\[-0.4cm]
68\% interval on $\mu$ & \ \ $[-0.41 , 3.0]$ \ \ & \ \ $[0.50, 1.6]$ \ \ & \ \ $[0.92, 1.1]$ \ \ \\[0.08cm]
\end{tabular}
\end{center}
Very similar results are obtained from an inclusive analysis without $m_{hh}$ categories, as one naively expects since $\mu$ accounts only for an
overall rescaling of the total cross section of the signal.

\begin{figure}
\begin{center}
\includegraphics[width=0.445\linewidth]{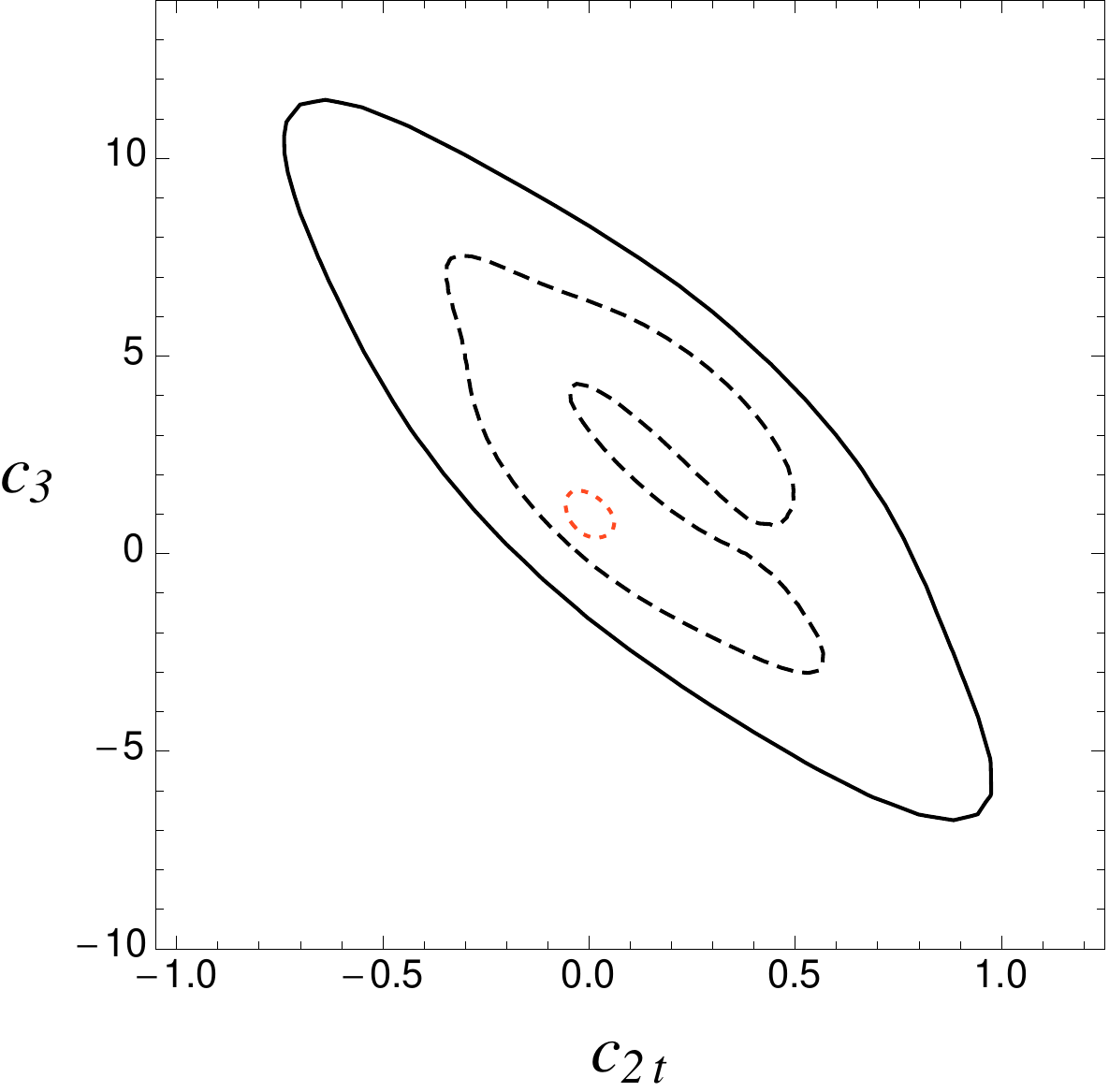}\qquad
\includegraphics[width=0.472\linewidth]{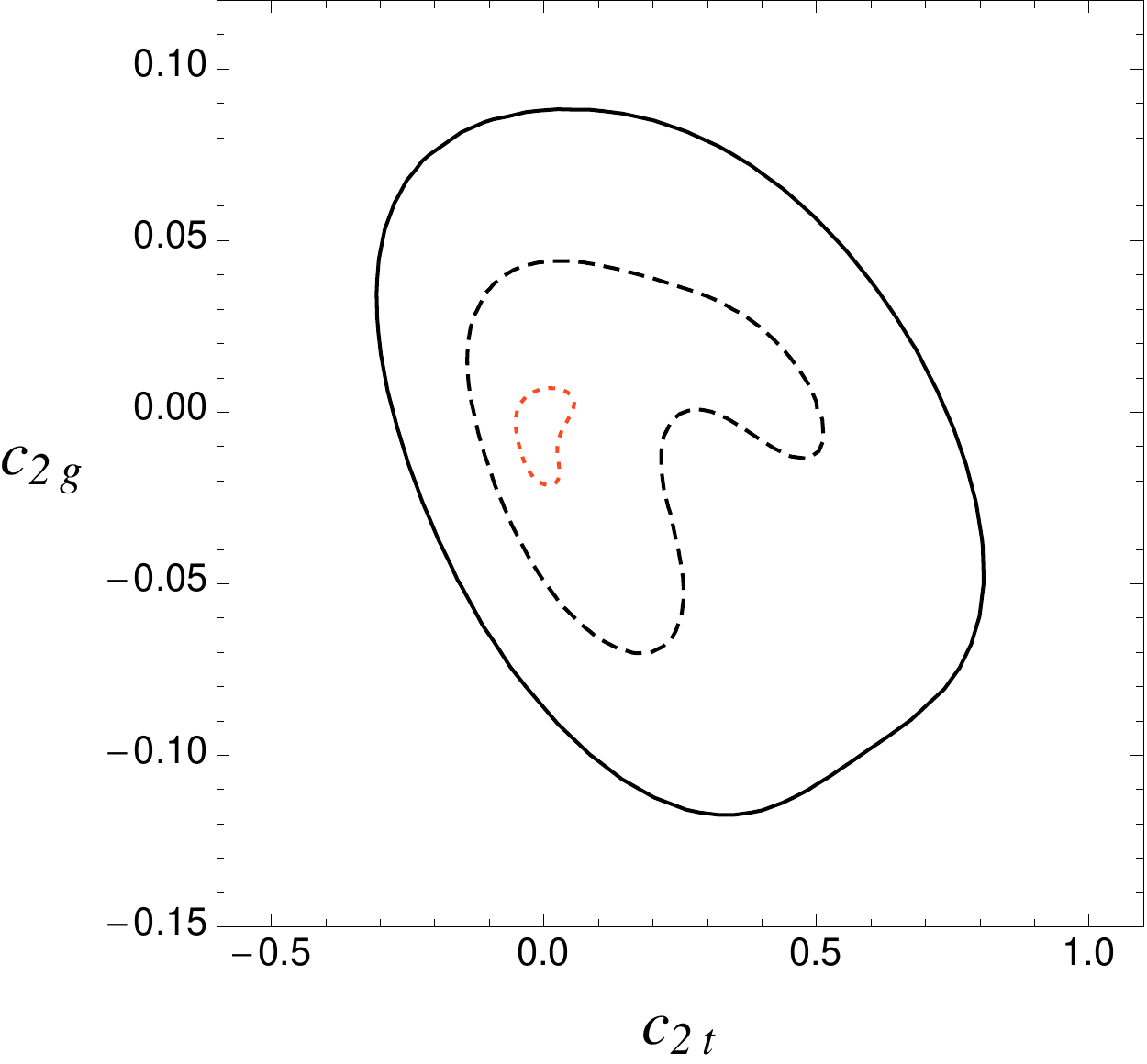}
\caption{$68\%$ probability contours in the planes $(c_{2t}, c_3)$ (left plot) and $(c_{2t}, c_{2g})$ (right plot). The different curves
refer to the three benchmark scenarios: \LHC (black continuous line); \HLLHC (black dashed line);  \FCC  (red dotted line).
}
\label{fig:summary-nonlinear}
\end{center}
\end{figure}

Turning to the determination of the  coefficients of the effective Lagrangian, we first consider the case of the non-linear parametrization of Eq.~(\ref{eq:nonlinearL}).
Figure~\ref{fig:summary-nonlinear} shows the 68\% probability contours in the planes $(c_{2t}, c_3)$ and $(c_{2t}, c_{2g})$ obtained through the procedure described above.
%
%
In this case the marginalization is performed over two parameters: $c_t$, with a prior obtained from single-Higgs measurements, and 
the branching ratio for the decay $hh\to b\bar b\gamma\gamma$.  For the latter parameter we assume a Gaussian distribution around
the SM value with standard deviation equal to 0.15 and 0.10 respectively for $L=300\,\text{fb}^{-1}$ and $L=3\,\text{ab}^{-1}$ (both at $14\,$TeV
and $100\,$TeV). For simplicity, the remaining couplings are set to their SM values: $c_g = c_{2g} =0$ in the plot on the left of Fig.~\ref{fig:summary-nonlinear};
$c_3 = 1$ and $c_{g} = 0$ in the plot on the right. We have checked that performing an additional marginalization over $c_g$ slightly decreases
the precision on the measured couplings, without changing the shape of the contours.

\begin{figure}
\begin{center}
\includegraphics[width=0.428\linewidth]{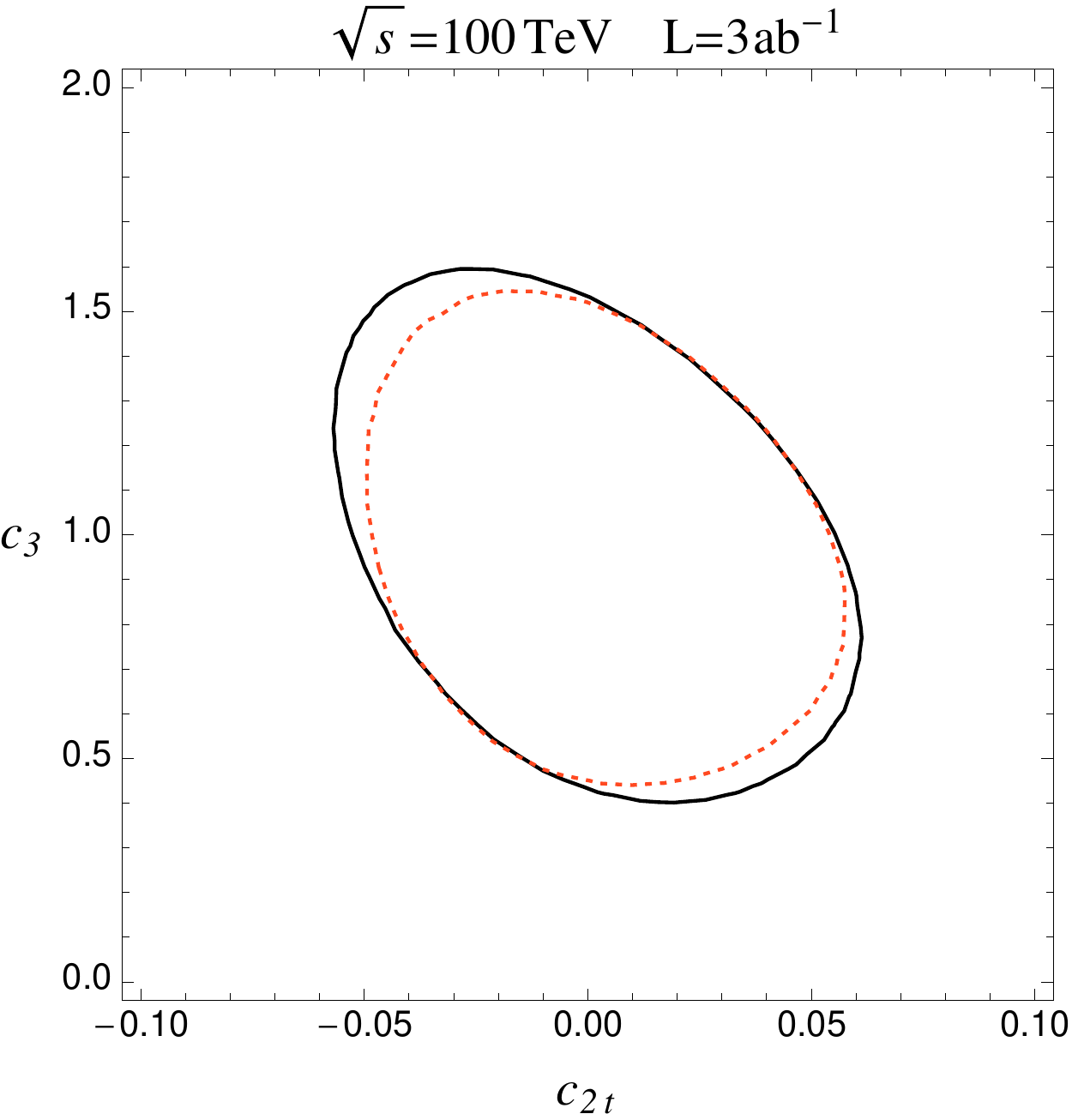}\qquad
\includegraphics[width=0.475\linewidth]{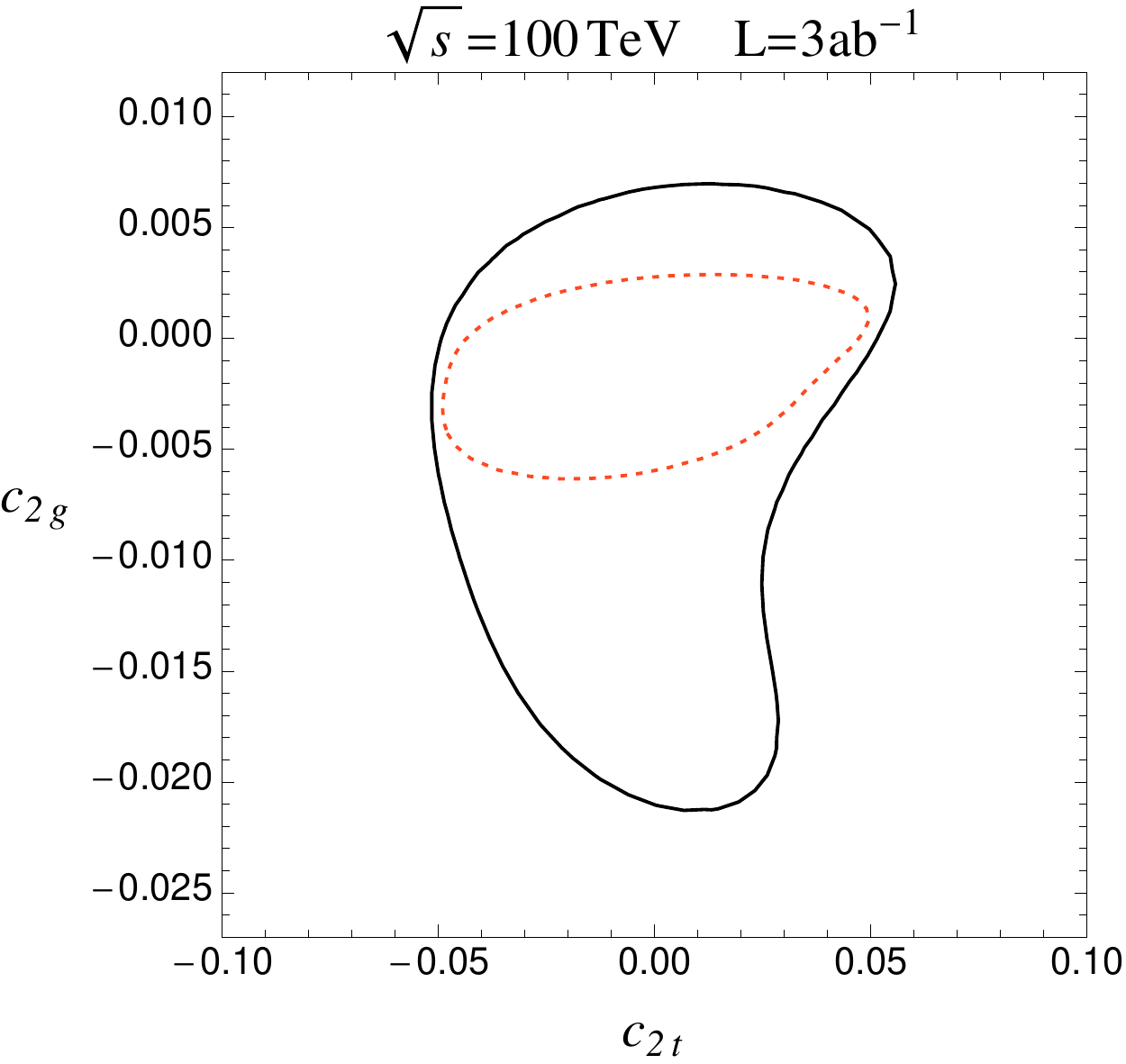}
\caption{$68\%$ probability contours for the \FCC
with a traditional analysis (black continuous line)
and one using jet substructure (red dotted line), see text. 
Left plot: plane $(c_{2t}, c_3)$; Right plot: plane $(c_{2t}, c_{2g})$.
}
\label{fig:boosted}
\end{center}
\end{figure}

We find that the couplings $c_3$ and $c_{2t}$ are strongly anti-correlated, and the precision expected on $c_{2t}$ is much 
higher than the one on the Higgs trilinear coupling $c_3$. This is in agreement with previous studies, which pointed out the strong sensitivity of the
double Higgs cross section on the $t\bar thh$ quartic coupling, see~\cite{Contino:2012xk} and references therein. 
The coupling $c_{2g}$ is determined even more accurately, although its naive estimate is suppressed, compared to that of $c_3$ and $c_{2t}$, 
by two powers of a weak spurion if the Higgs is a pseudo NG boson, see Eqs.~(\ref{eq:dictionary}),~(\ref{eq:estimatedim6}).
Although the Higgs trilinear coupling $c_3$ is the less accurately measured parameter, its precision can be highly increased at a $100\,$TeV collider,
as shown by the left plot of Fig.~\ref{fig:summary-nonlinear}. This is mainly due to the higher number of signal events which can be produced at this machine.
Using boosted jet techniques does not seem to improve significantly the accuracy on $c_3$ and $c_{2t}$, while it increases dramatically that on $c_{2g}$.
This is expected, since jet substructure techniques are most relevant to reconstruct highly boosted events with large $m_{hh}$,
which are  especially important to determine $c_{2g}$.
The situation is illustrated by
Fig.~\ref{fig:boosted}, where the red dotted curve is obtained by including the first 5 categories ($250\,\text{GeV} < m_{hh} < 1000\,\text{GeV}$) of the
traditional jet-based analysis and the last 5 categories ($m_{hh} > 1000\,\text{GeV}$) of the ``Substructure II'' analysis, see 
Tables~\ref{tab:mhh:100TeV} and~\ref{tab:mhh:100TeV:jetsub}.
%
%

\begin{figure}
\begin{center}
\includegraphics[width=0.435\linewidth]{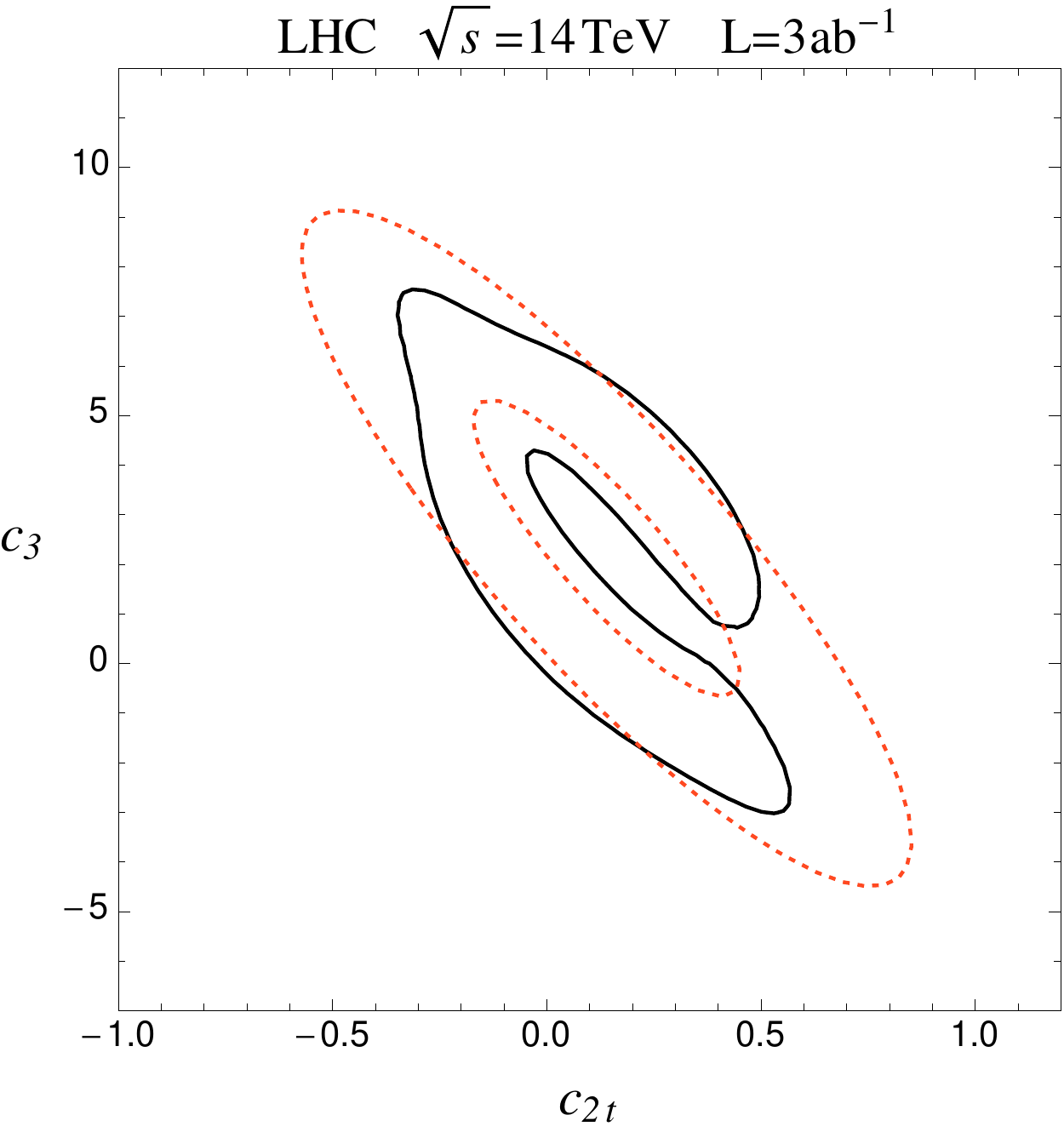}\qquad
\includegraphics[width=0.435\linewidth]{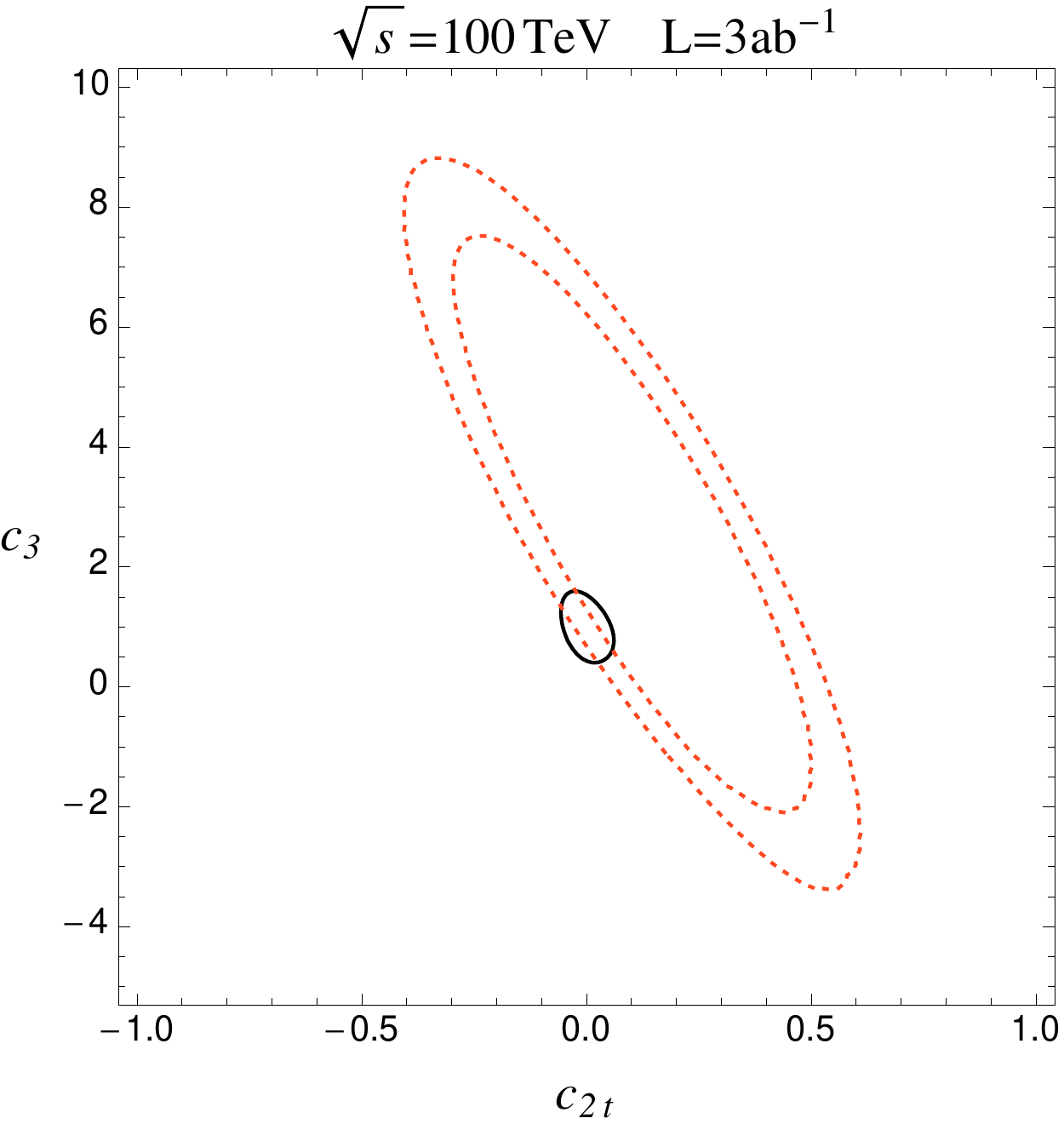}
\caption{$68\%$ probability contours in the plane $(c_{2t}, c_3)$  with the exclusive analysis (black continuous line) and an inclusive one
without $m_{hh}$ categories (red dotted line), see text. 
Left: plot for the \HLLHC; Right: plot for the \FCC.
}
\label{fig:inclusive}
\end{center}
\end{figure}

\begin{figure}
\begin{center}
\includegraphics[width=0.45\linewidth]{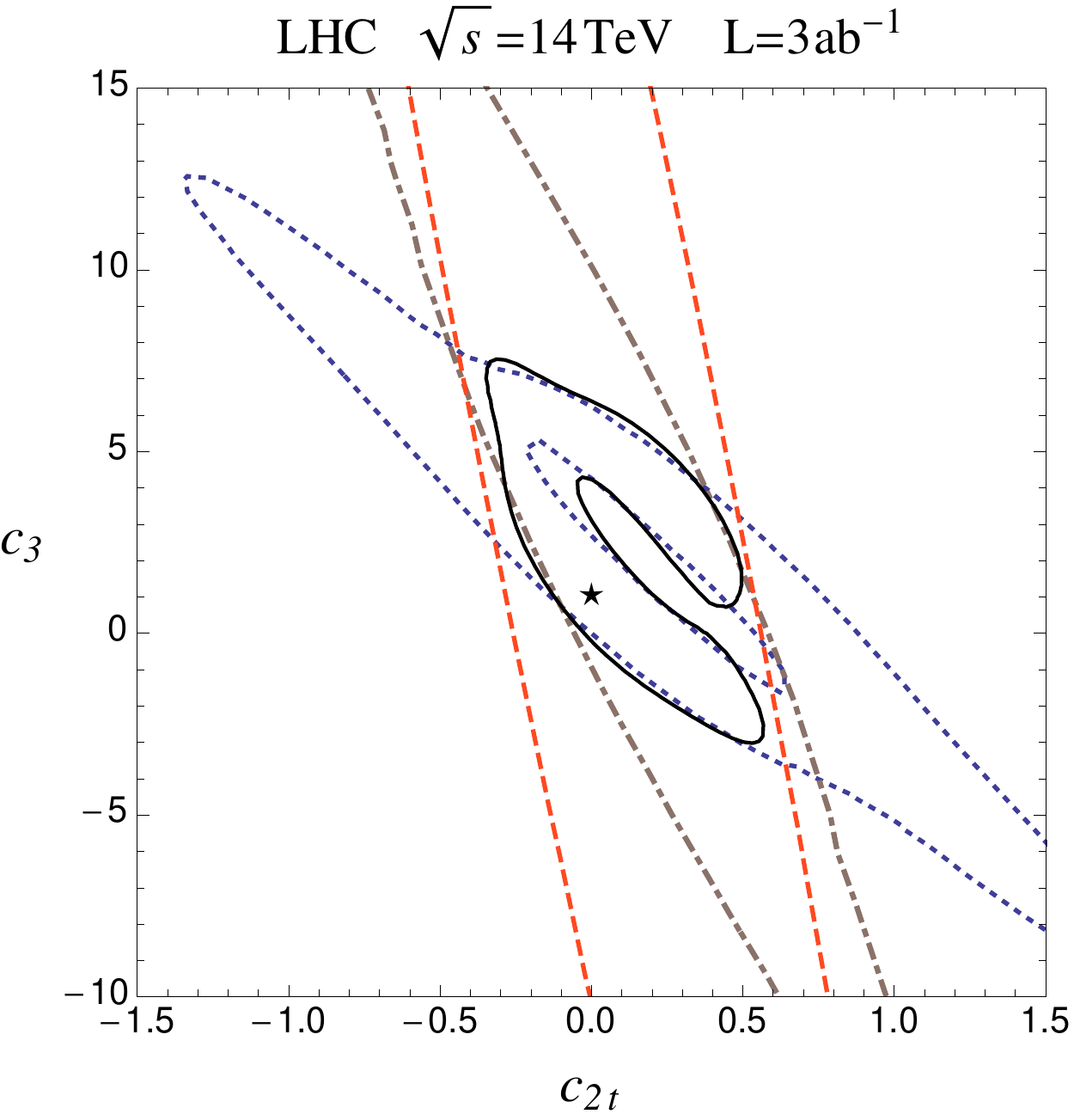}\qquad
\includegraphics[width=0.47\linewidth]{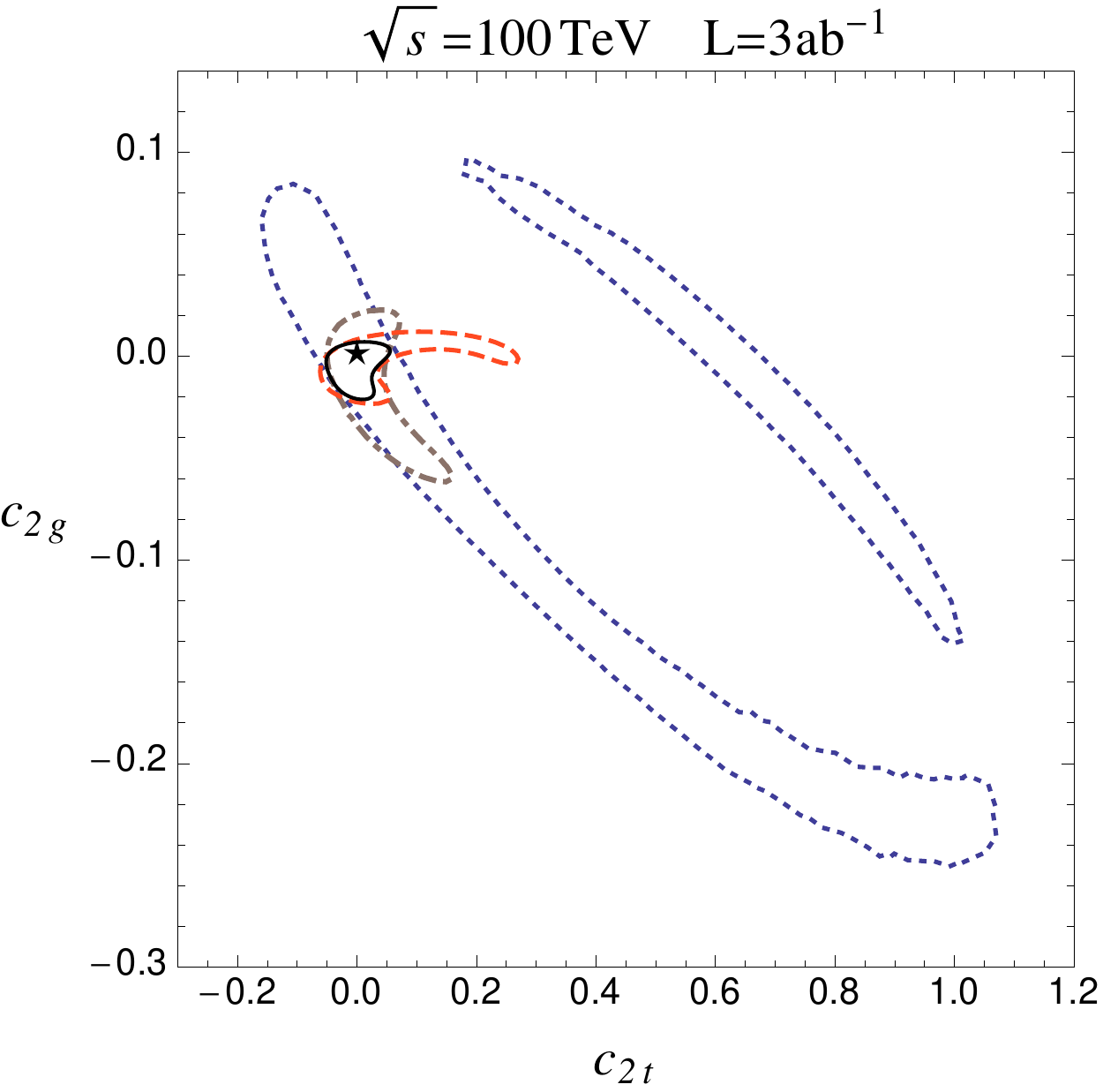}
\caption{Breakdown of $m_{hh}$ categories in the plane $(c_{2t}, c_3)$ for the \HLLHC
(left plot), and in the plane
$(c_{2t}, c_{2g})$ for the \FCC
(right plot). The various curves indicate the $68\%$ probability contours for the following
pairs of categories of Tables~\ref{tab:mhh:14TeV} and~\ref{tab:mhh:100TeV}: 1~and~2 (dotted blue line); 3~and~4 (dot-dashed brown line); 5~and~6 (dashed red line).
}
\label{fig:breakdown}
\end{center}
\end{figure}

\begin{figure}
\begin{center}
\includegraphics[width=0.476\linewidth]{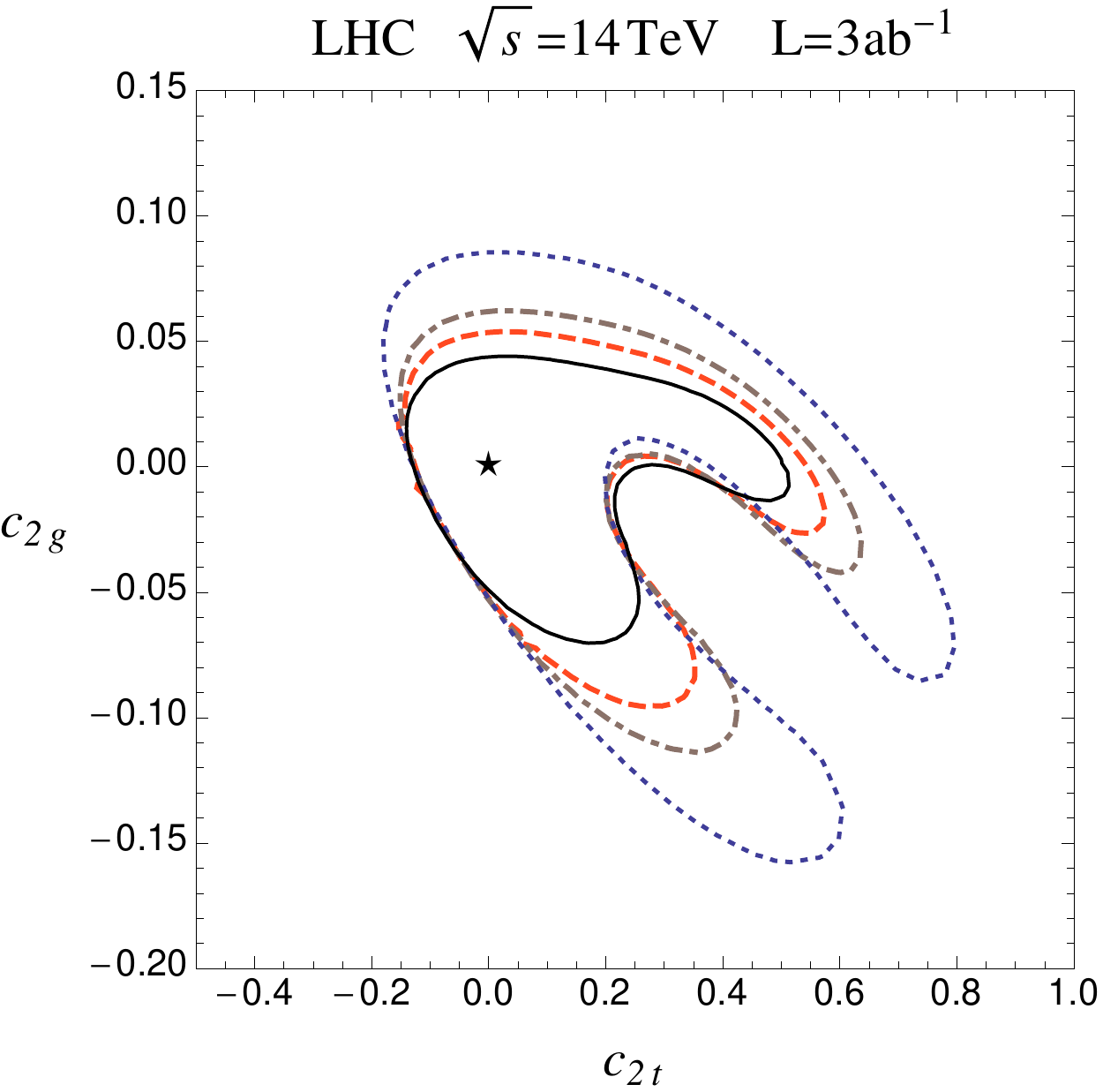}\qquad
\includegraphics[width=0.476\linewidth]{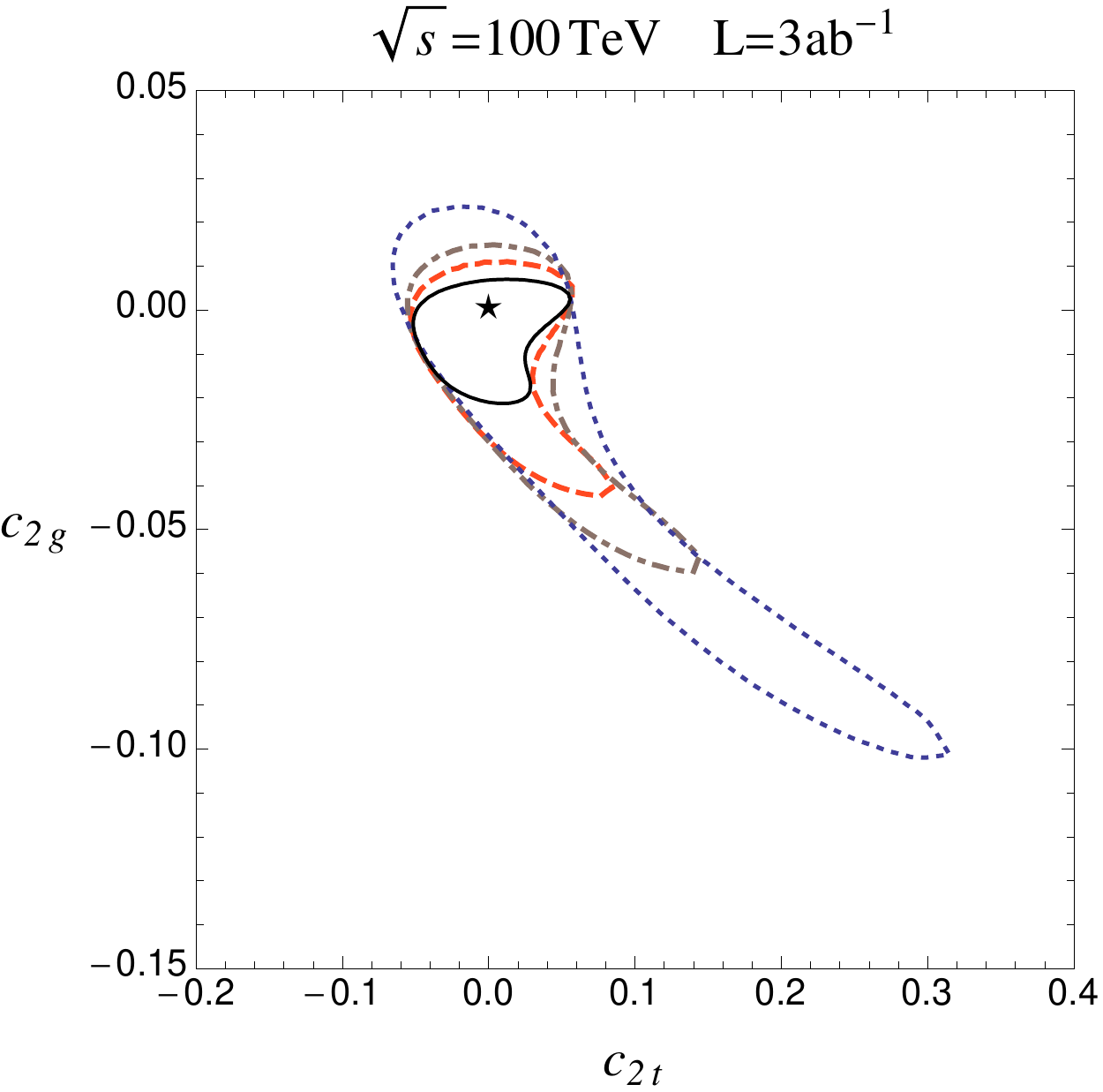}
\caption{$68\%$ probability contours in the plane $(c_{2t}, c_{2g})$ for the \HLLHC
(left plot) and the \FCC
(right plot). The different curves have been obtained by removing the following $m_{hh}$ categories
of Tables~\ref{tab:mhh:14TeV} and~\ref{tab:mhh:100TeV} from the fit: 6 (dashed red line); 6 and 5 (dot-dashed brown line); 6, 5 and 4 (dot blue line).
The continuous black contour is obtained by including all the categories in the fit.
}
\label{fig:fewercat}
\end{center}
\end{figure}

In order to break the degeneracy among the various parameters and extract them precisely, it is crucial to make use of the information on $m_{hh}$.
We find that an inclusive analysis, where events are not classified in the six $m_{hh}$ categories of Tables~\ref{tab:mhh:14TeV} and~\ref{tab:mhh:100TeV}, is much less
powerful in constraining the Higgs couplings, especially in the case of a $100\,$TeV collider. This is illustrated by the plots
of Fig.~\ref{fig:inclusive} in the plane $(c_{2t}, c_3)$. 
%
%
It is evident how at $100\,$TeV only an exclusive analysis can break the degeneracy between 
$c_{2t}$ and $c_3$.  As expected from the discussion of Section~\ref{sec:crosssection}, categories with larger $m_{hh}$ most strongly constrain $c_{2t}$ and $c_{2g}$,
while $c_3$ is mainly determined by events at threshold, see Fig.~\ref{fig:breakdown}.
%
%
Given the relevance of the categories with large $m_{hh}$ in determining the Higgs couplings, it is important to give an assessment on the
validity of the effective field theory approach in our analysis. Following the discussion of section~\ref{sec:higherorder}, we evaluate the minimum coupling strength
$g_{min}\sim \sqrt{\delta} (\bar E/v)$ by estimating the maximal invariant mass $\bar E$ of the events which lead to a precision $\delta $ on $c_{2g}$ 
(i.e. the parameter controlling the term  which grows quadratically with the energy in the scattering amplitude).
To this aim we re-derive the $68\%$ probability contours in the plane $(c_{2t}, c_{2g})$ by removing one or more of the categories with large $m_{hh}$, as a way
to estimate their impact on the determination of $c_{2g}$. We show such plots in Fig.~\ref{fig:fewercat} for the \HLLHC  and the \FCC.
%
%
We find \\[-0.3cm]
\begin{center}
\setlength{\tabcolsep}{15pt}
\begin{tabular}{cccc}
\LHC & \HLLHC & \FCC & \FCC \!(boosted) \\[0.05cm] 
\hline &&& \\[-0.4cm]
$\delta \simeq 0.15$ & $\delta \simeq 0.1$ & $\delta \simeq 0.05$ & $\delta \simeq 0.016$ \\[0.08cm]
$\bar E \simeq 1.0\,$TeV & $\bar E \simeq 1.0\,$TeV & $\bar E \simeq 1.0\,$TeV & $\bar E \simeq 1.8\,$TeV\\
$g_{min} = 1.6$ & $g_{min} = 1.3$ & $g_{min} = 0.9$ & $g_{min} = 0.9$ 
\end{tabular}
\end{center}
\vspace{0.3cm}
where the last column refers to the analysis including jet substructure at the \FCC (first 5 categories of traditional analysis + last 5 categories of ``Substructure II'' analysis).
As discussed in Section~\ref{sec:higherorder}, the value of $g_{min}$ determines the extension of the region which can be probed
under the validity of the effective field theory, see Fig.~\ref{fig:cartoon}.
In particular, $g_{min} < y_{t}$ only in the case of a $100\,$TeV collider, which suggests that an analysis including only dimension-6 operators
at the LHC (even at its high-luminosity upgrade)  is not sensitive to the case of a pseudo NGB Higgs with fully or partially composite $t_R$,
although it can probe the case of a generic composite Higgs.

We now turn to the effective Lagrangian for a Higgs doublet, Eq.~(\ref{eq:linearL}), and show the constraints on its coefficients. 
The production cross section, in this case, depends on  $\bar c_H$, $\bar c_u$, $\bar c_g$ and $\bar c_6$, while
the branching ratio for $hh \to b\bar b\gamma\gamma$ depends on $\bar c_H$, $\bar c_u$, $\bar c_g$ as well as on $\bar c_d$ and $\bar c_\gamma$.
The latter two coefficients parametrize respectively the modification of the down quark Yukawa couplings, in particular that of the bottom quark, and the contact interaction 
between the Higgs boson and two photons. Their precise definition can be found in Ref.~\cite{Contino:2013kra} and it is analogous to that of $\bar c_u$ and $\bar c_g$ 
in Eq.~(\ref{eq:linearL}). We will set $\bar c_\gamma =0$ for simplicity in the following.

\begin{figure}
\begin{center}
\includegraphics[width=0.46\linewidth]{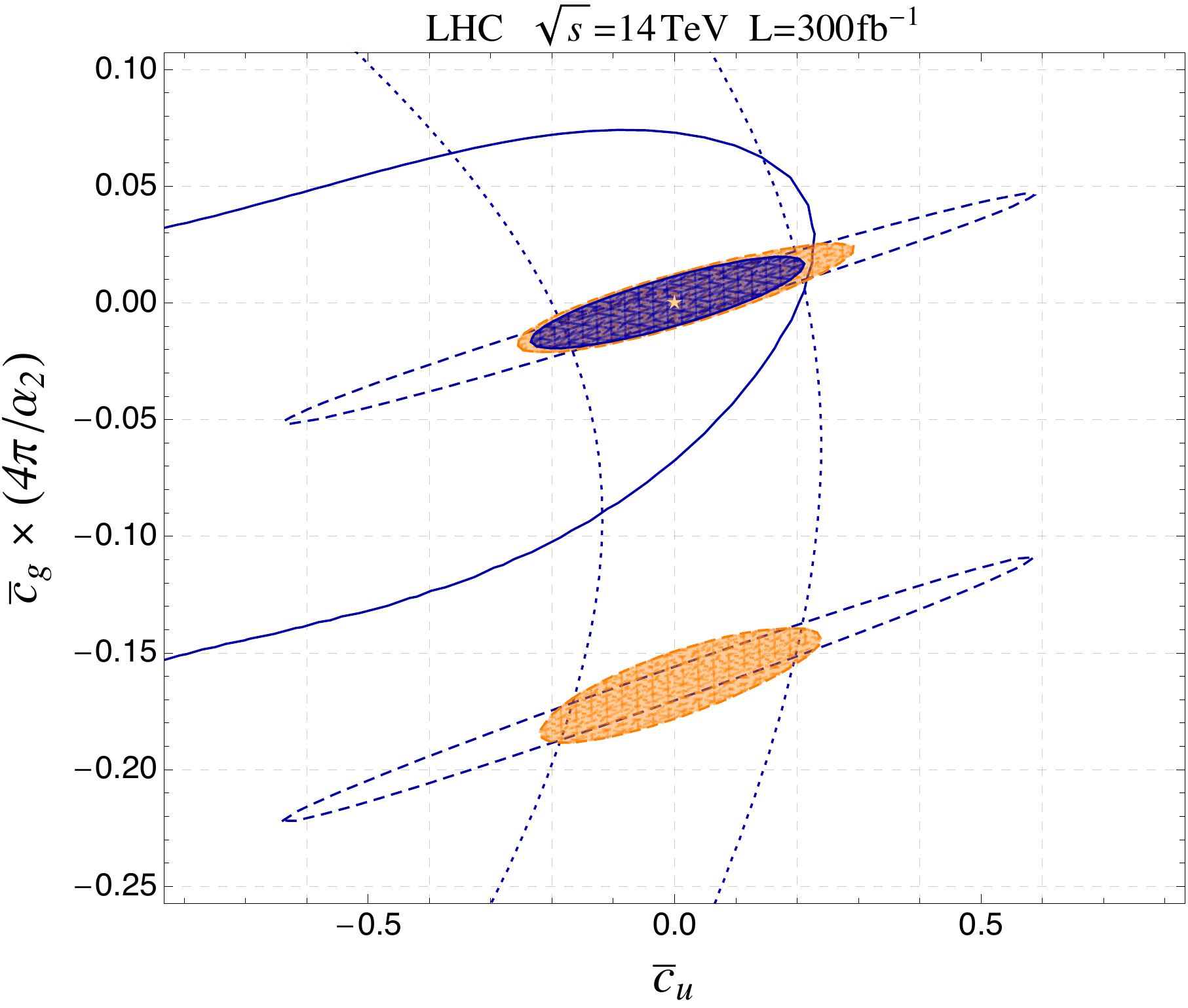}\qquad
\includegraphics[width=0.46\linewidth]{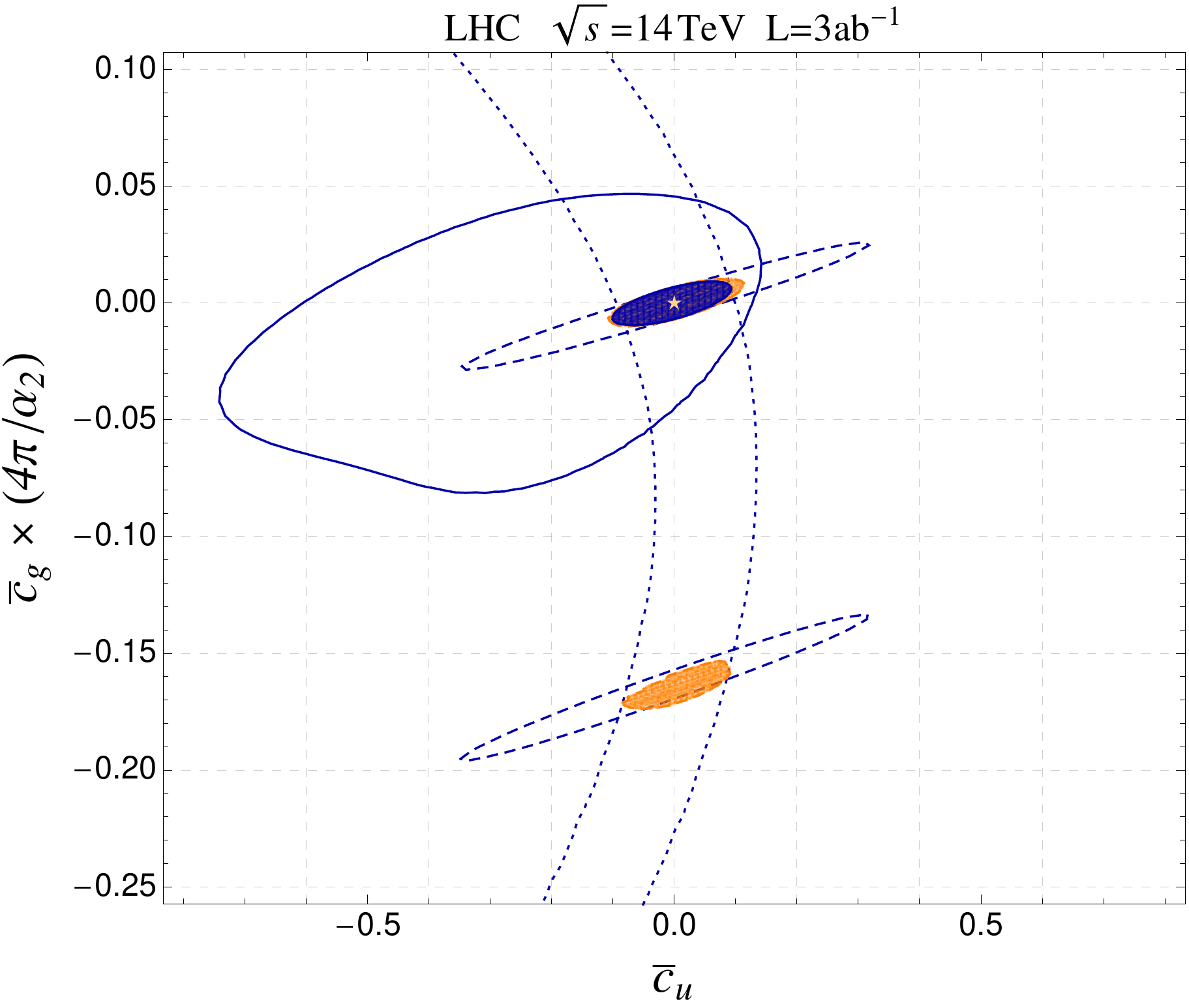} \\[0.3cm]
\includegraphics[width=0.46\linewidth]{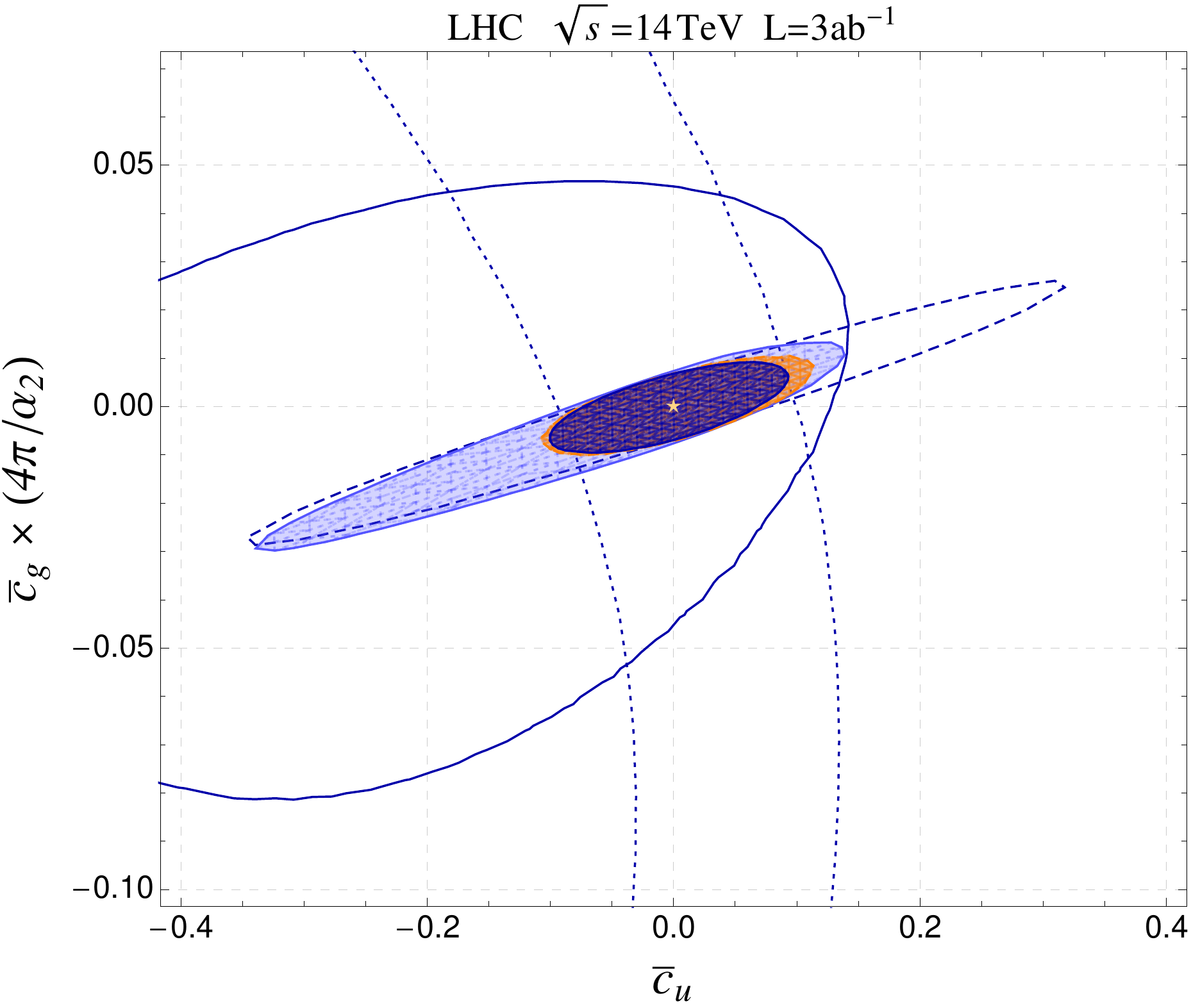}\qquad
\includegraphics[width=0.46\linewidth]{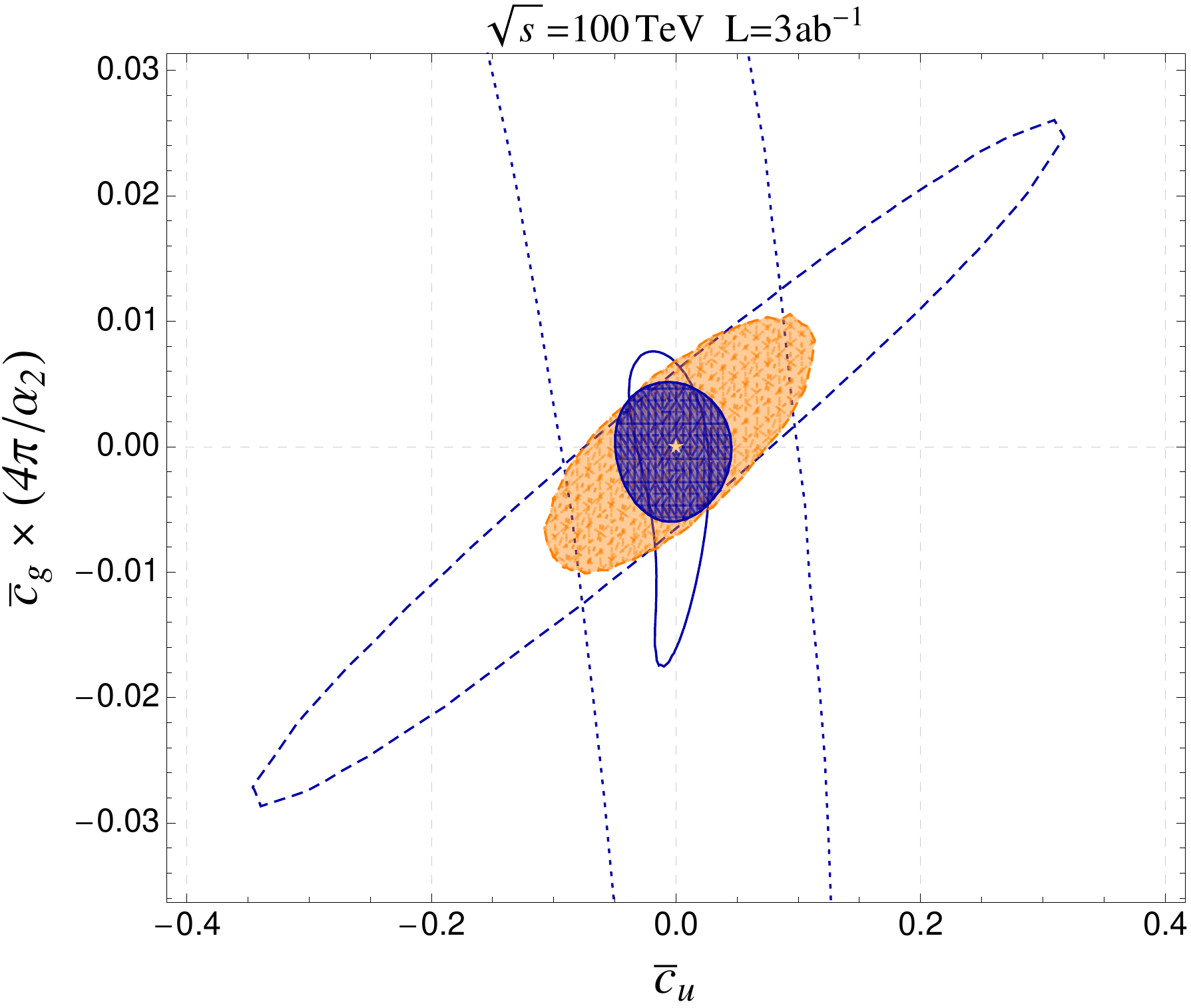}
\caption{$68\%$ probability contours in the plane $(\bar c_u,\bar c_g)$ for the three benchmark scenarios: \LHC (upper left plot);
\HLLHC (upper right and lower left plots); \FCC (lower right plot).
Different curves correspond to:  double Higgs production in the $b\bar b\gamma\gamma$ final state (solid blue line); all single-Higgs processes except $t\bar th$
(dashed blue line); $t\bar th$ alone (dotted blue line). Filled areas correspond to:  all single Higgs processes including $t\bar th$ (orange region);
all single Higgs processes plus double Higgs production (dark blue region); all single Higgs processes except $t\bar th$
plus double Higgs production (light blue area in the lower left plot).
}
\label{fig:cucg}
\end{center}
\end{figure}

We start considering the constraints on $\bar c_u$ and $\bar c_g$ that follow from both single and double Higgs production.
The  $68\%$ probability contours are shown in Fig.~\ref{fig:cucg} for the three benchmark scenarios considered in this section. 
%
%
The solid blue curve refers to double Higgs production in the $b\bar b\gamma\gamma$ final state (our analysis), while the dashed and dotted blue curves correspond 
to the constraint  from respectively all single-Higgs processes except $t\bar th$, and $t\bar th$ alone~\footnote{Notice that the dotted curves are not vertical in the 
plane $(\bar c_u, \bar c_g)$ due to the dependence of the total Higgs width on $\bar c_g$. A further dependence on $\bar c_g$ would follow from the contribution of 
diagrams with an insertion of $O_g$ to the $t\bar t h$ cross section; this effect has not been included in our fit. We thank C.~Grojean for drawing our attention on this point.} 
(the ATLAS projections from Ref.~\cite{ATL-PHYS-PUB-2013-014}).
They have been obtained by fixing $\bar c_H = \bar c_d =0$ and marginalizing over $\bar c_6$ with flat prior (in the case of double Higgs production).
The two dashed ellipses in the plots in the upper row correspond to the two degenerate solutions 
that follow from the interference between $\bar c_g$ and $\bar c_u$ in the $gg\to h$  cross section.
The filled areas denote instead the $68\%$ probability regions obtained by combining the following processes: all single Higgs processes including $t\bar th$ (orange region);
all single Higgs processes plus double Higgs production (dark blue region); all single Higgs processes except $t\bar th$
plus double Higgs production (light blue area in the bottom left plot of Fig.~\ref{fig:cucg}). They have been obtained by marginalizing over $\bar c_H$, $\bar c_d$
and  $\bar c_6$.  One can see that double Higgs production breaks the degeneracy on $\bar c_g$ that remains even after including $t\bar th$ among single-Higgs
processes. It also helps increasing the precision on both $\bar c_u$ and $\bar c_g$ compared to that following from single Higgs processes alone, especially at
a $100\,$TeV collider. In this respect, it is interesting to notice that $\bar c_u$ and $\bar c_g$ are strongly correlated when considering single-Higgs measurements 
without~$t\bar th$, and that the extension of the dashed ellipses  is much larger than the error obtained on each individual parameter by setting the other to its SM value 
(in fact, the ellipses become infinite bands if one lets $\bar c_\gamma$ vary, as a consequence of the degeneracy in the $h\to \gamma\gamma$
decay rate)~\footnote{By  fixing all the parameters to their SM value but one, 
the likelihood obtained from Ref.~\cite{ATL-PHYS-PUB-2013-014} is approximately Gaussian. 
By including all single Higgs processes except $t\bar th$ we obtain the following standard deviations: $\sigma(\bar c_u) = 0.06$, $\sigma(\bar c_g) = 0.005$, 
$\sigma(\bar c_H) = 0.08$ for $L = 300\,\text{fb}^{-1}$;  $\sigma(\bar c_u) = 0.05$, $\sigma(\bar c_g) = 0.004$, $\sigma(\bar c_H) = 0.05$ for $L = 3\,\text{ab}^{-1}$.
From $t\bar th$ alone we find: $\sigma(\bar c_u) = 0.2$ for $L = 300\,\text{fb}^{-1}$; $\sigma(\bar c_u) = 0.08$ for $L = 3\,\text{ab}^{-1}$.}.
The process $t\bar th$ mainly constrains $\bar c_u$ and is crucial to reduce the overall experimental uncertainty.  In comparison, 
$gg\to hh \to b\bar b\gamma\gamma$ is less powerful but still competitive in constraining $\bar c_u$. For example, by removing $t\bar th$ from the combination
of single and double Higgs at the \HLLHC,
one obtains the light blue region (instead of the orange one) which is sensibly
smaller than the dashed contour.  One must also notice that the $t\bar th$ dotted curve in Fig.~\ref{fig:cucg} corresponds to the combination of several decay channels
(all those studied by Ref.~\cite{ATL-PHYS-PUB-2013-014}: $h\to \gamma\gamma, ZZ, \mu\mu$), while the continuous blue curve denotes double Higgs
production in the $b\bar b\gamma\gamma$ final state alone, i.e. the one studied in this work.
Combining with other final states (e.g. $b\bar b\tau\tau$ and $b\bar bWW$) will make double Higgs production even more competitive for the determination
of $\bar c_u$, as well as of $\bar c_g$.

\begin{figure}
\begin{center}
\includegraphics[width=0.45\linewidth]{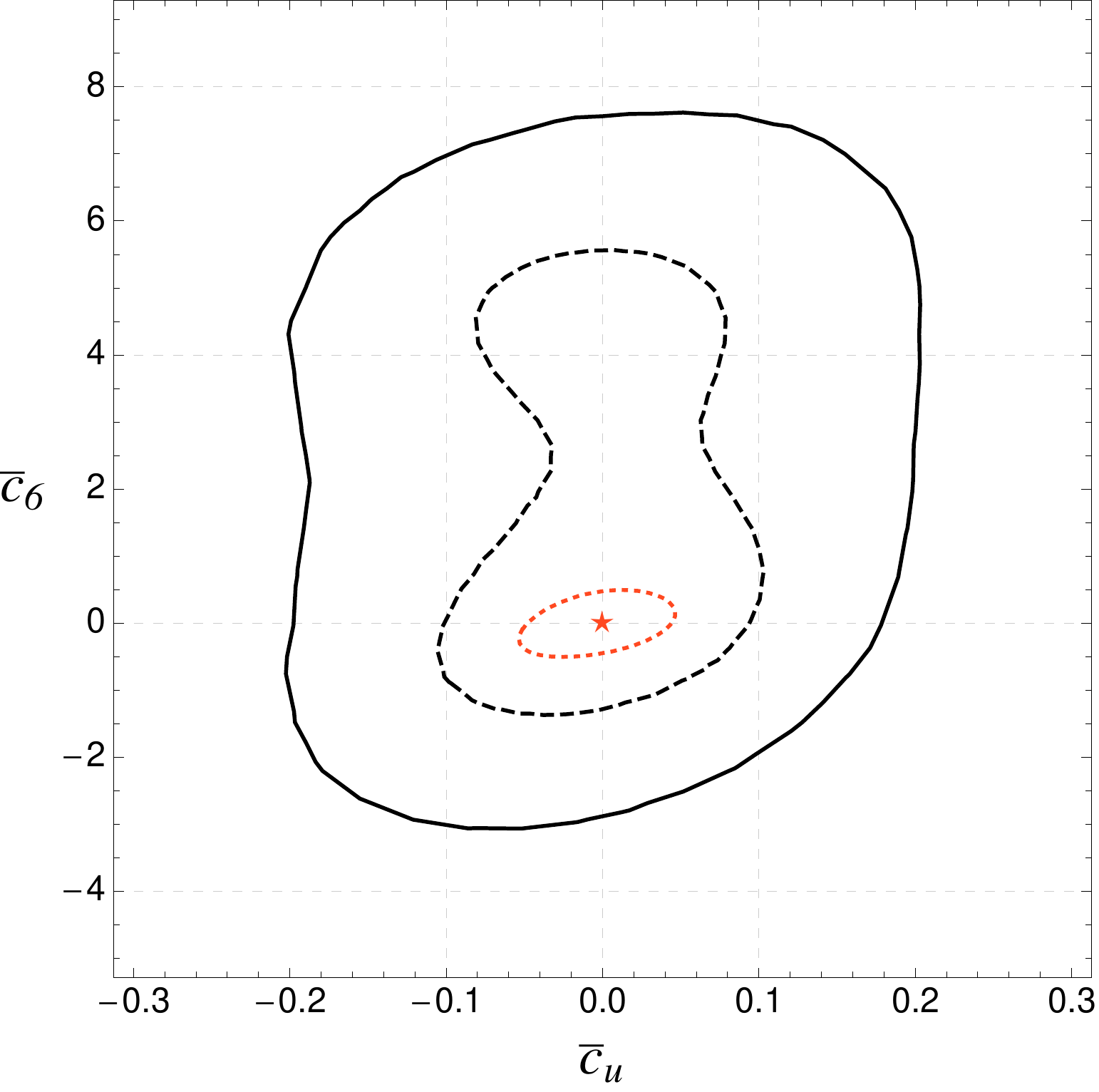}\qquad
\includegraphics[width=0.45\linewidth]{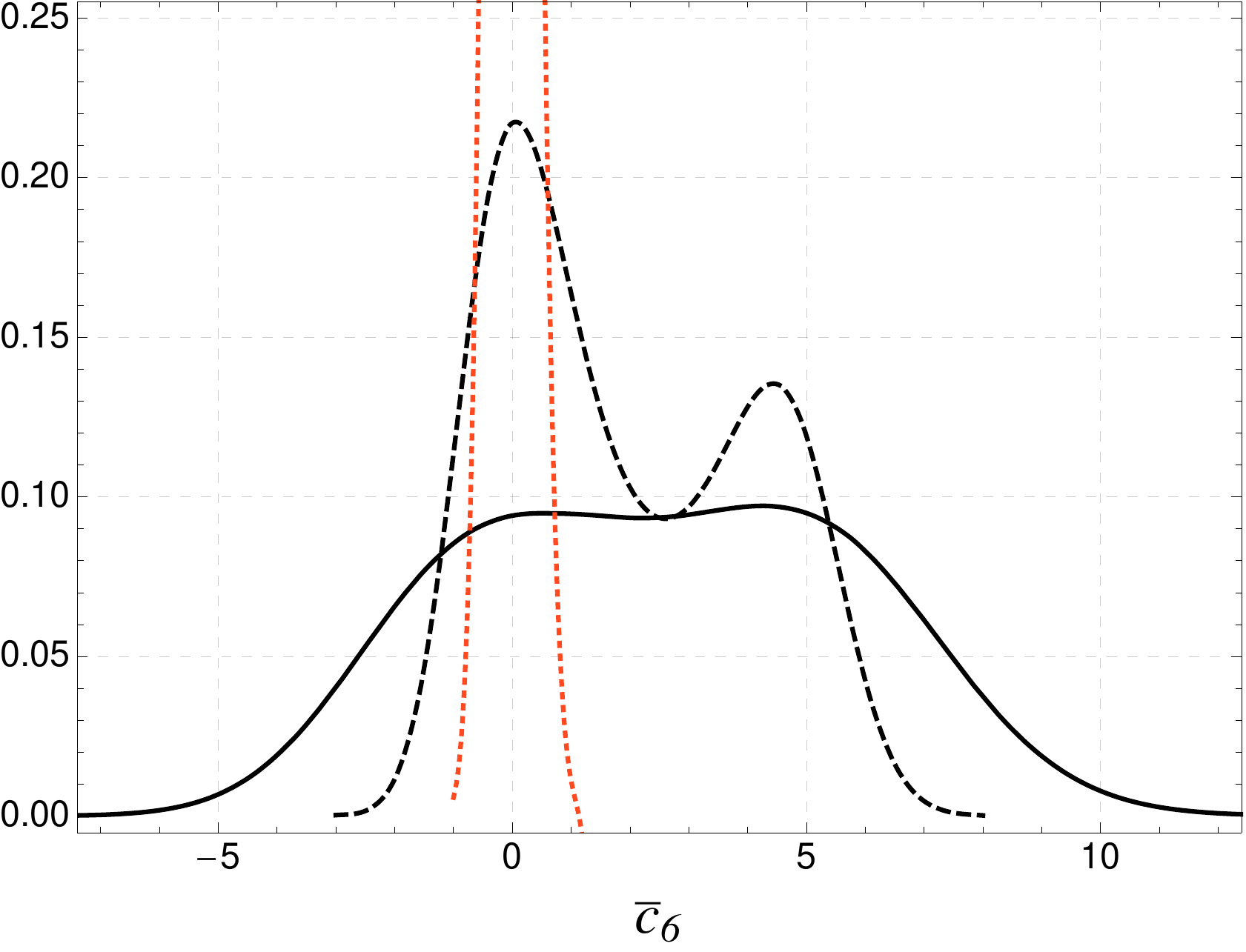} \\
\caption{Left plot: $68\%$ probability contours in the plane $(\bar c_{u}, \bar c_6)$; 
Right plot: probability distributions as functions of $\bar c_6$.
The different curves refer to the following benchmark scenarios: 
\LHC (black continuous line); \HLLHC (black dashed line); \FCC (red dotted line).
}
\label{fig:cu-c6}
\end{center}
\end{figure}

\begin{figure}
\begin{center}
\includegraphics[width=0.45\linewidth]{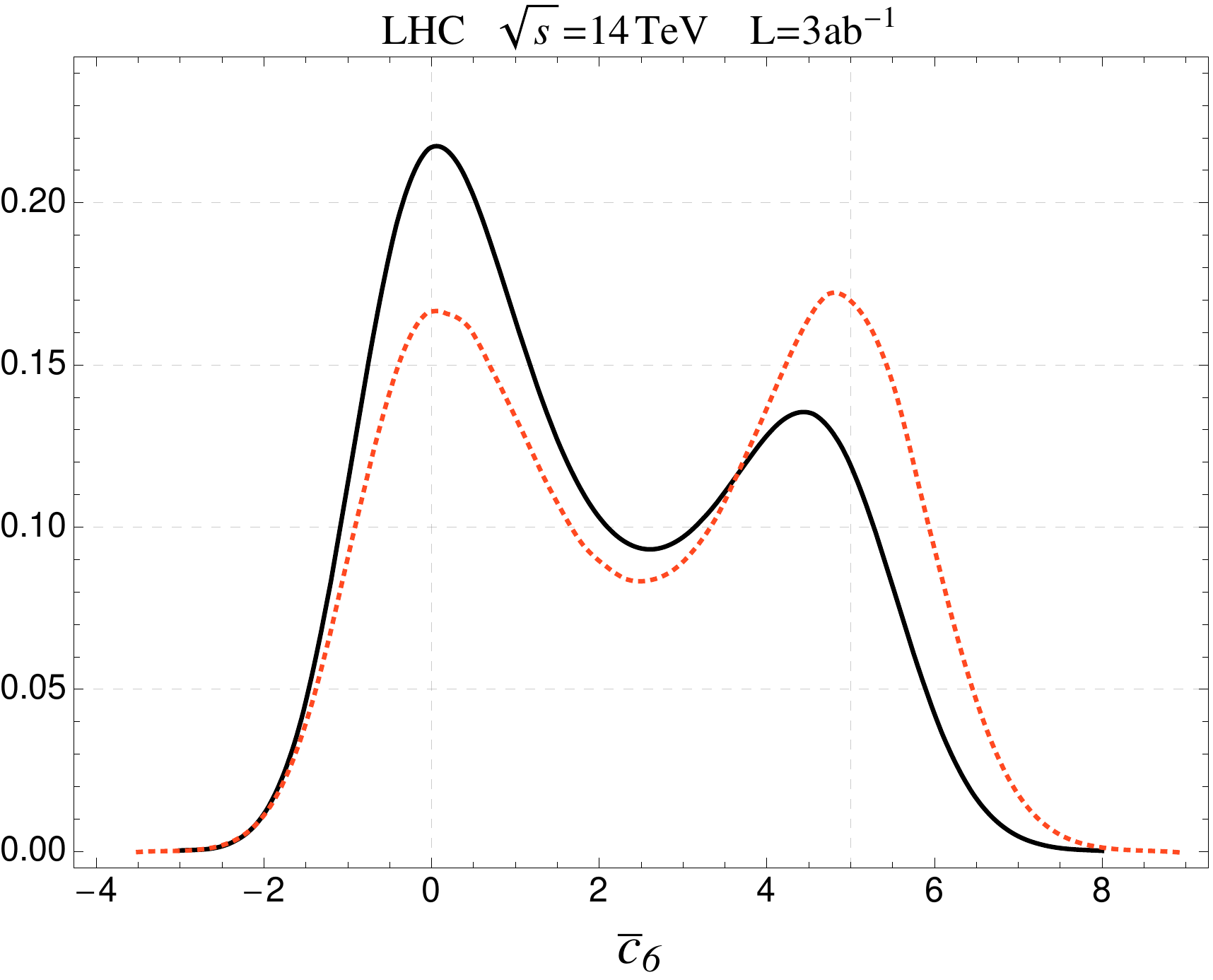}\qquad
\includegraphics[width=0.45\linewidth]{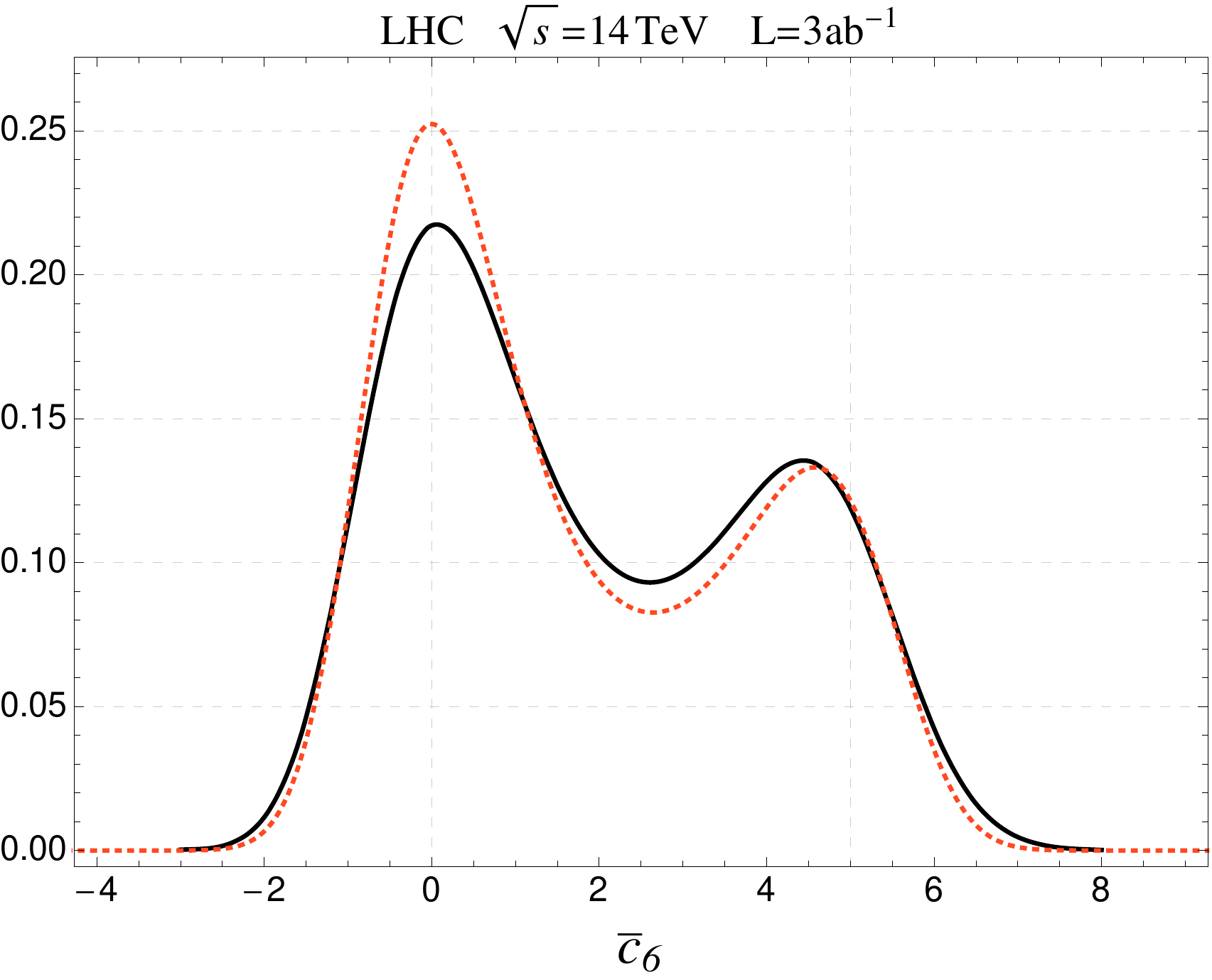} \\
\caption{Probability distributions as functions of $\bar c_6$ at the \HLLHC. The black continuous curve refers to the standard exclusive analysis
discussed in section~\ref{sec:analysis}. The dotted red curve refers to an inclusive analysis in the left plot, and to an analysis
without marginalization on $\bar c_H$, $\bar c_d$, $\bar c_g$ and $\bar c_u$ in the right plot. See text.
}
\label{fig:probabilityfunctions}
\end{center}
\end{figure}

The plot on the left of Fig.~\ref{fig:cu-c6} shows the $68\%$ probability contours in  the plane $(\bar c_u, \bar c_6)$, obtained by marginalizing over 
$\bar c_H$, $\bar c_d$ and $\bar c_g$. 
%
%
As already suggested by Fig.~\ref{fig:summary-nonlinear}, the precision on $\bar c_6$ (i.e. the parameter which controls the modification of the Higgs trilinear
coupling) is much smaller than that on $\bar c_u$. At $\sqrt{s} = 14\,$TeV, in particular, the constraint is on values of $\bar c_6$ larger than~1.~\footnote{See
the discussion of Section~\ref{sec:modifiedpowercounting} on the validity of the effective field description in this case.}
Further marginalizing over $\bar c_u$ leads to the probability functions shown in the right plot of Fig.~\ref{fig:cu-c6} for the three
benchmark scenarios. Notice that even at the \HLLHC
the likelihood is far from being Gaussian, and a second
maximum is present at $\bar c_6 \simeq 4.5$. We find  the following $68\%$ probability intervals on~$\bar c_6$:
\begin{center}
\setlength{\tabcolsep}{15pt}
\begin{tabular}{ccc}
\LHC & \HLLHC & \FCC \\[0.05cm]
\hline \\[-0.4cm]
$[-1.2, 6.1]$ & $[-1.0, 1.8] \cup [3.5, 5.1]$ & $[-0.33, 0.29]$
\end{tabular}
\end{center}
\vspace{0.3cm}
It is interesting to compare the probability  functions of Fig.~\ref{fig:cu-c6} with those obtained with an inclusive analysis
without $m_{hh}$ categories. For the \HLLHC
for example, an inclusive analysis gives the function shown in
the left plot of Fig.~\ref{fig:probabilityfunctions} (dotted red curve).
%
%
The two solutions are in this case almost degenerate, and in fact the second (non-SM) maximum is slightly higher than the one at $\bar c_6=0$ as an
effect of the marginalization on $\bar c_u$~\footnote{This ``unwanted'' feature can be avoided by profiling, instead of marginalizing, over $\bar c_u$. The
second maximum in this case would be lower than the one at $\bar c_6=0$, since the highest peak of the 2-dimensional likelihood is indeed the one at the SM point.}.
A fully exclusive analysis thus removes the degeneracy and helps reduce the significance of the unphysical solution.

\begin{figure}
\begin{center}
\includegraphics[width=0.45\linewidth]{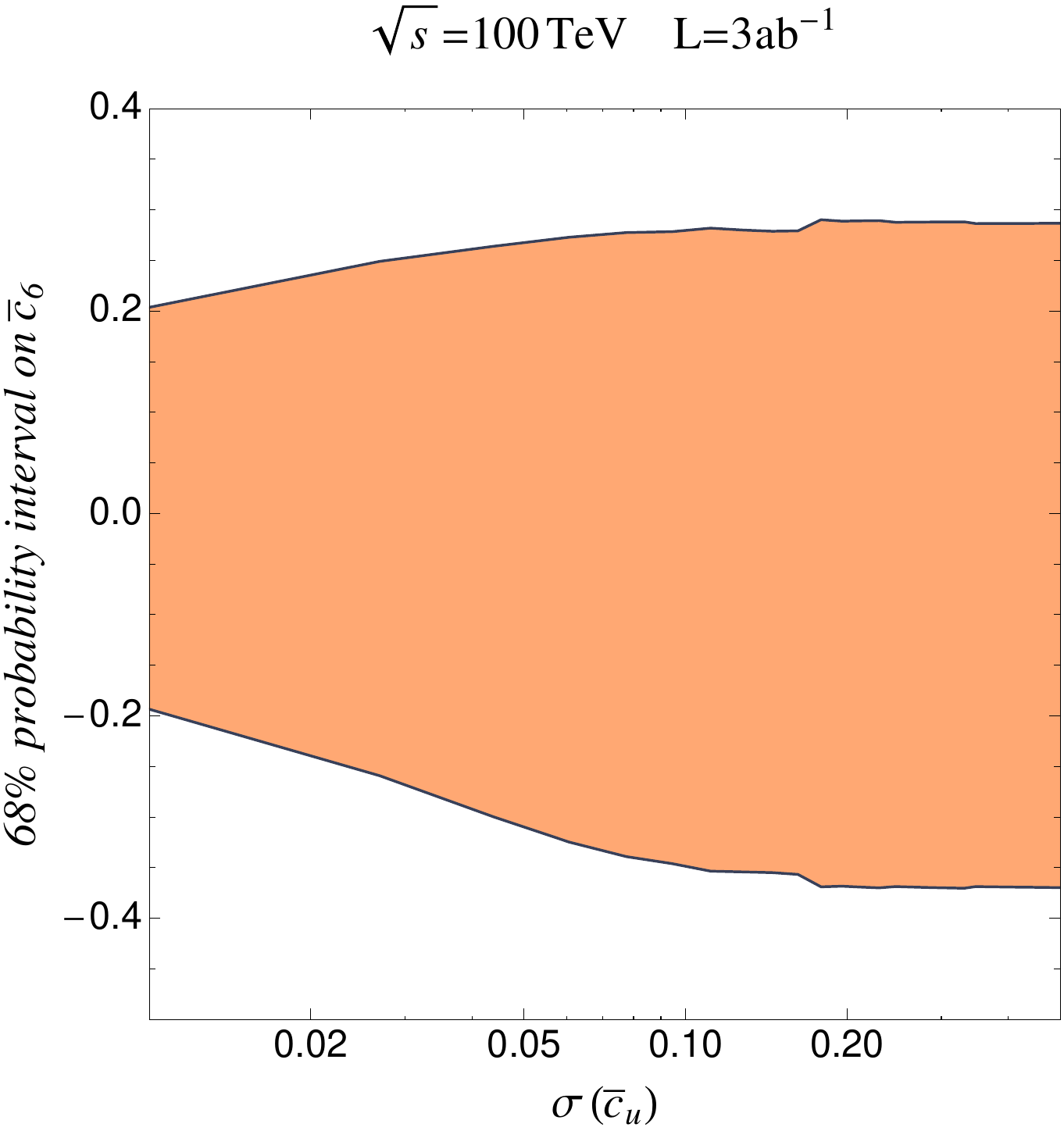}\qquad
\caption{Width of the $68\%$ probability interval on $\bar c_6$ (orange region) as function of the uncertainty on $\bar c_u$
for the \FCC. See text.
}
\label{fig:c6vssigmacu}
\end{center}
\end{figure}

Another interesting question concerns the relevance of the marginalization over $\bar c_H$, $\bar c_d$, $\bar c_g$ and $\bar c_u$ 
in determining the Higgs trilinear coupling.  The right plot of Fig.~\ref{fig:probabilityfunctions} shows  the probability function which is obtained  by switching off
the marginalization in the \HLLHC scenario  
(dotted red curve). 
The effect of marginalization is that of flattening the SM solution, although in a marginal way at $14\,$TeV. 
The effect is on the other hand much more important at $100\,$TeV. On a more quantitative side, the $68\%$ probability intervals which follow without
marginalization are: $[-1.4, 5.8]$ (\LHC),
$[-1.0, 1.6]\cup [3.8, 5.1]$ (\HLLHC),
and  $[-0.18, 0.18]$ (\FCC).
It is clear that a precise measurement of the Higgs trilinear coupling, as it follows at $100\,$TeV from the $b\bar b\gamma\gamma$ channel or as it might
be obtained at the high-luminosity LHC from a combination of several decay modes,  relies of the accurate extraction of the other couplings.
In particular, a significant uncertainty on $\bar c_6$ could follow from a poor determination of $\bar c_u$. This is illustrated in Fig.~\ref{fig:c6vssigmacu} for the $100\,$TeV
scenario. The plot shows the extension of the $68\%$ probability interval on $\bar c_6$ as a function of $\sigma(\bar c_u)$, where $\bar c_u$ is marginalized with 
a Gaussian distribution with standard deviation $\sigma(\bar c_u)$.
%
%
The uncertainty on $\bar c_6$ grows  with $\sigma(\bar c_u)$ for small values of this latter parameter, while it becomes constant
for $\sigma(\bar c_u) \gtrsim 0.2$. In this  limit double Higgs production alone determines $\bar c_u$ with a smaller error, independently of
single-Higgs measurements.

The impact of the uncertainty on the top Yukawa coupling in a precise determination of the Higgs trilinear was already discussed in Ref.~\cite{Goertz:2013kp},
although in a scenario where the only modifications with respect to the SM are in the values of these two couplings.
Such scenario of New Physics, however, is a disfavored one.
If the Higgs boson is part of a weak doublet and the new states are heavy,
one gets the effective Lagrangian (\ref{eq:linearL}) at low energy, where the same operator $O_u$ that gives a modification to the top Yukawa coupling also
generates the quartic $t\bar t hh$ interaction, whose contribution must thus be included. On the other hand, it is possible to get a modification of the top Yukawa without 
having a direct $t\bar t hh$ interaction if the new states are light, as for example in two-Higgs doublet models. In that case, however, the new states will give
an extra contribution to the production cross section, which is in fact required to match that from the $t\bar t hh$ vertex in the decoupling limit.
One can envisage a situation where the top Yukawa coupling $c_t$ is modified by new heavy physics while $c_{2t} =0$ if the Higgs does not belong to 
a weak doublet, like for example in the case of a dilaton. This class of models is however disfavored by current data.
These considerations suggest that any New Physics which modifies the top Yukawa coupling is likely to have an additional impact on the double Higgs cross section.
More specifically,  if the new states are heavy and the Higgs is part of a doublet, then any uncertainty on the value of the top Yukawa coupling will reflect
on a  contribution from the $t\bar thh$ vertex which must be properly included. Of course one can restrict oneself to the SM case and ignore the 
$t\bar thh$ anomalous coupling; in that case however, the top Yukawa coupling can be more precisely determined from the top quark mass, and its uncertainty is
sufficiently small to have a negligible impact on double Higgs production.

%% file: comparison.tex
\label{sec:comparison}

The results of the previous section suggest that an analysis of double Higgs production in the  $b\bar b\gamma\gamma$ final state can determine
the Higgs trilinear coupling with a $\sim 30\%$ precision at the \FCC.~\footnote{Here and in the following, by sensitivity/precision on some Higgs coupling or 
coefficient of the effective Lagrangian we mean the $68\%$ error on its measured value for injected SM signal.}
At the \HLLHC, on the other hand, the estimated uncertainty is at the level of~$200\%$,
and in fact another solution of the fit is present for a trilinear coupling approximately equal to 5.5 times its SM value. These results are more pessimistic
than other estimates that appeared in the literature, and it is thus useful to compare our analysis with  previous works in order to highlight
the different strategies and assumptions.

The first detailed analysis of the $b\bar b\gamma\gamma$ final state was performed by Baur, Plehn and Rainwater in Ref.~\cite{Baur:2003gp}.
They retained events with one or more $b$-tags at the LHC, while two $b$-tags were required at its high-luminosity phase.
An analysis with only one $b$-tag seems much more involved due to the larger background and the systematic error
from combinatorics. We will thus compare the results for the \HLLHC obtained with two $b$-tags.
A first difference with our analysis is in the value of the $b$-tagging efficiency and the corresponding light jet mis-tag rate, 
since Ref.~\cite{Baur:2003gp} assumes $\epsilon_b = 0.5$ and $\epsilon_{j\rightarrow b} =1/23$, to be compared
with $\epsilon_b = 0.7$ and $\epsilon_{j\rightarrow b} =0.01$ used in the present work (for both the LHC and its high-luminosity upgrade). 
While the choice of Ref.~\cite{Baur:2003gp} was based on early Montecarlo simulations, more recent results based on LHC data
have shown that larger $b$-tagging efficiencies can be achieved while maintaining acceptable levels of rejection 
rates~\cite{Chatrchyan:2012jua,CMS:2013vea,ATL-PHYS-PUB-2013-009}. As a consequence of the larger value of $\epsilon_{j\rightarrow b}$, the 
 background $jj\gamma\gamma$ is found to be the largest one in Ref.~\cite{Baur:2003gp}, differently from our estimates.
A second difference concerns the estimate of the background $b\bar b\gamma\gamma$, which we found to be enhanced at NLO by a large k-factor 
$k_{b\bar b\gamma\gamma} \simeq 2$.  Reference~\cite{Baur:2003gp} instead computes all the backgrounds at LO and
rescales their cross sections by  an ad-hoc common factor 1.3
to account for NLO effects, thus underestimating~$b\bar b\gamma\gamma$.
Finally, a narrower mass window for the photon pair, $m_{\gamma\gamma}^\text{reco} = m_h \pm 2.3\,$GeV, is assumed in the study of  Ref.~\cite{Baur:2003gp}.
This leads to a reduction by a factor $\sim 2$ compared to our estimates of all the non-resonant backgrounds, since their $m_{\gamma\gamma}^\text{reco}$
distribution is nearly flat around the Higgs mass. We have chosen instead to make a more conservative cut on  $m_{\gamma\gamma}^\text{reco}$  and
to leave the issue of optimizing the width of the mass window to a fully-realistic experimental analysis.
As a consequence of the above different assumptions, the  total background in Ref.~\cite{Baur:2003gp} is rather smaller than in our analysis,
which led the authors to a more optimistic estimate of the precision on the trilinear coupling.

A more recent analysis of the $b\bar b\gamma\gamma$ channel has been carried out in Refs.~\cite{Baglio:2012np,Barger:2013jfa,Yao:2013ika}. 
Reference~\cite{Baglio:2012np} by Baglio et al. estimates 47 SM signal events after all cuts with $L = 3\,\text{ab}^{-1}$ at the high-luminosity LHC, which 
should be compared with our $\sim 13$ events. 
Although our smaller rate can be in part explained by the inclusion of the photon efficiency factor $0.8^2 = 0.64$ (not included in Ref.~\cite{Baglio:2012np}), 
we were not able to fully identify the origin of such difference. 
%
Furthermore, the $b\bar b\gamma\gamma$ continuum is found to be negligible in Ref.~\cite{Baglio:2012np} due to a limited MC statistics, thus
leading to an underestimation of the background rate.
The comparison with Ref.~\cite{Barger:2013jfa} is instead difficult, since the number of signal and background events is not reported.
The  same kinematic cuts of Ref.~\cite{Baglio:2012np} are however applied, and it is said that an agreement is found with the efficiencies there reported.
This suggests that the same underestimation of the background also plagues Ref.~\cite{Barger:2013jfa}.

Finally, we compare with the analysis by Yao in Ref.~\cite{Yao:2013ika}, upon which the Snowmass study~\cite{Dawson:2013bba} is based.
This is a sophisticated study including a full-fledged simulation of showering, hadronization and detector effects.
The estimated number of SM signal and background events at the LHC after all cuts is respectively 16.6 and 53.4 
(40.1 of which from $b\bar b\gamma\gamma$) for $L = 3\,\text{ab}^{-1}$, with a corresponding statistical significance $S/\sqrt{B} = 2.3$. 
This is compatible with the
numbers reported in Table~\ref{tab:cutflow:14TeV}, which give $S/\sqrt{B} = 2.1$. Our analysis thus agrees with the  signal and background rates estimated by Yao.
Based on these numbers, this latter calculates a precision on the trilinear coupling equal to $50\%$ and $8\%$ respectively 
for the \HLLHC and the \FCC.
This is much more optimistic than our corresponding estimates $200\%$ and $30\%$ reported in Section~\ref{sec:results}.
The discrepancy follows because the uncertainty on the trilinear is derived in Ref.~\cite{Yao:2013ika} from the dependence of the signal cross section
on this parameter \textit{before} cuts. In particular, a linear approximation is used around the SM point with a slope $d(\sigma/\sigma_{SM})/dc_3\simeq -0.8$
taken from Ref.~\cite{Baglio:2012np}. However, the slope decreases substantially (in absolute value) after imposing all cuts: at the end of our analysis we find 
$d(\sigma/\sigma_{SM})/dc_3\simeq -0.58$ and $-0.50$  respectively for 
the \HLLHC and the \FCC.~\footnote{In fact, the dependence on the trilinear
coupling also depends on the $m_{hh}$ category considered, the categories with lowest $m_{hh}$ having the largest absolute slope, as expected.}
By using these numbers, the uncertainty on the trilinear coupling that follows from the rates of Ref.~\cite{Yao:2013ika} is equal to $90\%$ and $16\%$
respectively for 
the \HLLHC and the \FCC.~\footnote{During the completion of this paper, an updated analysis by Yao was presented
in a conference talk~\cite{talkbyYao}, where a similar precision is obtained at $100\,$TeV after taking into account the correct slope.}
The difference between these improved values and our results is finally due to the 
fact that the linear approximation  is not accurate for $\sqrt{s} = 14\,$TeV (a second solution exists), 
and to the marginalization on $\bar c_u$, $\bar c_H$, $\bar c_d$, $\bar c_g$  (not included in Ref.~\cite{Yao:2013ika}) which has a large effect at $\sqrt{s} = 100\,$TeV
as explained in Section~\ref{sec:results}.

Besides the above theoretical studies, the ATLAS collaboration has recently completed a Montecarlo analysis of the $b\bar b\gamma\gamma$ decay mode
for the  \HLLHC~\cite{ATL-PHYS-PUB-2014-019}. The simulation of the $b\bar b\gamma\gamma$ background is done by 
matching up to 1 extra jet at the matrix-element level, thus properly including the  bulk of the NLO corrections. 
The reported signal and background rates are 
consistent with those obtained in our analysis, although backgrounds with fake photons ($b\bar b\gamma j$) and fake $b$-jets from charm quarks 
($c\bar c \gamma\gamma$) are found to be sizable, differently from what assumed in this work. It is interesting to see if further experimental analyses will 
confirm this preliminary study and strategies will be found to reduce the size of these reducible backgrounds.

%% file: Conclusions.tex
\label{sec:conclusion}

In this work we presented an analysis of  double Higgs production via gluon fusion.
The novelty of our approach with respect to most of the previous studies is the use of an
effective field theory perspective that allows one to encode all possible effects from heavy New Physics
into a small set of deformations of the SM Lagrangian.
Although the leading contributions to observables are in general expected to arise from dimension-6 operators, when these latter are suppressed due
to selection rules higher-dimensional operators can become important. We pointed out that this occurs in double Higgs production, where the
dimension-6 operator~$O_g$ violates a Higgs shift symmetry and is thus suppressed if the Higgs is a  pseudo Nambu-Goldstone boson.
In this case dimension-8 operators become relevant in the high invariant-mass tail of the kinematic distributions. 
A careful assessment of the range of validity of the effective field theory description is thus required, like the one provided in Section~\ref{sec:parametrization}.
There we analyzed in detail the implications of the standard SILH power counting, which predicts $O(v^2/f^2)$ deviations
in a large set of observables, including the Higgs trilinear coupling. We  discussed alternative scenarios, including a Higgs-portal model, that
imply a modified power counting 
and lead to large deviations in the  trilinear coupling while predicting small modifications to the other
Higgs couplings. These will be presumably the first scenarios to be probed by early measurements of double Higgs production at the LHC,
where the precision on the trilinear coupling is expected to be limited.

We provided an explicit application of the effective field theory approach by assessing the experimental
sensitivity on the coefficients of the relevant local operators. For definiteness we focused on the
$gg \rightarrow hh \rightarrow \gamma \gamma b \bar b$ process, which has been recognized as one of the
cleanest channels to exploit double Higgs production.
Although this process was previously analyzed in several works, our study provides some important
improvements which led to significantly different results (see Section~\ref{sec:comparison} for a full comparison).
One of the most relevant observables that can be extracted  is the Higgs
trilinear coupling $c_3$. We found that the $\gamma\gamma b \bar b$ channel allows the determination of $c_3$
with a fair precision ($\sim 30\%$ with $L= 3\,\text{ab}^{-1}$) only at a future $100\ \mathrm{TeV}$ hadron collider. The prospects for the
LHC, instead, are much less optimistic and only an $O(1)$ determination seems possible even with $3\ \mathrm{ab}^{-1}$ of
integrated luminosity.
This result is significantly worse than what usually claimed in the
literature. The origin of the discrepancy is in large part due to a more careful determination of the background processes.
In particular we found that the irreducible $b\bar b\gamma\gamma$ background is 
enhanced by a sizable NLO k-factor ($k_{b\bar b\gamma\gamma} \sim 2$), not included in most of the previous works, and is thus
larger than previously thought.
We also showed that the determination of $c_3$ is affected by the choice of the statistical treatment. In particular
a full fit including possible corrections to all relevant dimension-6 operators leads to significant differences with respect
to the procedure, often used in the literature, where only the Higgs trilinear coupling is allowed to vary.
An important role is played by the uncertainty on the coupling of the Higgs boson to the top quark. The reason is that, in the context of the effective
Lagrangian for a Higgs doublet, the same operator $O_u$ which controls the modification to the top Yukawa coupling also generates a quartic
$t\bar t hh$ anomalous interaction which leads to a new contribution to the double Higgs production cross section.
This latter effect must be properly included when estimating the precision on the Higgs trilinear coupling. 
We find that an uncertainty of order $10\%$ on the measurement of $\bar c_u$ can nearly double the error on
$c_3$ at a future $100\,\mathrm{TeV}$ collider (see Fig.~\ref{fig:c6vssigmacu}). 
One can in fact turn the argument around and use double Higgs production to determine $\bar c_u$. Our results show that 
the accuracy one can obtain in this way is potentially competitive with the one from the $t\bar th$ process (see Fig.~\ref{fig:cucg}), especially
if several final states are combined.
On more general terms, other couplings can be extracted from the $gg\to hh$ process.
In particular, producing a pair of Higgses gives unique access to the $t\bar thh$ and $gghh$ quartic interactions,  parametrized respectively by $c_{2t}$ and $c_{2g}$.
These should be regarded as independent parameters from the wider perspective of a non-linear effective Lagrangian.
We found that these couplings can be determined much more accurately that $c_3$, as shown in Fig.~\ref{fig:summary-nonlinear}, in agreement with previous works.

Another important point of our study is the effectiveness of an exclusive analysis exploiting the $m_{hh}$ differential
distribution. This procedure is particularly useful to disentangle the contributions arising from different effective
operators since each of them leads to very specific deformations of the $m_{hh}$ distribution.
This analysis strategy can be relevant for the LHC mainly in its high-luminosity phase, and can have a dramatic impact
at a future $100\ \mathrm{TeV}$ collider (see Figs.~\ref{fig:inclusive} and~\ref{fig:breakdown}).
The other kinematic variable characterizing  double Higgs events, namely the angular separation between the two Higgses,
was instead found to have a marginal role in our analysis. The expected distribution, indeed, is almost flat in the SM and nearly unchanged
by New Physics effects. A possible exception to this result are the deformations due to dimension-$8$ operators.
These are however unlikely to be accessible at the LHC, especially in the rare $b\bar b \gamma\gamma$ decay mode,  
although they are expected to be relevant at future colliders.
We also explored the possibility of using jet substructure techniques to
improve our analysis strategy. We found that these methods can be relevant in the high invariant-mass tail
of the $m_{hh}$ distribution, namely for $m_{hh} \gtrsim 1\ \mathrm{TeV}$. The impact of such improved analysis
seems to be marginal at the LHC, whereas at future higher-energy colliders it can lead to a significant increase in the sensitivity on
$O_g$ (see Fig.~\ref{fig:boosted}) and, possibly, on dimension-$8$ operators.

There are a few directions in which our analysis could be improved, which we leave for future investigation.
In the present work, for example, we did not try to optimize the mass windows for the reconstruction of the $b\bar b$ and $\gamma\gamma$
pairs. The choice of these windows has a strong impact on the non-resonant backgrounds, as they scale roughly linearly with the window size,
and it is not unreasonable to expect that
some improvement could be achieved with a more sophisticated analysis. Another point that we did not fully explore is
the estimate of the backgrounds with fake photons. The analogy with single-Higgs production suggests that it should be possible
to reduce these backgrounds to a subdominant level, but a thorough quantitative analysis is required to clarify this issue.

As previously remarked, in this paper we were mostly interested in showing how double Higgs searches can be performed and interpreted in
the general framework of effective field theory. We thus adopted a very simple analysis strategy avoiding unnecessary
complications. It is clear, however, that the use of a more advanced procedure, as for instance a multivariate analysis, could
be useful to improve the final sensitivity. In  view of future high-energy colliders, it would also be interesting to
investigate how to extract information about dimension-$8$ operators. An possible strategy, in this case,
could be the use of angular distributions and a more systematic analysis of the high-energy tails of  kinematic distributions.

Finally, a major point that is still missing in the literature is a full analysis combining the results obtained in the
various double Higgs channels. Although $\gamma\gamma b\bar b$ is undoubtedly the easiest final state to look for,
other channels with larger rates can be competitive and might  lead to a higher sensitivity on the BSM coefficients.
The estimates vary significantly in the literature depending on the details of the simulation and of the statistical analysis.
The  highest  sensitivity is expected to come from the $b\bar b\tau^+\tau^-$ final state, which has been claimed to lead to a 
determination of $c_3$ at the $40-60\%$ level at the high-luminosity LHC with $\sqrt{s} = 14\,$TeV and 
$L = 3\,\text{ab}^{-1}$~\cite{Dolan:2012rv,Barr:2013tda,Goertz:2013kp,Goertz:2014qta}.
A slightly worse sensitivity on $c_3$ has been estimated for the $b\bar bWW^*$ channel~\cite{Papaefstathiou:2012qe,Goertz:2013kp}, while
extracting the trilinear coupling from the $b\bar bb\bar b$ channel will be much harder. For the latter case it has
been estimated that the signal can be distinguished from a background-only hypothesis with 95\% probability if $c_3 < 1.2$~\cite{deLima:2014dta},
as smaller values of $c_3$ lead to larger cross sections. 
For a fair comparison with our results, it should be noted that none of the above estimates, with the exception of those of Ref.~\cite{Goertz:2014qta}, includes the uncertainty 
implied by the presence of other effective operators besides the one controlling the trilinear coupling. Our analysis suggests that this will have an important
impact in all cases in which a high precision can be reached on the trilinear coupling.

All analyses of double Higgs production, including the one illustrated in this work, agree in concluding that the observation of double Higgs production will be
a challenge for the LHC, even at its high-luminosity upgrade. Extracting the Higgs trilinear coupling with sufficient precision will thus most certainly require 
combining as many different channels as possible. This will also open up the possibility of testing precisely New Physics by following the strategy outlined in this paper.

\vspace{0.5cm}
\noindent Note added: \ \ During the completion of this work Ref.~\cite{Barr:2014sga} appeared, where an analysis of $gg\to hh \to b\bar b\gamma\gamma$ is performed at 
a future $100\,$TeV collider in the context of the Standard Model. 
The authors estimate a $\sim 40\%$ sensitivity on the Higgs trilinear coupling with $3\,\text{ab}^{-1}$ of integrated luminosity,
which roughly agrees with our result. Their simulation differs in several aspects compared to ours: on the one hand backgrounds with fake photons 
are included and their importance is pointed out; on the other hand backgrounds are computed at LO (including the $b\bar b\gamma\gamma$ continuum, whose
simulation does not include extra jets at the ME level)
and jet substructure techniques are not used.

%% file: Appendix.tex

\section{Angular decomposition}
\label{app:angular}

By angular momentum conservation,
the scattering amplitude can be decomposed into two contributions, $M_0$ and $M_2$, mediating respectively $J_z =0$ and $J_z=\pm 2$
transitions~\cite{Glover:1987nx}:
\begin{equation}
A(g(p_a) g(p_b) \to h(p_c) h(p_d)) = \left(P_0^{\mu\nu} M_0+P_2^{\mu\nu} M_2\right) \epsilon^\mu(p_a) \epsilon^\nu(p_b)\, .
\end{equation}
The
$P^{\mu\nu}_{0,2}$ are (orthogonal) projectors
\begin{equation}
\begin{split}
P_0^{\mu\nu} & = \eta^{\mu\nu}-\frac{p_a^\nu p_b^\mu}{(p_a p_b)} \\[0.1cm]
P_2^{\mu\nu} &= \eta^{\mu\nu}+\frac{p_c^2 p_a^{\nu}p_b^\mu}{p_T^2(p_a p_b)}-\frac{2(p_bp_c)p_a^\nu p_c^\mu}{p_T^2 (p_a p_b)}
-\frac{2(p_a p_c)p_b^\mu p_c^\nu}{p_T^2 (p_a p_b)}+\frac{2p_c^\mu p_c^\nu}{p_T^2}\, ,
\end{split}
\end{equation}
and $p_T^2 = 2 (p_a p_c)(p_b p_c)/(p_a p_b) - p_c^2$ is the transverse momentum of the Higgses.
The expression of $M_0$ and $M_2$ is given by
\begin{equation}
\begin{split}
M_0 & = \frac{G_F \alpha_s \hat s}{2\sqrt{2} \pi}  
\left[ c_t^2 F_{\Box} +2 c_{2t} F_{\Delta}+8 c_{2g}+\left( c_t F_{\Delta} +8 c_g\right) c_3\frac{3 m_h^2}{\hat s-m_h^2} 
 -\frac{8 c_{gD0} }{m_W^2}\, \frac{\left(\hat s-2 m_h^2\right)}{2} \right]\\[0.2cm]
M_{2}& =\frac{G_F \alpha_s \hat s}{2\sqrt{2} \pi} 
 \left[   c_t^2 G_{\Box}  +\frac{8 c_{gD2} }{m_W^2}\,\frac{\left(\hat u \hat t-m_h^4\right)}{\hat s} \right]\, ,
\end{split}
\end{equation}
where the form factors $F_{\Box}$, $F_{\Delta}$ and $G_{\Box}$ are given in Ref.~\cite{Plehn:1996wb}, and
$c_{gD0}$, $c_{gD2}$ are the coefficients of the higher-derivative operators 
\begin{equation}
\frac{g_s^2}{8\pi^2 v^2 m_W^2} \Big[ 
  c_{gD0} \, \big(\partial_\rho h  \partial^\rho h\big) G^a_{\mu\nu} G^{a\,\mu\nu} 
 + c_{gD2} \, \big(\eta^{\mu\nu} \partial_\rho h^\dagger \partial^\rho h - 4\, \partial^\mu h \partial^\nu h\big)
G^a_{\mu\alpha} G_\nu^{a\,\alpha} \Big]\,,
\end{equation}
which give a next-to-leading correction to the non-linear Lagrangian (\ref{eq:nonlinearL}). 
They are related to the coefficients of the dimension-8 operators in (\ref{eq:dim8L}) 
by the following simple relations:  $c_{gD0} =   (4\pi/\alpha_2) \bar c_{gD0}$, $c_{gD2} =   (4\pi/\alpha_2) \bar c_{gD2}$.

\section{Simulation}
\label{app:simulation}

\subsection{Cross sections of signal and backgrounds}
\label{app:xsec:sig:bkg}
The signal and background cross sections at the generation level  are summarized in Table~\ref{tab:xsection:signal:bkg}. We simulate one extra jet from 
the matrix-element in addition to $b\bar{b}\gamma\gamma$ system in all backgrounds except $t\bar{t}h$ and $\gamma\gamma jj$.
No generation-level cuts have been applied to the resonant backgrounds $t\bar{t}h$ and $Zh$.
The $b\bar{b}h$ samples are restricted to satisfy $p_T(b)>20\,$GeV at $14\,$TeV (increased to $30\,$GeV at $100\,$TeV) within $|\eta(b)|<3$ 
as well as $\Delta R(b,\bar{b})>0.3$. 
All extra partons are required to be within $|\eta|<5$ and their $p_T$ threshold are set to xqcut in {\sc MadGraph5}. The size of $\gamma\gamma jj$ is reduced by demanding 
$p_T(j)>15\,$GeV at $14\,$TeV ($25\,$GeV at $100\,$TeV) within $|\eta(j)|<5$ and $p_T(\gamma)>20\,$GeV ($30\,$GeV at $100\,$TeV) within $|\eta(\gamma)|<2.5$. 
Jets and photons are required to be separated in $\Delta R$ by 0.3. Appropriate branching fractions of $Z\rightarrow b\bar{b}$ (15.12\%) and $h\rightarrow \gamma\gamma$ 
(0.228\%), $h\rightarrow b\bar{b}$ (57.7\%) are folded in the cross sections in Table~\ref{tab:xsection:signal:bkg}. 
\begin{table}[tbp]
\centering
\begin{tabular}{r|c|ccccc}
  & $hh$ ($\gamma\gamma b\bar{b}$)  & $\gamma\gamma b\bar{b}$  & $\gamma\gamma jj$  & $t\bar{t}h$ ($t\bar{t}\gamma\gamma$)  & $b\bar{b}h$ ($b\bar{b}\gamma\gamma$) 
 &  $zh$ ($b\bar{b}\gamma\gamma$)  \\
\hline \hline
14 TeV & 9.71 $\times 10^{-2}$ & 192 & 40.3 $\times 10^2$ & 1.42 & 0.14 &0.23 \\ \hline
100 TeV & 4.04 fb & 579 & 10.5 $\times 10^3$ & 86.4  & 1.47  & 2.67 
\end{tabular}
\caption{Cross sections (in units of fb) for signal and backgrounds at $14\,$TeV and $100\,$TeV after generation-level cuts (see text for details).
}
\label{tab:xsection:signal:bkg}
\end{table}

\subsection{NLO estimate of $\gamma\gamma b\bar b$ background and k-factor}
\label{app:NLOestimate}

As mentioned in Section~\ref{sec:analysis}, we found that the cross section of the  $\gamma\gamma b\bar{b}$ background is significantly enhanced (roughly by a factor 2) after
including diagrams with one extra parton at the matrix-element level. Considering that this is the most important background included in our analysis
it is worth exploring more in detail the origin of such enhancement. No analytic NLO calculation is available in the literature, so any study should currently rely on the use
of numerical Montecarlo computations. 
In addition to the matched sample used for our analysis, we generated two additional ones. The first is obtained through a full NLO simulation performed with
\textsc{MadGraph}5\_aMC$@$NLO v2.1.1,  including both real emissions and virtual corrections; it has been showered through {\sc PYTHIA8}.
The second is a LO simulation without extra partons at the ME level, showered through {\sc PYTHIA6} (with the choice of $p_T$-ordered shower scheme).
The three sets of  samples are compared in Table~\ref{tab:cutflow14TeV:NLO}. 
While the same version of \textsc{MadGraph}5\_aMC$@$NLO has been used to generate all three samples, different sets of generation cuts had to be imposed
(partly as a consequence of different matching procedures), making the comparison subtle.
In fact, cuts have been imposed on ``jets'' (defined to be  the clustered partons before parton shower) in the case of the NLO sample, while they apply to partons for the LO ones.
For the NLO sample we imposed
(see Eq.~(3.5) of Ref.~\cite{Frixione:1998jh} for a definition of $\epsilon_\gamma$ and $n$)~\footnote{We thank Marco Zaro for helping us to impose appropriate cuts on $b$-jets 
in our NLO simulation.},
\begin{equation} \label{eq:cutsNLO:14TeV}
\begin{split}
 &R_{\rm anti-kt} = 0.3~,\quad p_T(j) > 15\ {\rm GeV}~, \quad |\eta(j)| < 5~,\\
 &p_T(b) > 20\ {\rm GeV}~, \quad |\eta(b)| < 3~,\\
 &p_T(\gamma) > 20\ {\rm GeV}~, \quad |\eta(\gamma)| < 2.5~,\quad R_{\rm iso}=0.3~,\quad \epsilon_\gamma =1~,\quad n=1\, ,
\end{split}
\end{equation}
whereas the following cuts (with $\text{QCUT}=7\,$GeV)
\begin{equation}\label{eq:gencutsLOaabb:14TeV}
\begin{split}
 &p_T(j) > 3\ {\rm GeV}~, \quad |\eta(j)| < 5~,\\
 &p_T(b) > 20\ {\rm GeV}~, \quad |\eta(b)| < 3~, \\ 
 &\Delta R(b,\, \gamma) > 0.3~, \quad \Delta R(b,\, b),\ \Delta R(\gamma,\, \gamma) > 0.3~,\\
 &p_T(\gamma) > 20\ {\rm GeV}~, \quad |\eta(\gamma)| < 2.5~,\quad R_{\rm iso}=0.3~,\quad \epsilon_\gamma =1~,\quad n=1~,\\
 &30 < m_{b\bar{b}} < 300\ {\rm GeV}~,\quad 30 < m_{\gamma\gamma} < 300\ {\rm GeV}\, ,
\end{split}
\end{equation}
have been imposed on the LO samples.
The prescription of Ref.~\cite{Frixione:1998jh} for the photon isolation is important especially for the NLO computation to guarantee the correct cancellation of IR divergences. We apply the same photon isolation to the tree-level matched samples for  consistency~\footnote{However, we find no practical discrepancy between the step-function type cone isolation, i.e. $\Delta R(\gamma,X)>0.3$ where $X$ denotes any object near the photon, and the prescription in~\cite{Frixione:1998jh} in the tree-level matched samples.} (choosing this option automatically inactivates the cut on $\Delta R(j,\gamma)$ in \textsc{MadGraph}5\_aMC$@$NLO). The low value of $p_T(j)$ in Eq.~(\ref{eq:gencutsLOaabb:14TeV}) is due to the low value of the 
matching scale.
The lower bound of the $m_{b\bar{b}}$ mass window  in Eq.~(\ref{eq:gencutsLOaabb:14TeV}) has been chosen so as to obtain an optimal matching
scale and, at the same time, sufficiently small to avoid possible distortions in the signal region.
\begin{table}[tbp]
\centering
\begin{tabular}{|l|c|c|c|}
\hline
 $\sqrt{s}=14$ TeV   & \multicolumn{2}{c|}{$\gamma\gamma b \bar b$ at LO, matched up to $n_j$ jet} & $\gamma\gamma b \bar b$ at NLO \\ \cline{2-3}
  & $n_j=0$ & $n_j=1$ &  \\ \hline 
$\sigma$ [fb] at generation          & 111 fb &  192 fb  & 287 (223) fb \\ \hline
Cuts in Eq.~(\ref{eq:acceptanceI})     & 3619  & 6919   & 6581  \\
Cuts in Eq.~(\ref{eq:acceptanceII})    & 557   & 1274   & 1164  \\
Cuts on $m^{\rm reco}_{b\bar{b}}$        & 148   &  334   & 301  \\
Cuts on $m^{\rm reco}_{\gamma\gamma}$     & 13.7    &  24.2  & 24.5 \\ \hline
\end{tabular}
\caption{Cut flow of the three $b\bar b\gamma\gamma$ samples described in the text at $14\,$TeV. The number in  parentheses corresponds to the NLO cross section 
when the parton-level events are restricted to the same mass window as in the last line of Eq.~(\ref{eq:gencutsLOaabb:14TeV}).}
\label{tab:cutflow14TeV:NLO}
\end{table}

\begin{figure}[h]
\begin{center}
\includegraphics[width=0.32\linewidth]{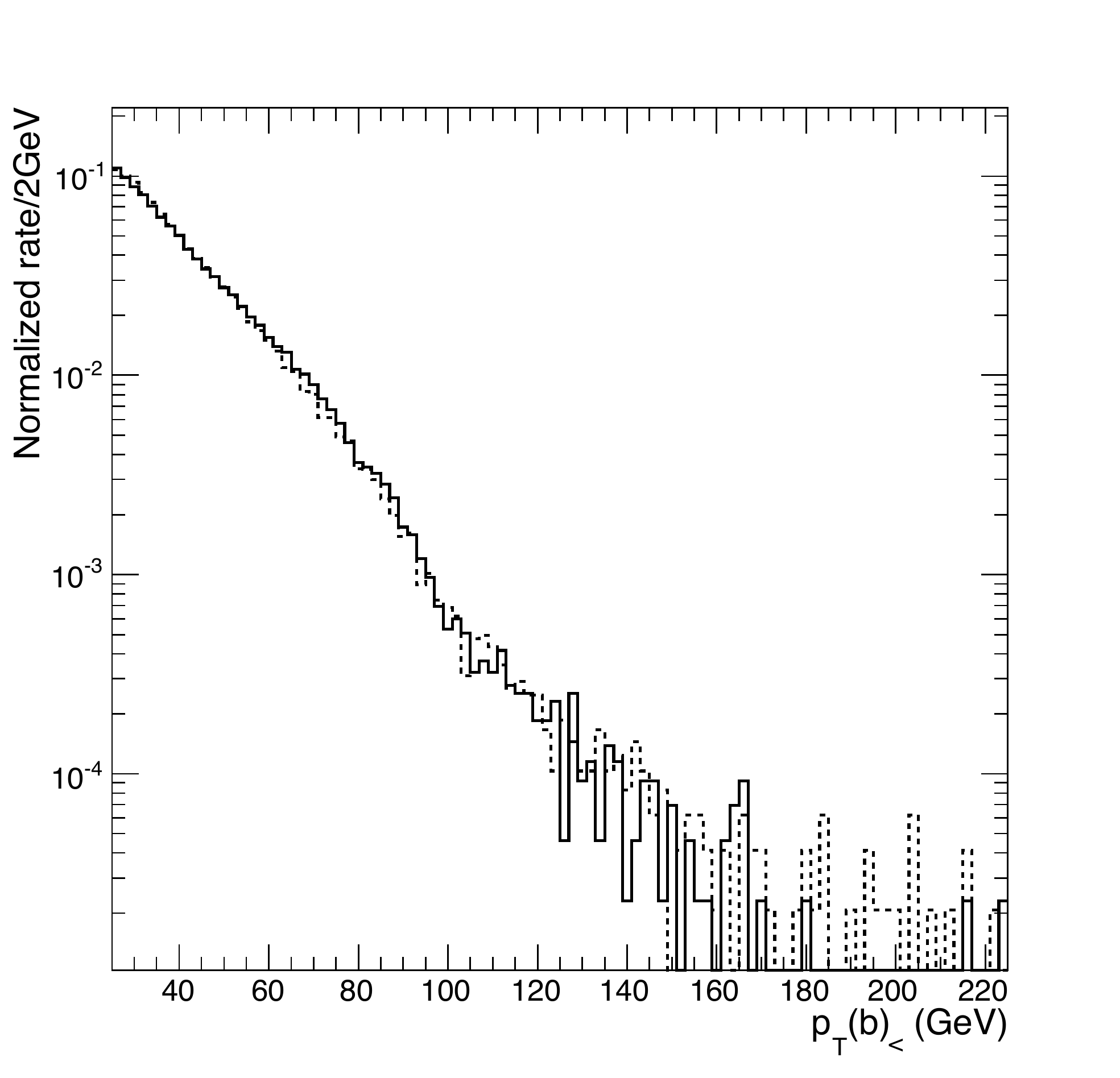}\quad
\includegraphics[width=0.32\linewidth]{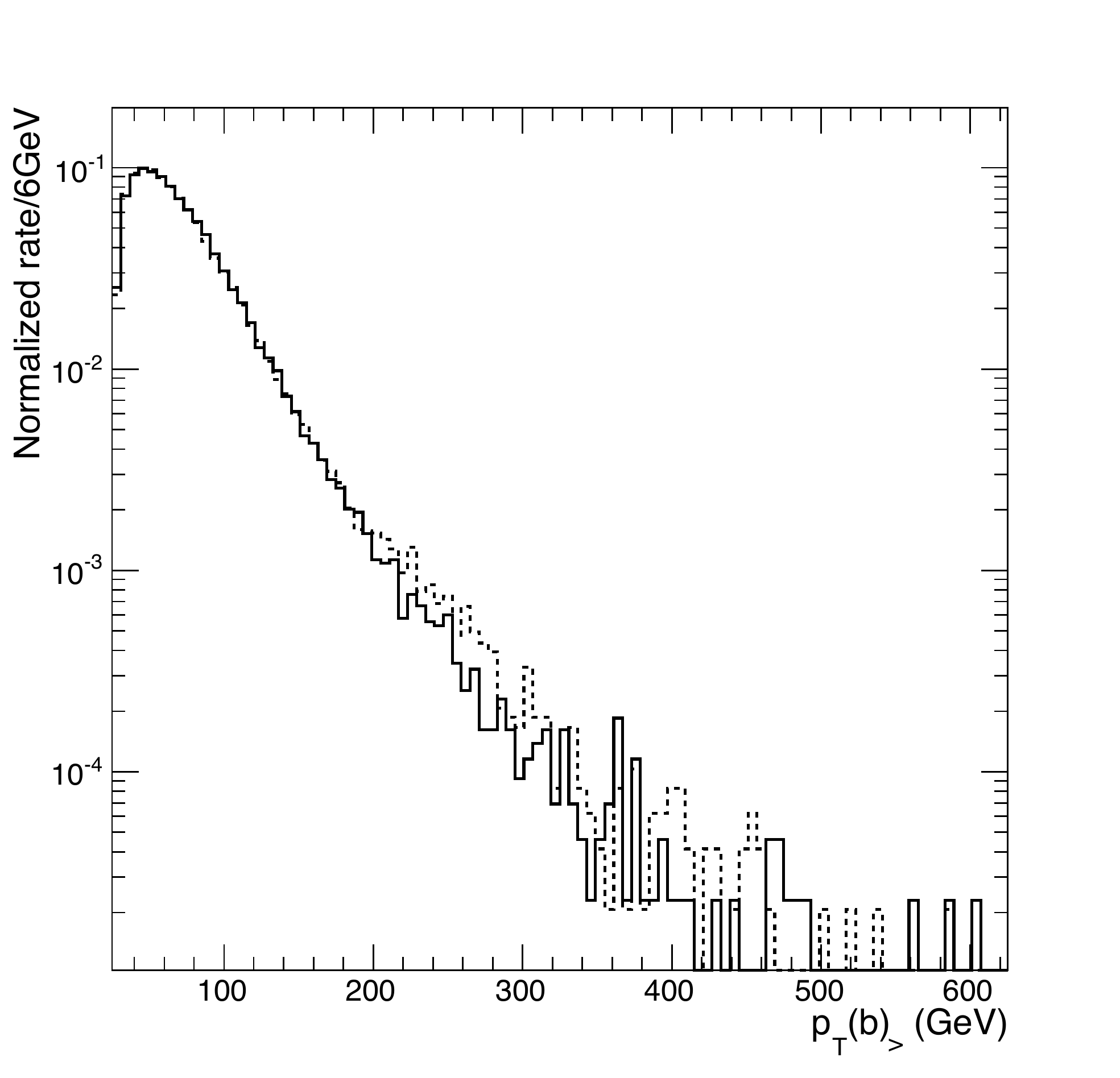}\\
\includegraphics[width=0.32\linewidth]{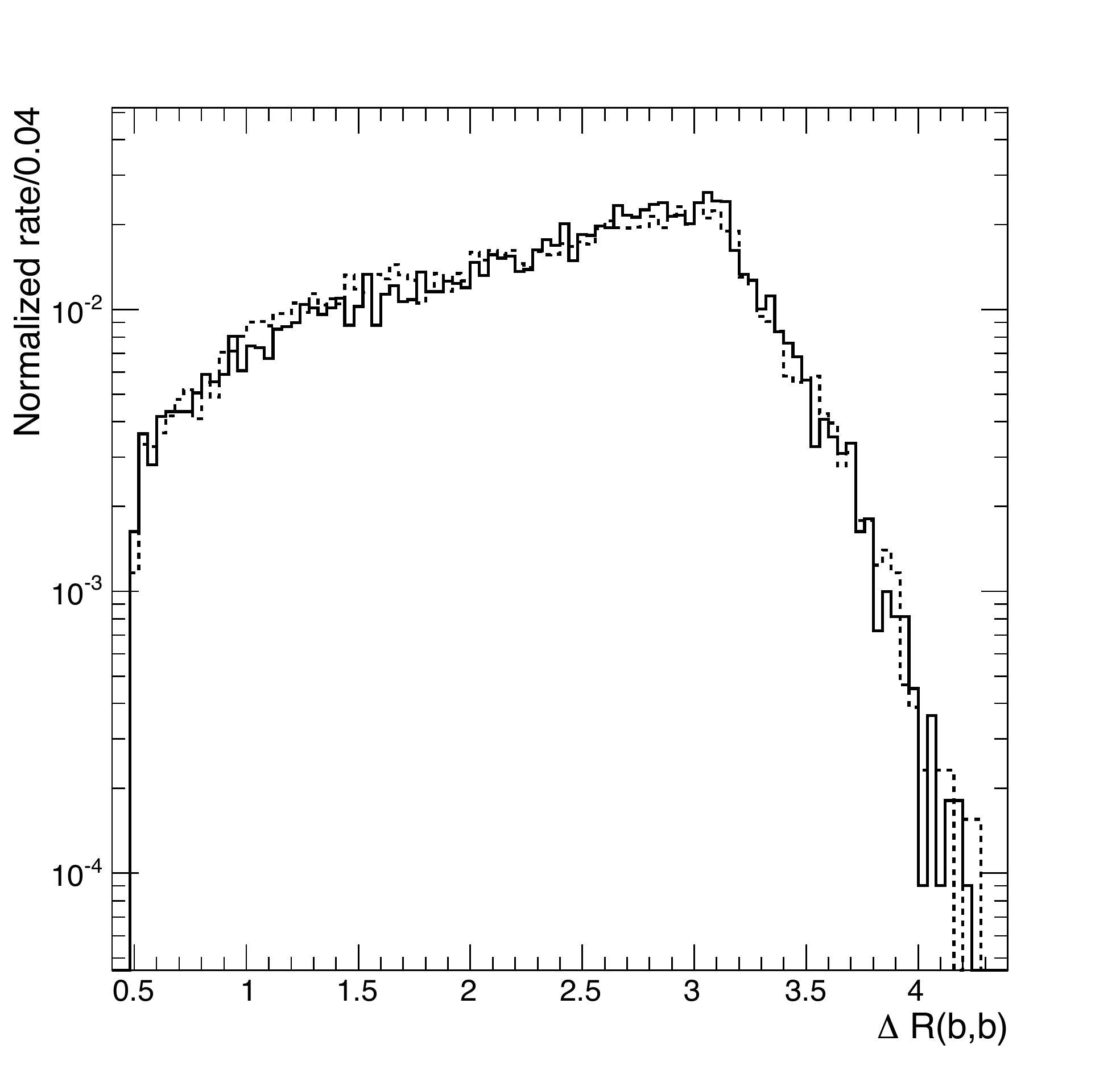}
\includegraphics[width=0.32\linewidth]{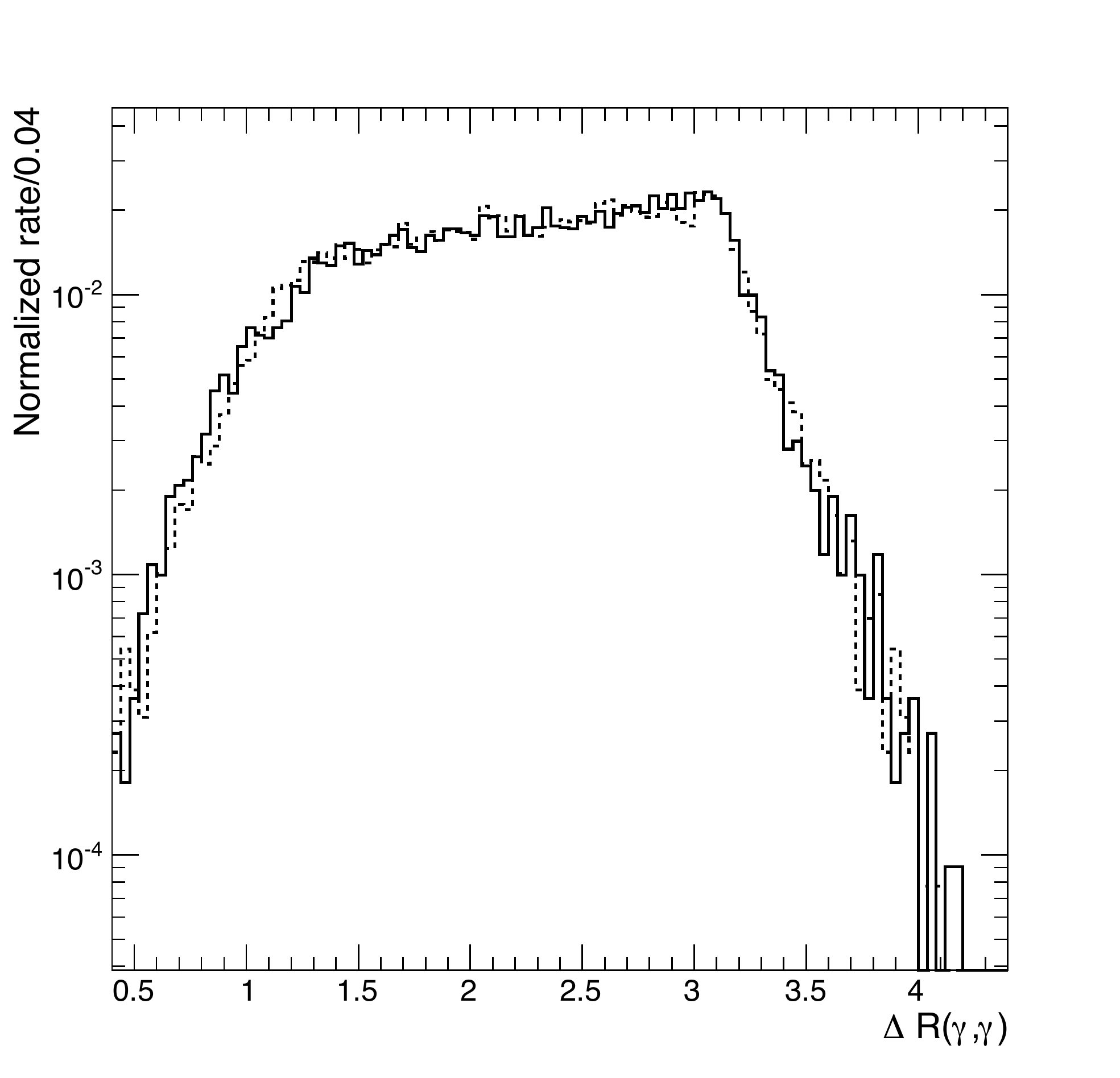}
\includegraphics[width=0.32\linewidth]{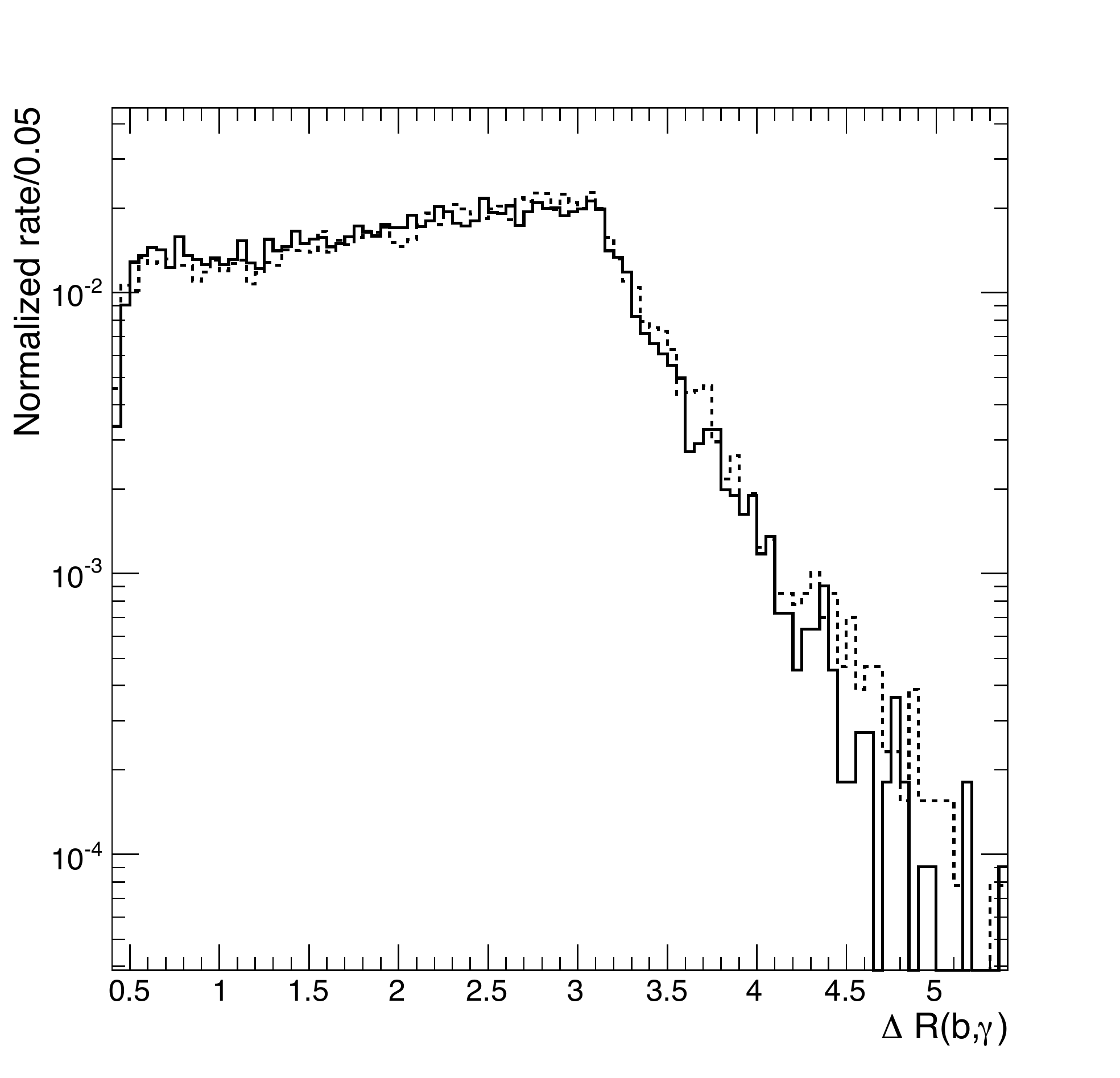}\\
\includegraphics[width=0.32\linewidth]{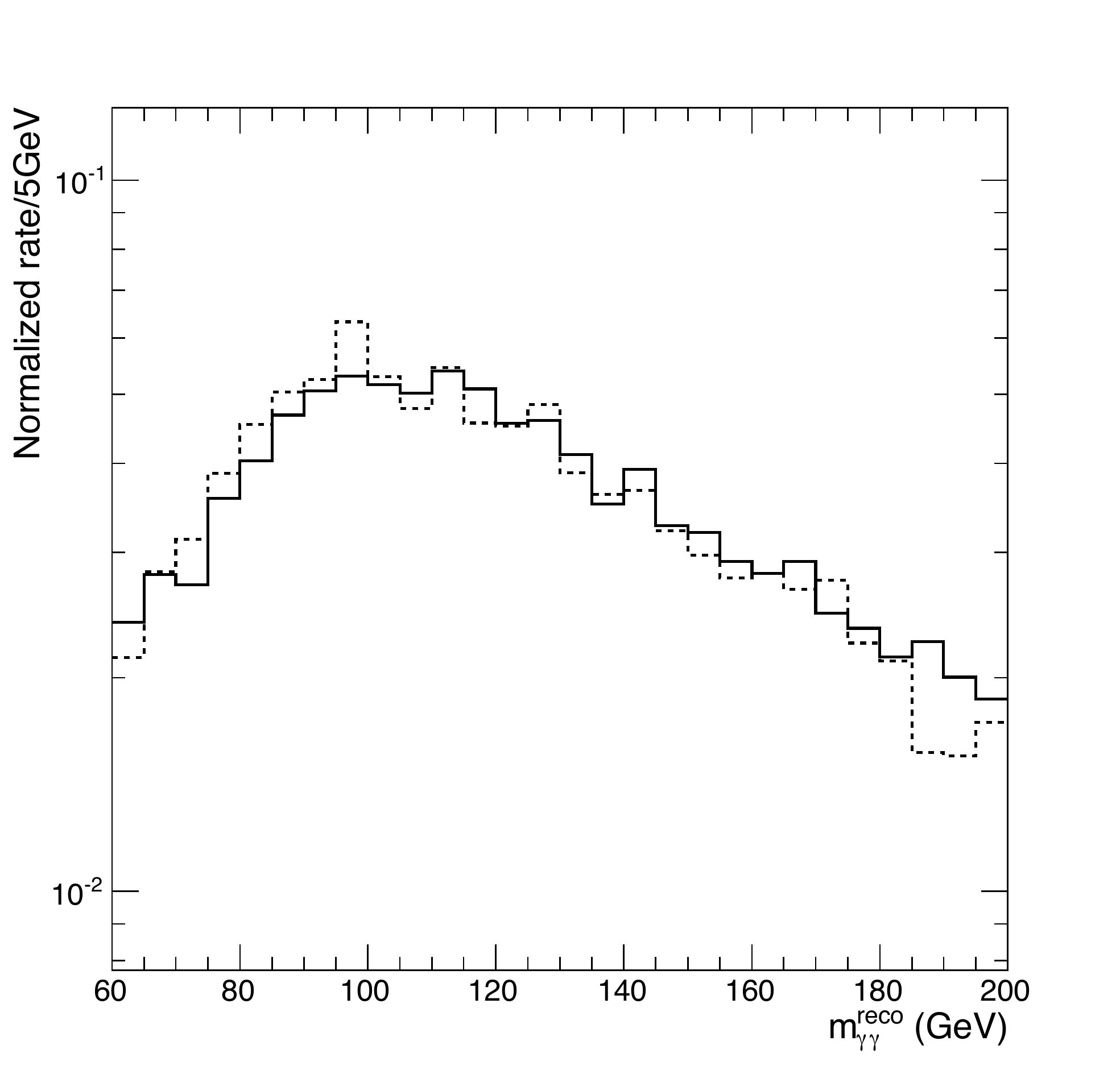}
\includegraphics[width=0.32\linewidth]{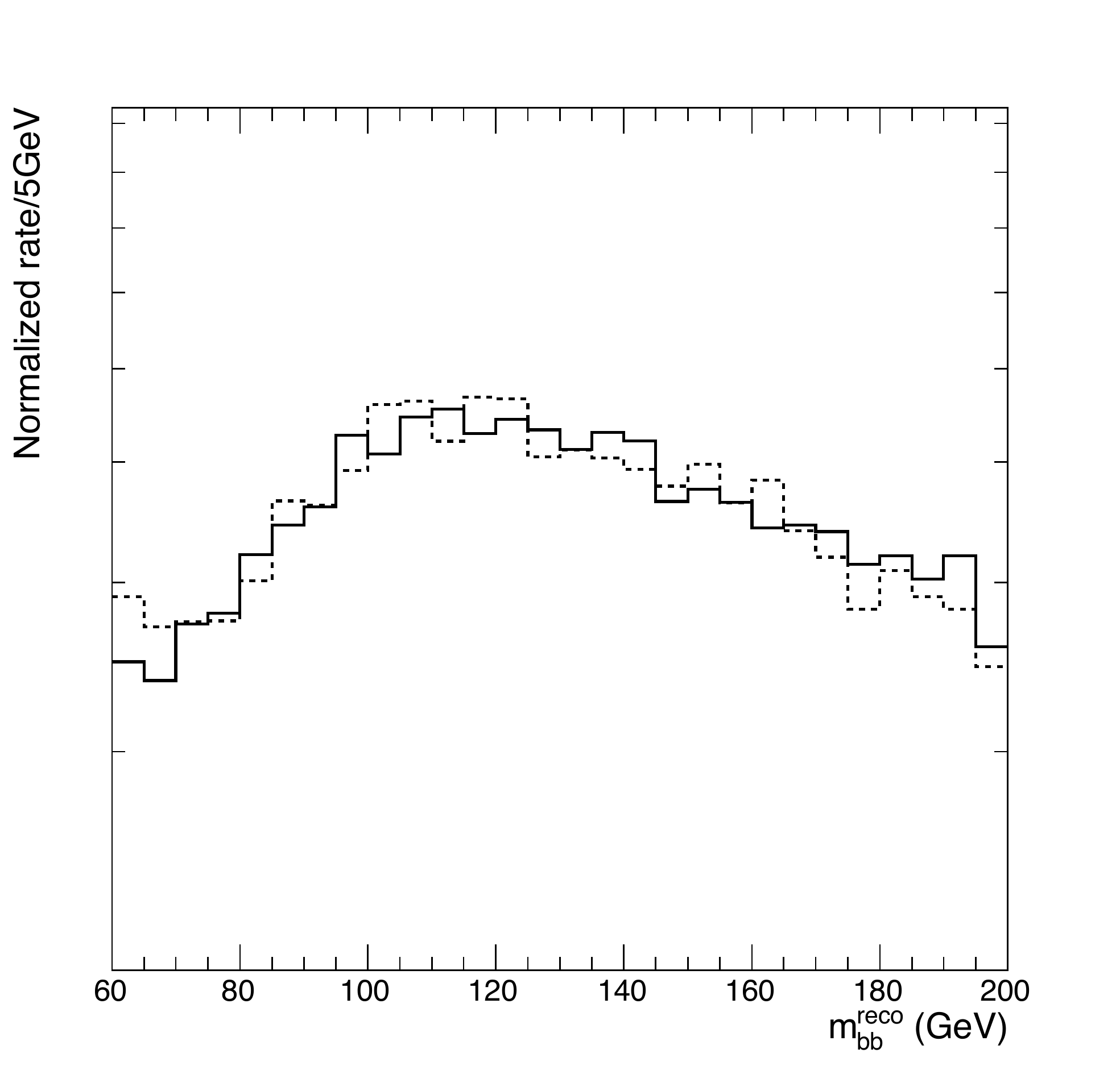}
\includegraphics[width=0.32\linewidth]{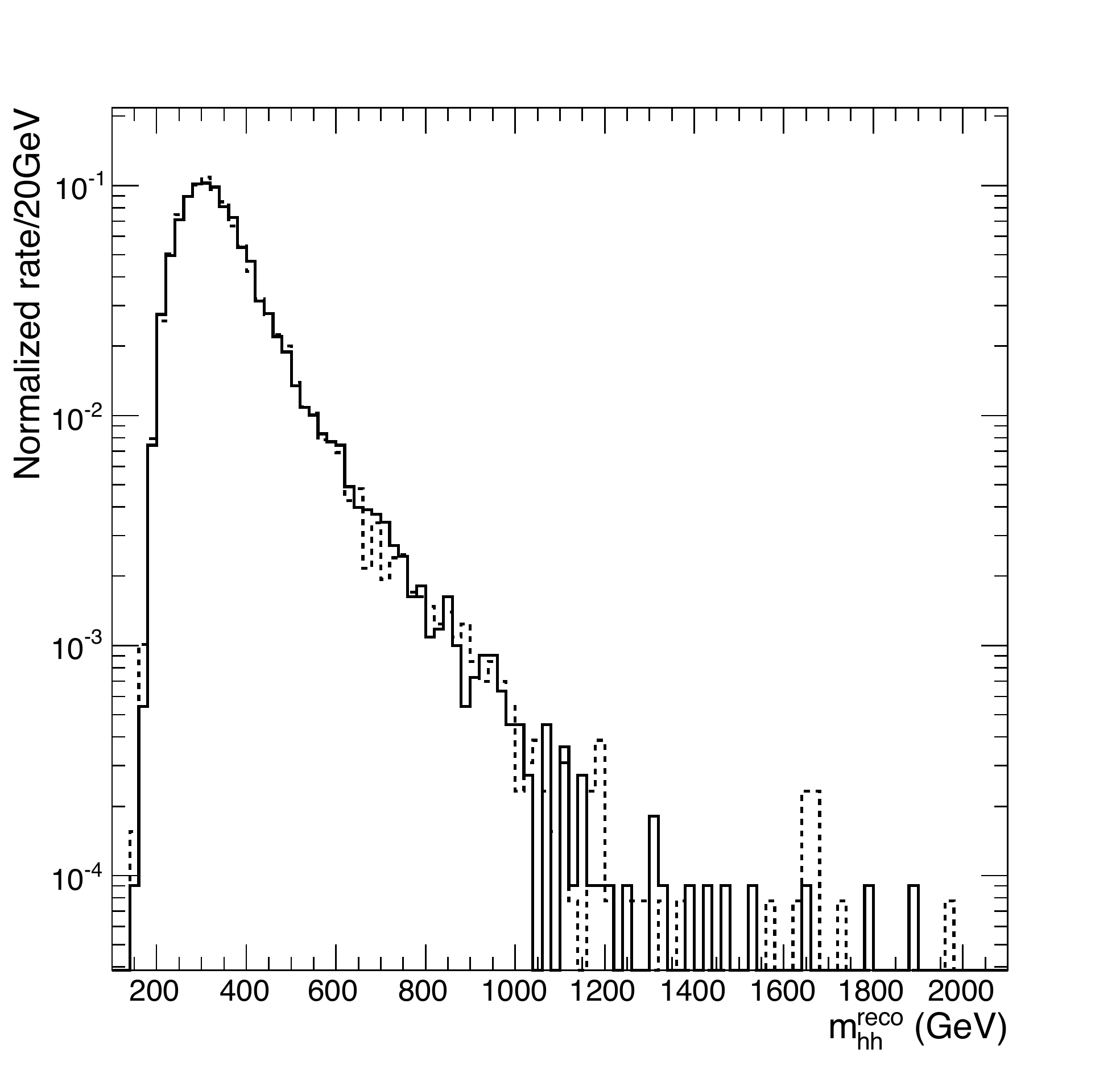}
\caption{Differential distributions for the $b\bar b\gamma\gamma$ background obtained with the tree-level matched sample (solid) and the full NLO sample (dotted) for several 
variables at the $14\,$TeV LHC. Due to the mass window cuts at the generation level  (see Eq.~(\ref{eq:gencutsLOaabb:14TeV})), the events are restricted to the mass window 
$60 <m^{\rm reco}_{b\bar{b}}<200\,$GeV and $60 <m^{\rm reco}_{\gamma\gamma}<200\,$GeV.}
\label{fig:LOvsNLO:14TeV}
\end{center}
\end{figure}

The value of the cross section for each of the three samples is reported in Table~\ref{tab:cutflow14TeV:NLO}.
The entries in the first row  should not be literally compared, since they refer to
samples generated with different sets of cuts and procedures. These generation-level differences, however, become irrelevant after imposing the first round of cuts 
leading to the number of events in the second row of  Table~\ref{tab:cutflow14TeV:NLO}.
The k-factor, i.e. the ratio of the fourth to second column, is approximately 2 in each step of cut-flow. 
A similar k-factor is also found by taking the ratio of the third to  second column, i.e. of LO samples with and without one extra parton.
This shows that, at least at the level of cross section, the matched sample gives a very good approximation of the NLO one. This indicates that virtual corrections 
give a small contribution and the bulk of the NLO correction comes from real emissions.
Another important aspect are the differential distributions. Although total rates agree well, the shape of distributions could differ between the NLO sample and the LO matched one,
leading to different cut efficiencies. 
Figure~\ref{fig:LOvsNLO:14TeV}  compares the distributions of several  kinematic variables relevant for  our analysis.
While the higher value bins are subject to  statistical fluctuations, overall the differential shapes of the two samples agree well.
This justifies our use of the LO matched $\gamma\gamma b \bar b$ sample --whose generation requires much less CPU  time-- throughout our analysis.
We expect that the same results will also apply in the case of a $100\,$TeV collider.